# Hate in the Time of Algorithms: Evidence on Online Behavior from a Large-Scale Experiment

Aarushi Kalra*

March 5, 2025

(Please click here for the most recent version)

**Abstract**

As social media usage reaches record highs, there are growing concerns that personalization algorithms can radicalize users by reinforcing their existing beliefs. However, quantitative evidence on how algorithms and user preferences jointly shape harmful online engagement is limited. I conduct an individually randomized experiment with 8 million users of a prominent TikTok-like platform in India, replacing the feed-ranking algorithm with random content delivery. I focus on hateful content targeting minority groups, given its prominence on Indian social media, and establish a trade-off: random post recommendation lowers exposure to harmful ("toxic") content by 27%, but at a substantial cost to the platform as overall usage falls by 35%. Strikingly, treated users share a larger proportion of the toxic posts they view, mitigating the decline in the number of toxic posts shared. Users with a higher interest in toxic content at baseline drive this result as they reduce overall platform usage, and seek out posts the algorithm does not show them. I collect survey evidence to trace users' behavior beyond the platform and show that the most affected users switch to other applications. I rationalize these findings with a dynamic model of a revenue-driven algorithm that faces users choosing which posts to consume according to their fixed preferences. Counterfactual simulations evaluate alternative interventions that cannot be implemented in the field, using elasticities estimated from the model. My results underscore the limits of piecemeal algorithmic regulation intended to moderate harmful content online, especially when user behavior is immalleable.

*Department of Economics, Brown University, Providence, RI, 02906 (email: aarushi_kalra@brown.edu).
I am grateful to Andrew Foster, Brian Knight, Daniel Björkegren, Pedro Dal Bo, Peter Hull, and Stelios Michalopoulos, for their continued guidance and support. This project has greatly benefitted from helpful discussions with Jesse Shapiro, Bryce Steinberg, Jesse Bruhn, Neil Thakral, Ro'ee Levy, Matthew Pecenco, Lorenzo Lagos, Elisa Macchi, Martin Mattsson and seminar participants at Brown University. I thank Ahad Bashir and Farrukh Zaidi for excellent research assistance. The experiment was preregistered on the AEA RCT registry, ID AEARCTR-0010933. Protocols for survey data collection were approved by the Institutional Review Board at Brown University. This project was generously supported by the NSF Dissertation Research Improvement Grant, and the Weiss Family Fund for Research in Development Economics, hosted at the Population Studies Training Center at Brown University.

# 1 Introduction

Social media platforms engage 64% of the world's population, raising concerns about the potential exposure of 5 billion users to harmful unmoderated content (Acemoglu et al., 2023). Personalization algorithms designed to maximize engagement can create radicalizing echo chambers (Sunstein, 2018) that have been linked with physical violence against minorities (Müller and Schwarz, 2021). These concerns have engendered proposals for a regulatory response to diversify content feeds by disabling algorithmic personalization.[1] While such content-moderation policies can reduce exposure to harmful content, little is known about their effectiveness which can be limited if users' online behavior is unresponsive to viewing diverse information, and is largely determined by fixed preferences (Hosseinmardi et al., 2024). Further, industry-wide policy adoption is uncertain due to costs imposed on revenue-maximizing platforms (Acemoglu, 2021; Kasy, forthcoming).

In this paper, I conduct an individually randomized experiment with 8 million social media users to study the causal effect of "switching-off" personalization algorithms on online behavior, focusing on interactions with anti-minority "toxic" content.[2] I develop a partnership with TipTop (pseudonym), an Indian social media platform that has a user base of about 200 million users to identify the impact of randomized content delivery that replaces algorithm-generated personalized feeds.[3] The recommendation algorithms employed by this TikTok-like platform, akin to the algorithms typically used by YouTube, Instagram, Netflix, Amazon or Spotify, optimize for the time users spend online based on their previous engagement with similar content. This feed-ranking algorithm remains unchanged for 3% of TipTop's user base, 6 million users who were randomly assigned to the control group. I intervene on the personalization algorithm in a 1% random sample–2 million users–diversifying content feeds by randomizing exposure to different kinds of posts. I find a reduction in the absolute number of harmful posts viewed and shared, but an increase in the relative quantities of toxic shares showing that users' baseline preferences can blunt the gains from the intervention, that too at a high cost to the platform in terms of overall usage.

Studying the causal effects of feed-ranking algorithms is challenging for several reasons. First, user behavior and personalized recommendations are endogenous as algorithms "learn" from past interactions to maximize engagement. Prior on-platform experimental research partly addresses these concerns, finding that algorithmic interventions are ineffective in re-ducing polarization (Guess et al., 2023a,b; Nyhan et al., 2023). However, these experiments may not identify users' demand for polarizing content, as content exposure in these interventions—such as displaying posts in reverse chronological order—depends on social networks shaped by user preferences. Second, opt-in requirements to participate in the studies likely introduce selection bias in the research cited above. Third, supply-side incentives to "self-regulate" are opaque as algorithms typically optimize over unknown objectives that may not align with harm reduction (Jiménez Durán, 2022). Further, platforms are reluctant to experimentally remove posts based on subjective assessments of harm as they want to avoid being accused of political bias or suppressing free speech (Kominers and Shapiro, 2024). This makes it hard to collect evidence that can inform design of content moderation policies.

[1] See `https://shorturl.at/WKWPU` for minutes of the Subcommittee on Privacy, Technology, and the Law convened under the US Senate Committee on the Judiciary.

[2] Toxicity, as defined by Google's Perspective API, measures a post's potential harm as it scores comments on a scale of 0 to 1 based on their likelihood to make someone leave a discussion.

[3] The DUA between Brown University and TipTop does not impose restrictions on the publication of the results of the study due to the anonymization of firm identity and relevant measures to protect user privacy.

My study offers the largest-scale experimental evidence to date on the causal effects of personalization algorithms. By replacing algorithmically curated feeds with randomly picked content, this experiment–conducted in collaboration with a platform that has a user base equal to the combined population of Germany, France, and the UK–provides clean identification of demand factors driving harmful engagement. My findings have important policy implications for regulating India's ICT sector, which contributes 13% of the country's GDP and operates under unique institutional constraints, yet remains an understudied context (ITA, 2024; Blair et al., 2024). The intervention has limited general equilibrium effects as the incentives of content creators remain unchanged, and interactions do not depend on users' on-platform networks, thus minimizing spillovers.[4] The experimental sample is not selected as users consented through the platform's terms of service. High-frequency administrative data characterize engagement decisions and survey outcomes evaluate behavior beyond TipTop. I draw on insights gained from my partnership with the platform to model the optimization problems faced by both the platform and its users, and to evaluate counterfactual moderation policies that are infeasible to implement in the field (such as direct censorship of posts and users), using behavioral parameters estimated within the model.

The first set of results tests for heterogeneous treatment intensities conditional on users' baseline exposure to toxic content, resulting from the exposure to randomly drawn content. The treatment had the strongest impact on users at both extremes—those who initially viewed very little toxic content and those who initially viewed a high amount of toxic content. Random content delivery enhanced the diversity of user feeds as treated users encountered fewer posts of the types they were used to consuming. That is, some users had, through prior platform-engagement, tuned their algorithm to receive large amounts of toxic posts ex-ante, and so were expected to receive more toxic content at baseline and under the control condition. However, users with the highest exposure to toxic content at baseline experienced the largest reductions in toxic exposure upon being treated, 49% below the 38 toxic posts viewed over a period of one month by control users in this group. On the other hand, control users with the lowest baseline exposure to toxic content viewed 10 toxic posts on average, whereas treated users in this group viewed 3 more toxic posts during the same time period. On average, the intervention reduced the number of toxic posts viewed in one month by 27%.

[4]Supplier incentives are unchanged because the treatment is assigned at the user level only for 1% of the platform's user base. Control users are unaffected by online behavior of treated users as the control algorithm's recommendations on TipTop minimally depend on social networks. This is because content is typically shared on other platforms like WhatsApp and users typically do not follow each other on TipTop.

Similarly, the intervention altered the feed composition with respect to other types of posts, such as religious or romantic content. I focus on the intervention's impact on engagement with toxic content as I show that the percentage reduction in exposure to this content category is most salient across various genres and topics, among top users of the platform. This focus on toxicity also addresses a key policy concern especially in India where online misinformation (Garimella and Eckles, 2017) has been linked to deaths fueled by anti-Muslim hate crimes (Banaji et al., 2019). I identify posts that verbally attack or threaten India's Muslim minority using posts deemed political as per TipTop's internal classification, and toxic or abusive as per the Google-developed Perspective API. The latter algorithm scores a "phrase based on the perceived impact the text may have in conversation."

My second set of results deals with overall platform usage. Random content delivery decreased the total time an average user spent online over a period of one month by 35%, around 5 minutes per day in a sub-sample of active users consuming content in the Hindi language. The average treated user viewed 35 fewer posts compared to control users who viewed 247 posts of any variety in the same time period, a 14% decrease in activity by this measure. Therefore, the intervention is costly for the platform. For every rupee lost in ad-revenues due to decreased overall usage, an average user is exposed to 2.5 fewer toxic posts, compared to the 19 toxic posts viewed by control users in a month on average.

The largest intensive margin decreases in usage are due to users who viewed the most toxic content at baseline, with a 23% reduction in the total number of posts viewed during the intervention period. These users also reported spending more time on other platforms in my endline survey, consisting of over 8,000 randomly selected users from the experimental sample. This means that while platforms are unlikely to self-regulate in the short run, such content moderation policies may be effective in driving out problematic users from one platform, but potentially causing harm on others. These extensive margin results may also raise concerns about differential attrition as treated users with higher affinity towards toxic content were more likely to leave the platform. However, I show that the Lee bounds on treatment effects are tightly estimated (Lee, 2009).

My third set of results considers user engagement with toxic content. The average treated user reduced the number of toxic posts shared by 20% over one month. Users with the highest interest in toxic content drive this decrease in toxic engagement as they reduced the number of toxic posts shared by 34%, where toxic posts constituted 3% of the control users' total shares in a month. However, this reduction is smaller than the decrease in the number of toxic posts viewed meaning that users are only partially responsive to diversified feeds. On the other hand, users with low toxic exposure at baseline are even more unresponsive, and do not share more toxic posts even as they are served a larger number of toxic posts.

My fourth set of results focuses on the degree of malleability in user behavior. Although random content delivery reduced the number of toxic posts viewed and shared, the sharing response to decreased toxic exposure is blunted by the increased rate of sharing such posts. That is, the intervention led to an 18% increase in the probability of sharing a toxic post conditional on viewing one. The implication is that the reduction in the number of toxic posts shared would have been larger if users were more responsive to diversified feeds. This means that users seek out posts that align with their tastes even when they are not readily served to them. This result is supported by survey evidence: political attitudes are indistinguishable between treatment and control users in the post-intervention period.[5]

To rationalize these results, I propose and estimate a dynamic model of users' online behavior. In this model, time spent on the platform is endogenous, users choose the posts to share in order to balance static preferences for content consumption and their social-image concerns, and platforms maximize engagement by choosing the proportion of toxic posts to show a user during her session. I predict observed behavior in an equilibrium where users receive both consumption utility from viewing and public recognition utility from sharing posts that are learned to be socially acceptable, where this perception of norms is informed by the algorithm's recommendations. The steady state condition provides an estimation strategy for key behavioral parameters that cannot be directly estimated with the experimental data, such as the elasticity of sharing with respect to exposure.[6] These estimates are then used to simulate the effects of counterfactual policy proposals.

Bringing the model to the data, I find that a 1% decrease in exposure to toxic content decreases toxic sharing only by 0.16%. This elasticity measure highlights the limited role of the influence of content exposure, and supports my finding that users' online behavior is largely immalleable. Online behavior is mainly driven by pre-existing user tastes that equal content exposure at baseline in equilibrium. Simulated policy counterfactuals suggest that interventions targeting toxic posts, such as diversifying feeds of users with a higher proclivity towards such content, are ineffective in changing behavior if user responses are inelastic. However, a combination of diversified and customized feeds is shown to lower the dissemination of toxic content while minimizing the risk of losing users.

[5]The effects are precisely estimated as the survey design enables me to rule out even modest changes in attitudes, as small as a 5 pp change.

[6]This is because in replacing the personalization algorithm with random content delivery, treated users are exposed to a random draw of posts from an average user's feed over time. By the law of large numbers, the average proportion of user feeds consisting of toxic posts is constant for treated users. This means that the variation in "random" exposure is insufficient to identify this influence parameter, especially in the presence of unobserved heterogeneity that impacts sharing behavior.

These results have important policy implications for content moderation on digital platforms especially in scarcely regulated environments. First, platforms will not willingly moderate content by diversifying feeds because they lose user engagement. Second, targeting this intervention to users who are likely to share toxic content can help reduce the spread of harmful content as these users disengage with the platform. Third, substitutability among platforms necessitates cross-platform regulation. Fourth, limited malleability in behavior suggests that blanket regulations targeting algorithms may not be as effective as hoped for.

The rest of the paper is organized as follows. Section 2 presents background of the study as well as the administrative, experimental, and survey data sources. Section 3 outlines the experiment's design and presents descriptive statistics. Section 4 presents the main empirical results. Section 5 introduces the theoretical framework, and the model parameters are estimated in Section 6. Section 7 concludes.

## 1.1 Relation to the Literature

This paper contributes to three strands of the literature. The first strand examines the role of new communication technologies in aggravating political divisions. In collaboration with Meta, Guess et al. (2023a,b) find that on-platform interventions were not effective in reducing polarization, and users seek out like-minded content sources (Nyhan et al., 2023). However, my experiment offers clean identification of the demand for toxic content due to random content delivery. While the results on overall platform usage are consistent with closely related studies (Beknazar-Yuzbashev et al., 2022), I complement the literature by showing that users with a higher exposure to toxic content at baseline are more likely to disengage with the platform. I study a period of "calm" in the second-largest market for digital platforms, as opposed to the election period in the US as in the Meta studies, further enhancing the generalizability of the results (Bagchi et al., 2024). Finally, my model-based counterfactuals simulate the effects of policies that are difficult to implement in the field.

Second, this paper contributes to a rich literature finding strong effects of media bias on political polarization (DellaVigna and Kaplan, 2007). Although social media differs from traditional media in some key aspects (lower entry costs and content personalization), existing research shows that supply factors can influence behavior (Chiang and Knight, 2011). I build on these studies to show that demand for slanted information drives outcomes (Gentzkow and Shapiro, 2010; Martin and Yurukoglu, 2017) even when supply is endogenous due to algorithms. In so doing, the paper joins a growing literature on the welfare effects of social media (Allcott et al., 2022; Bursztyn et al., 2023; Brynjolfsson et al., 2024). This literature shows that social media usage can positively impact political outcomes, and encourage persistent civic engagement, especially among the underrepresented (Bursztyn et al., 2021; Zhuravskaya et al., 2020). However, direct evidence on the drivers of behavioral responses due to media bias is scarce, especially on TikTok-like platforms (Aridor et al., 2024). My findings are policy relevant because social media platforms like TipTop often introduce some randomly drawn posts in personalized feeds to expose users to a more diverse set of content, and users may be less resistant to algorithmic interventions like partial feed-diversification than a complete blackout of social media applications (Allcott et al., 2020).[7]

Finally, this paper adds to a literature on the consequences of AI adoption in the economy (Acemoglu et al., 2022). Despite fairness concerns (Rambachan et al., 2020), algorithms are widely applied across government and industry (Goldfarb and Tucker, 2019; Obermeyer et al., 2019; Aridor et al., 2022). This literature finds that algorithmic decision-making has unanticipated consequences due to misaligned values of platforms, users and regulators (Björkegren et al., 2020; Kasy, 2024). I show that consumers positively value personalization algorithms, and yet seek out the content that aligns with their pre-existing biases.

# 2 Background and Data

## 2.1 The Harms of Social Media and AI

With an average user spending 151 minutes daily on social media platforms, increased social media usage has tightened the scrutiny of the harms these platforms can potentially cause (GWI, 2023). Recent work also shows that social media can adversely affect users' mental health and can encourage the spread of misinformation (Allcott and Gentzkow, 2017; Braghieri et al., 2022). Content recommendation algorithms are often accused of boosting engagement with misinformation and hate speech by pushing users into echo chambers of radicalizing content (Pariser, 2011; Van Bavel et al., 2021). On the other hand, Gentzkow and Shapiro (2011) show that the internet has reduced the cost of accessing diverse viewpoints, and new communication technologies can also have remarkable societal benefits (Manacorda

[7]TipTop typically randomizes a small proportion of posts in a user's feed to maximize learning in an algorithm operating on an exploration-exploitation frontier (Dimakopoulou et al., 2017). While this helps the platform to continuously learn about user preferences, I show that it also diversifies the content that users consume (Kleinberg et al., 2022).

and Tesei, 2020; Gonzalez and Maffioli, 2024).

However, the link between social media and violence due to misinformation is widely discussed, especially in the case of the Capitol Hill riots on January 6, 2021 in Washington DC, as well as widespread violence in Myanmar and Ethiopia (Narayanan and Kapoor, 2024). Academic research has also linked exposure to radicalizing content on social media to hate-crimes and politically motivated violence, for example Müller and Schwarz (2021, 2023) show that exogenous reductions in social media usage led to a decrease in anti-immigrant hate crimes. This makes the study of social media platforms important as policymakers discuss optimal regulation that prevents the spread of misinformation and hate speech, while preserving users' right to free speech. The implications of the study are global, as the problems of misinformation and hate speech are universal (Avalle et al., 2024).

## 2.2 Social Media in India

More than 600 million people in India use social media platforms (GWI, 2023). This makes it one of the largest markets for online platforms in the world, second only to China. On average, 40.2% of the Indian population uses social media, and 67.5% of internet users have used at least one social networking platform (GWI, 2024). Therefore, social media usage is a significant part of the daily lives of a large number of Indians, who spend 141.6 minutes on various platforms every day (Wong, 2024). With a mobile phone penetration rate of 83%, social media users primarily use handheld devices to access the internet. This is not surprising as India has added more than 500 million mobile broadband connections in the last six years (Waghmare, 2024). While the Demographic and Health Survey (NFHS-5) reports that vulnerable populations have lower access to mobile phones and internet (IIPS and ICF, 2021), social media has become the most widely used platform for public discourse in India, and has been used to communicate underrepresented views (Wankhede, 2021).

However, social media usage has also proven to be harmful in this setting as it has been linked to instances of violence, in the form of mob lynchings, riots, and hate crimes (Arun, 2019). Threats to minority communities, stemming from social media usage in India, are speculated to be bolstered by content recommendation algorithms, which are customization algorithms that employ machine learning technologies. This is because in optimizing content engagement, social media is predicted to generate political filter bubbles or echo chambers (Barberá et al., 2015). Such echo chambers are likely to increase user exposure to more extreme and polarized view points, possibly leading to radicalization (Huszár et al., 2022).[8]

[8] Facebook whistleblower, Frances Haugen, has alleged that the company's personalization algorithms promote extreme content (Haugen, 2021). She also leaked the company's internal documents to show that the company is aware of the harms that algorithms have caused, not just in the US, but also in India.

India's regulatory framework has struggled to keep pace with the rapid proliferation of social media, leaving significant gaps in addressing the spread of harmful content and its consequences. Despite the introduction of the Information Technology (Intermediary Guidelines and Digital Media Ethics Code) Rules in 2021, enforcement has been inconsistent, and the regulatory mechanisms lack the teeth to effectively curb the influence of recommendation algorithms that amplify harmful content.[9] If anything, this regulatory framework can be used to stifle dissenting voices (Kumar and Jha, 2022), which is in direct contravention of the Santa Clara Principles on Transparency and Accountability in Content Moderation.[10]

Consequently, there is an urgent need for comprehensive policy interventions to tackle the challenges posed by personalization algorithms. However, it is unclear if the continued use of recommendation algorithms is the primary driver of user engagement with extreme content, or if user preferences play a bigger role. This has important implications for the design of regulatory measures in the future, which need to take a multi-pronged approach because behavioral responses may dampen the societal benefits of such policies.

## 2.3 The Platform

I partner with TipTop, a prominent social media platform in India, to understand the effects of exposure to extreme content via recommendation algorithms. I study how the nature of online interactions changes with my intervention in TipTop's rich content-based network, comprising almost 200 million monthly active users. TipTop's user interface resembles that of TikTok, and the platform made massive gains in market share when TikTok was banned in India due to escalating geo-political tensions with China in 2020.

The interactions on TipTop are mostly conducted through the algorithmically generated 'trending' feed shown in Figure E.1, which is also the landing page when a user logs onto the platform. TipTop posts, comprising text, images and videos, are created by influencers on this platform, as most users do not create content themselves. TipTop is a content-based social network, meaning that users interact with content rather than with other users, unlike X (formerly, Twitter), where users frequently engage with users they 'follow,' and unlike Facebook, where users can interact with content from 'Friends,' or from the 'Groups' they join. Connectedness with other users is of little consequence, as is evidenced by the distribution of the number of accounts a user follows in Figure E.2. This minimizes spillovers from the intervention to the control group, as I argue in Section 3.

See documents on internally conducted experiments, providing concrete evidence of the problem in India `https://www.nytimes.com/2021/10/23/technology/facebook-india-misinformation.html`

[9]`https://www.freelaw.in/legalarticles/Regulation-of-Social-Media-Platforms-in-India-`. Accessed on September 29, 2024.

[10]See Foundational Principles `https://santaclaraprinciples.org/`. Accessed on January 15, 2025.

Content based social networks, such as TipTop, are centered around topics like Politics, Religion, and Good Morning (or Greetings) messages.[11] Religious posts (both relating to Islam and Hinduism) are by far the most popular genre on the platform. Politics is the least favored genre on the platform, but a large part of political posts is seen to be harmful. Posts are automatically classified into broad genres in the data, potentially using the user generated hash-tags associated with each post. The algorithms used to classify content were not disclosed by the platform.

Due to the new (TikTok-like) features this platform offers, and its multi-lingual interface, TipTop attracts a large proportion of voters among the urban and rural poor in India. This makes such analysis especially important as little is known about political behavior of this demographic in India or about the users of this massive platform (Aridor et al., 2024).

### 2.4 Toxicity Classification

To measure the main outcome variable, i.e. toxicity of shared posts, I further process the text from posts in the form of images (using OCR). I problematize posts that may be a direct threat to the safety of a group or individual. Specifically, I use Perspective's machine learning algorithms, developed by Jigsaw at Google, to identify toxicity in the Hindi text extracted from about 20 million posts.[12] Perspective offers functionality in various languages, including Hindi, and is therefore able to preserve the context of the text in the classification process which could potentially be lost in a translation to English. Toxic content is defined as "a rude, disrespectful, or unreasonable comment that is likely to make one leave a discussion." Engagement with such content has been shown to create positive affect among some social media users (Schmid et al., 2024), which can potentially make social networking sites more addictive (Jo and Baek, 2023).

Perspective is a widely recognized machine learning solution for toxicity detection. It leverages transformer-based models, trained on millions of comments annotated by multiple human raters who evaluate contributions on a scale ranging from "very toxic" to "very healthy" (Fortuna et al., 2020). Such models, like GPT and BERT, use self-attention mechanisms to process entire sentences at once, enabling them to capture long-range dependencies in text (Vaswani, 2017). Perspective's Machine Learning models are being widely adopted to identify and filter out abusive comments on platforms like New York Times webpage, and are also being frequently used in academic research (Jiménez Durán et al., 2024).

[11] For context on the enormity of the 'Good Morning!' genre, see `https://www.wsj.com/articles/the-internet-is-filling-up-because-indians-are-sending-millions-of-good-morning-texts-1516640068`. Accessed on May 1, 2024.

[12] Jigsaw is a research unit within Google that builds technology to address global security challenges. For more information, see `https://jigsaw.google.com/`.

I construct a binary variable, labelled "toxic," which takes value 1 when Perspective's toxicity score on a post is higher than 0.2. The 0.2 threshold is chosen to maximize the criterion of true positive rate in the classification because toxicity is a rare outcome and can make automatic detection difficult (Banerjee et al., 2023). In Figure E.3, I show that 0.2 satisfies this threshold selection criterion, where the true labels for a random set of posts in the confusion matrix were determined by human raters, who were Hindi-reading undergraduate students at Brown University. While a 0.1 threshold mechanically has a higher rate of correctly classifying toxic speech, the F1 score is higher at 0.2 making it a more appropriate measure because there is imbalance in the dataset.[13] I will show that the results are robust to the choice of threshold for toxicity classification.

I further validate this threshold in Appendix H, by comparing the performance of this cut-off with other methods of detecting harmful content, and providing examples of Hindi posts from the platform along with the continuous toxicity scores. This paper focuses on toxic posts that target India's minority Muslim population. Since Perspective also correctly classifies homophobic and sexist content as toxic (for instance), I replicate the analysis on a subset of political posts, where a majority of the toxic posts are anti-Muslim.

## 2.5 Administrative Data

### 2.5.1 User-Post-Level Data

The administrative data provides information on each post that is viewed or engaged with (by way of sharing, liking, or downloading) by any given user. The precise time of exposure and engagement is also recorded in the data, which helps identify distinct patterns in usage according to time of the day or day of the week. This allows me to trace the posts a user was exposed to, and whether the user chose to engage with the post or not.

### 2.5.2 User-Level Data

I observe user characteristics, like their location, gender, age, date of account creation and language. These static user characteristics, along with users' exposure to and engagement with different types of content during the baseline period allow me to analyze heterogeneous treatment effects. The dimensions of heterogeneity used in this analysis were pre-registered with the AEA RCT Registry (Kalra, 2023). I provide a descriptive summary of user characteristics, as well as engagement at baseline in Table 1.

[13]The F1 score is a metric used to evaluate the performance of a model, particularly in tasks like classification. It is the harmonic mean of precision (how many of the predicted positive results are actually correct) and recall (how many of the actual positive results the model successfully identified) (Dell, 2024).

### 2.5.3 Post-Level Data

The platform characterizes posts by broad tag genres, based on user generated hashtags using an undisclosed algorithm. Further, the text on the images/ videos in the user generated posts is a rich source of information. I tokenized the text data and applied various methods to analyze it, including generating word clouds using TF-IDF and performing LDA topic modeling to identify broad themes in the posts (Ash and Hansen, 2023). This helps in measuring political slant of more than 20 million posts in the Hindi language that users engaged with during the course of the experiment.

## 2.6 Survey Data

I supplement the main outcome measures on platform usage, that are available in the administrative data, with an online survey that was sent out in three waves between May 2023 and July 2024. The protocol involved sending out a survey to users' registered WhatsApp numbers through the platform's WhatsApp business account during the intervention period. This received a low response rate, despite the survey being heavily incentivized. The second and third protocols, in the post period of the experiment (starting in January 2024), were telephonic surveys with a higher response rate, closer to 40%.

These data were especially useful in understanding how treated users spend their time off the platform if the intervention caused them to disengage from TipTop. For instance, the survey asked users about their time spent on other social media platforms, on the phone interacting with their friends and family, or watching TV. Other off-platform survey outcomes include user attitudes towards redistribution, and attitudes towards out-groups.

# 3 Experimental Design

## 3.1 Sample

I collaborated with TipTop to design and conduct a large-scale, long-term experiment that exposed users to content that was randomly drawn from the corpus of 2 million posts generated in the Hindi language each day. Out of the 200 million users on the platform, about 2 million users, or 1% of the user base, were randomly treated at the start of the experiment in February, 2023. Approximately 6 million users, or 3% of the user base, were randomly selected to be in the control group. This is done to make sure that the comparisons are not biased by other experiments running on the platform that may contaminate outcomes of the untreated users.

I limit the analysis to Hindi language users, who constitute about 45% of the total user base, to ensure the accuracy of the text analysis, given my native proficiency in this language. Further, I only include active users in the sample, defined as individuals who viewed at least 200 posts during the baseline period of December 2022 (5% of the experimental sample). These cuts reduce the sample size to 231,814 users, with 63,041 in the treatment group and 168,773 in the control group. Table 1 shows that 70% of users in the sample are men. This aligns with the gender distribution of social media users in India as reported in the Demographic and Health Survey data collected between 2019 and 2021 (NFHS-5), which found that 41% of women in India do not use the internet (IIPS and ICF, 2021). Further, the average user in the experimental sample created their account in 2022, which is almost two years after TikTok was banned in India. This is consistent with the estimated growth in internet penetration in India according to the World Bank, when the proportion of internet users increased from 20% in 2018 to 46% in 2022 (World Bank, 2022).

Table F.1 shows that the average user spent close to 7 hours on TipTop during the baseline period of 31 days, as opposed to the 141.6 minutes spent on social media in India each day, indicating that the average user was also actively consuming content from other social media platforms (GWI, 2024). The platform's integration with WhatsApp, a unique feature, indicates that its users are also very active on WhatsApp, the most popular social networking application in the country (GWI, 2023). As a result, the platform is representative of an average internet user in India who largely uses TipTop to consume content that is suitable for sharing as WhatsApp forwards in private conversations. This also has implications for the type of content TipTop users consume. For example, the average user viewed 1,087 posts during the baseline period, with most this content falling under the categories of "greetings" and "devotion."

## 3.2 Randomization

Treatment was randomly assigned at the user level to 1% of TipTop's user base. User IDs were picked randomly at the start of the experiment, and selected users were assigned to the treatment group for the entire duration of the intervention. Similarly, control users were also selected at the start of the intervention to ensure that their outcomes were not subject to contamination due to other AB tests/RCTs running on the platform. Therefore, users who joined the platform after the start of the experiment were not included in the sample.

Since users must opt into being randomly assigned to treatments for market research and AB tests when they agree to the platforms terms and conditions to create their account, they were unaware of their participation in the experiment, reducing the likelihood of biases due to selection, experimenter demand or Hawthorne effects. This is noteworthy as previous work on social media algorithms may suffer from selection as users were asked to participate in the research studies, were therefore, aware of the experiments and may have changed their behavior accordingly (Guess et al., 2023a).

I verify the validity of randomization in the treatment assignment across the sample by assessing balance in observable user characteristics across the treatment and control groups as I cannot reject the hypothesis that the treatment assignment was uncorrelated with user characteristics, either individually or jointly. I consider pre-registered user attributes, including gender, state and city of residence, and the week in which a user first created their account, as well as various measures of baseline usage, such as the total number and proportion of posts viewed. Table 1 provides estimates for a randomly selected set of attributes.

## 3.3 Control

Users in the control group continued to receive these standard algorithmic recommendations, while I intervene on the personalization algorithm for treated users. TipTop's personalization algorithm customizes user feeds based on their engagement history, using Field Aware Factorization Machines (Aggarwal et al., 2016). This algorithm generates a vector of preference weights for each user with respect to various post attributes, which are calculated using matrix factorization methods. These vector weights in the space of certain post features are referred to as embeddings (Athey and Imbens, 2019). This generates a ranking of posts for each user, and new posts are recommended daily according to this order.

The personalization algorithm generates these preference weights, or embeddings vectors, based on various (latent) features, which might represent a user's or post's affinity for characteristics such as humorous or toxic content, for instance. In general, these features are not human-interpretable but are learned by the algorithm to maximize user engagement with the platform. Appendix A provides a general overview of how these algorithms function, using simulated data on recent user-post engagements and a simple example. This example helps fix ideas and simulate the general properties of the personalization algorithm.

## 3.4 Treatment

Treated users were shown posts that were not ranked according to user preferences but randomly drawn from the entire corpus of content in their chosen language.[14] Figure E.4

[14]Appendix A discusses how the intervention was engineered in greater detail. The random draw of posts for treated users were generated by replacing the algorithmically generated embedding vectors with randomly picked multidimensional embeddings for each treated user. That is, for each treated user, the vector of preference weights is just a random draw of numbers. The "random embeddings" for treated

shows that treated user feeds changed along a variety of dimensions characterized by different topics. However, Figure E.5 shows that users with high interest in toxic content witnessed the largest decreases in exposure to such content, compared with the reduction in exposure among top users in other genres. I focus on toxicity because this is also the policy-relevant dimension due to regulatory concerns about the spread of anti-minority hate speech on social media platforms especially in India (Saha et al., 2021). Therefore, this intervention is a blunt instrument for a precision job as targeting specific types of problematic content is not feasible to implement on the field because the platform does not want to take a partisan stance, nor does it want to be accused of suppressing free speech.

I demonstrate the key properties of the content distribution in the treatment group by simulating a simple recommender system that generates the probability of the personalization algorithm assigning a post to a user. First, Figure A.3 shows that the distribution of content assignment probabilities among treated users over a period of one month is centered around the average assignment probability observed in the control group. This is predicted by the Law of Large Numbers (LLN), as the assignment probabilities are randomly picked from the set of control probabilities for treated users on each day of the intervention period. Second, the LLN also predicts a smaller spread in assignment probabilities for the treatment group given that the variance of these probabilities is divided by the number of control group users.

Crucially, the treatment has a greater effect on users with more extreme preferences regarding the toxicity of content. Figure A.4 shows that users with preferences closer to the average did not experience large differences in their assignment probabilities, and therefore the content feed, when treated. This important characteristic of content distribution is formally discussed as treatment intensity in Section 4.

Since interactions with content do not depend on user-to-user networks as shown in Figure E.2, the risk of spillovers from user-level treatment assignment is minimized. Such second order effects, including those on the incentives of content creators, are negligible as the intervention affected only 1% of the user base. The intervention began on February 10, 2023 and continued until the end of the year. Administrative and survey data on relevant outcomes were gathered for the baseline period (December 2022), the intervention period (February to December 2023), and the post-intervention period (January to March 2024).

---

users were uniformly sampled from an epsilon ball whose centroid was given by the mean embedding in the control group and the radius was twice the sum of variances in that vector. In particular, $\boldsymbol{\mu} = \frac{1}{N}\sum_{i=1}^{N} \boldsymbol{x}_i$, $\sigma^2 = \frac{1}{\nu}\sum_{i=1}^{\nu}(\boldsymbol{x}_i - \boldsymbol{\mu})^2$, where $\boldsymbol{x}_i$ represents the embedding with bias for user $i$ and $\nu$ is the total number of users. Formally, for each user embedding, the "random algorithm" uniformly sampled a point from an epsilon ball with the centroid $\boldsymbol{\mu}$ as the center and radius 2 $\times$ variance of control embeddings. Then, $\boldsymbol{\rho} \sim \mathcal{U}(\text{Ball}(\boldsymbol{\mu}, 2\sigma^2))$, where $\boldsymbol{\rho}$ is the newly sampled embedding for the user and $\mathcal{U}(\text{Ball}(\boldsymbol{\mu}, 2\sigma^2))$ represents a uniform distribution within the epsilon ball centered at $\boldsymbol{\mu}$ with radius $2\sigma^2$.

## 3.5 Descriptive Statistics

**The algorithm keeps users engaged online.** Users value the personalization algorithm because it decreases the cost of searching for preferred content. By making content discovery more difficult, the treatment reduces overall engagement with the platform, as reflected in the total number of posts viewed and shared in Table 2. Table F.1 shows disengagement in all aggregate measures of platform usage in the first month of the intervention. There are negative and statistically significant treatment effects on the number of logins per month but positive effects on the probability of leaving the platform. The average treated user reduced the total time spent on the platform during one month by 35% or 2.5 hours compared to the average control user, who spent close to 7 hours in the same time period.

This suggests that users gain value from the personalization algorithm and disengage from the platform when it is turned off, meaning that the intervention was costly for TipTop, as the platform generates revenue from the time users spend on the app and their attention to advertisements. In particular, Table 2 shows that on average, treated users viewed 35 fewer posts, while control users viewed about 250 posts in a month. Back-of-the-envelope calculations suggest that if the intervention is upscaled to the entire platform, TipTop loses about INR 4 million in advertising revenue in the first month of the intervention.[15] In fact, for every rupee lost in ad-revenues, an average treated user is exposed to 2.5 fewer toxic posts, compared to the 19 toxic posts viewed by control users in a month.

**Treated users view less toxic content.** The direction of the average treatment effect on the number of toxic posts viewed, or the treatment intensity for the average user, is unclear a priori. It is expected to be positive for users who do not prefer toxic content and negative for those who do.[16] I plot the raw data on number of toxic posts viewed using the dichotomous toxicity measures in Figure E.6. Panel (a) shows that between February 10 and March 10, 2023, the treatment group was exposed to fewer toxic posts on average due to the random content delivery.

The average reduction in exposure to toxic content is borne out in the average treatment effect of -27% in Table 2, as treated users viewed 5 fewer toxic posts on average compared to the control group's average of 19 toxic posts viewed in a month. Table F.2 shows that there is an average reduction of 14% in the toxicity of posts viewed when continuous toxicity score is not dichotomized in column (1), and that the result persists when different thresholds are used to construct the binary variable.

[15] The estimate was obtained using the price of INR 0.55 that TipTop charges advertisers per 1,000 impressions. The exchange rate on October 6, 2024 was 1 USD = 84.03 INR.

[16] The probability of being assigned toxic content does not necessarily equal the inverse of the number of toxic posts in the corpus, as it is determined by the cross-product of user and post embeddings (detailed in Appendix A). As a result, the average treatment intensity does not equal zero in this experiment.

**Online sharing behavior is inelastic.** I expect a reduction in the total number of toxic posts shared, as treated users face positive search costs in seeking out toxic posts to share that are not recommended to the average user. Additionally, exposure to fewer toxic posts on average may change their attitudes toward toxicity, potentially reducing their engagement with harmful content. Indeed, Table 2 shows that the number of toxic posts shared reduced by 20% over a month. However, this decrease in toxic engagement is smaller than the 27% reduction in the number of toxic posts viewed. While treatment reduced the percentage of toxic posts viewed by 0.64 pp, Table F.1 shows that the intervention increased the percentage of toxic posts shared by 0.12 pp over a control mean of 7% of the shares that are toxic in the first month of the intervention.

Table 2 also shows that the probability of sharing a toxic post conditional on viewing one, as seen by the ratio of toxic shares to toxic views, increased by 18%. Therefore, the responsiveness of toxic sharing with respect to toxic viewing—defined as the ratio of the percentage change in the number of toxic posts shared to the percentage change in the number of toxic posts viewed—is less than 1. I reject the null hypothesis that users behave mechanically with respect to the toxic content they are exposed to (or that this ratio equals 1) with a p-value of 0.002. Unresponsiveness in user behavior is corroborated by survey evidence in Figure E.7 that shows no differences in political attitudes of treated and control users in the post-intervention period. These differences in attitudes are precisely estimated in a random sample of over 8,000 surveyed users, so that that effect sizes as small as 5 pp can be ruled out using the survey data.

**Treatment induces behavioral responses to seek out content.** On average, the treatment effect on the ratio of toxic shares to toxic views is positive. Thus, the average treated user changed their behavior in response to the intervention, sharing toxic content at a higher rate and offsetting the negative effect on the number of toxic posts shared. To illustrate this, I decompose the treatment effect on the number of toxic posts shared

$$\text{Num of Toxic Posts Shared} = \frac{\text{Toxic Views}}{\text{Posts Viewed}} \cdot \text{Posts Shared} \cdot \frac{\text{Proportion Toxic Shares}}{\text{Proportion Toxic Views}}, \quad (1)$$

where the first term in the decomposition corresponds to the mechanical change in exposure to toxic posts due to the intervention, the second term corresponds to the disengagement effect that reduced platform usage, and the third term corresponds to the change in behavior upon viewing diverse content.

On average, the exposure and disengagement drive the reduction in the number of toxic posts shared in this empirical decomposition, as the behavior change goes in the opposite direction. Figure E.14 shows that the exposure effect dominates the effect on number of toxic shares, and the behavior change is larger than the disengagement effect in absolute terms. This implies that if the behavioral change component had been less than or equal to zero, the treatment effect on the number of toxic posts shared would have been more negative, yielding greater benefits for society.

**Treated users search more.** Table F.1 shows that treated users were more likely to use the search feature on the platform. This complements the evidence on the stickiness of sharing behavior, and that users seek out preferred content, as my measure of shares includes posts accessed through both the trending feed and search tabs. This finding also aligns with the fact that treated users were more likely to view fewer posts during the intervention period, as my measure of views excludes posts accessed through the search tab. While searching offers an intuitive channel for the positive effect on the ratio of toxic posts shared to toxic posts viewed, it is less likely to be driving the main treatment effects. This is because searched posts constitute 0.01% of viewed posts and 0.004% of shared posts.

**An Illustration of User Behavior.** While the treatment significantly decreased the number of toxic posts viewed by the average user, the reduction in the number of toxic posts shared was not as large. This suggests that user behavior is not malleable or elastic, as the change in sharing behavior did not correspond proportionally to the change in views.

For example, consider a user who is served 15 posts in a day, out of which they share 9 during the intervention period. If they are served 5 toxic and 10 non-toxic posts and share 2 toxic and 7 non-toxic posts, then the percentage of toxic posts shared is $\frac{2}{9} \times 100 \approx 22\%$. Now, consider a treated user who views 2 toxic posts and 7 non-toxic posts, and suppose they share 1 toxic post and 3 non-toxic posts. Thus, they are disengaged from the platform, sharing a total of 4 posts instead of the 9 they would have shared if they had not been treated. Note that they also view a smaller number of posts. This example illustrates that even though the average user views and shares fewer toxic posts upon being treated, the percentage of toxic shares increase. This occurs because 1 out of 4 shares is toxic under treatment, meaning the proportion of toxic shares is 25% which is greater than the 22% toxic posts shared in the control.

The stickiness in user behavior is reflected in the ratio of toxic shares to toxic views. In this example, control users shared 2 out of 5 toxic posts they viewed (40%), while treated users shared 1 out of 2 (50%). Therefore, the percentage change in toxic shares is -50%, while the percentage change in toxic views is -60%. Thus, the responsiveness of toxic shares with respect to toxic views is 0.83 which is less than 1.

# 4 Results: Four Facts

In this section, I present four key findings from the empirical analysis of the experiment.

## Fact I: Disabling the algorithm has heterogeneous effects

The intervention assigned average content from the full library of posts, reducing the amount of particular content for users who typically consumed above-average amounts and increasing it for those with below-average consumption. This results in heterogeneous treatment intensity (Duflo et al., 2011) which is higher for extreme users, as their baseline exposure to toxic content differed more significantly from the average feed.

To examine potential heterogeneous effects, I rank users based on the percentage of toxic content in their feed at baseline. This approach allows me to compute the effects on users with low, medium, and high levels of baseline toxic exposure. Using baseline exposure to represent user types is accurate as the personalization algorithm recommends posts based on users' past behavior. Figure 1 shows that the treatment effect on the proportion of toxic views is negative for users with a high degree of toxic exposure at baseline (Q3–Q5) and is positive for those with low baseline exposure (Q1 and Q2). The left panel in Figure E.8 replicates this result for the percentage of views that are toxic.

## Fact II: Extreme users disengage more

Personalization likely makes it easier for users to find the content they prefer, especially for those with more extreme preferences, as average content is further from their favored material. Overall platform usage changes based on users' baseline toxic exposure when personalization is removed, as shown in Figure 2. Users with the highest degree of baseline toxic exposure (Q5) reduced the number of posts viewed the most, by 23%. Views also declined for users in the middle of the distribution (Q2–Q4). However, those with the lowest toxic exposure at baseline (Q1) did not reduce their overall usage, even though the treatment changed their feed more than it did for those in the middle of the distribution.

Although there is a decline in the number of logins for the average user, the left panel of Figure E.9 shows that this effect is not heterogeneous by baseline exposure. This suggests that Q5 users largely disengaged on the intensive margin (measured with time spent in the right panel of Figure E.9) rather than on the extensive margin.

This may raise concerns about differential attrition across the treatment and control groups. Specifically, the main estimates may be biased if the type of treated users who left the platform differed from those who stayed. In fact, Figure 3 shows that users with higher exposure to toxic content at baseline reported spending a larger amount of time on other platforms, but no such pattern was found for users with lower baseline exposure to toxic content in the survey conducted after the intervention. However, controlling for baseline engagement with toxic content and treatment status, Table F.3 shows that leavers and stayers were balanced on observable characteristics. In Appendix G, I show that the Lee bounds for the main outcomes are tightly estimated as the magnitude of attrition is small (Lee, 2009).

## Fact III: Sharing of toxic content is nearly unresponsive but more responsive for users initially exposed to more toxicity

Figure 4 shows that Q5 users, who had higher baseline exposure to toxic content, reduced the number of toxic posts they shared when personalization was removed. Further, Figure E.10 also shows the largest decrease in the number of toxic posts shared comes from Q5 users when I focus on anti-Muslim content by considering the intersection of political and toxic content. This is also corroborated by the effect on other types of engagement with toxic content, for example, liking behavior in Figure E.11.

In contrast, Figure 4 shows that Q1 users, with the lowest baseline exposure to toxic content, did not increase the number of toxic posts they shared, even though their exposure to toxic posts increased. This lack of responsiveness among Q1 users helps explain why these users did not experience a decrease in the total number of posts viewed during the intervention period, as was expected for extreme users, despite the high and positive treatment intensity in Figure 2.

To assess how users' decision to share was affected by exposure, I measure the ratio of the percentage change in toxic posts shared to the percentage change in toxic posts viewed. If this ratio equals 1, then decreasing the toxic content a user is exposed to reduces the amount of toxic content they share by the same proportion. If the ratio is less than 1, users decrease their sharing by less than their exposure. I find that the measure of responsiveness is below 1 for all users, ranging from 0.08 for Q1 users to 0.69 for Q5 users. Figure E.12 shows that while treated users were more likely to use the platform's text search feature, the treatment effect on the number of times a user searched for any content (per post viewed on the landing page) is not heterogeneous. Moreover, this evidence does not identify the mechanisms as searched posts constitute 0.01% of viewed posts and 0.004% of shared posts.

### Fact IV: Behavioral responses dampen benefits of regulations

The intervention makes it more difficult for treated users to discover content that would have been recommended by the personalization algorithm. This may lead them to share fewer posts of the type they are initially inclined toward especially if they change their behavior. Alternatively, if users do change their behavior, they may seek out the content they like and share a higher proportion of the preferred content they view. Although Q5 users saw the largest decrease in the number of toxic posts shared upon treatment, Figure 5 shows that these users actually saw an increase in the proportion of toxic posts shared per toxic post viewed.

This suggests behavioral change counter to the direction of the effect on toxic exposure, as the probability of sharing toxic posts conditional on viewing such content strikingly, increases for Q5 users. The evidence that behavior is immalleable, and that users seek out the kind of content they prefer is also robust to choice of threshold for toxicity classification as Figure E.13 shows the effect on the ratio of toxic shares to toxic views is positive and the largest for Q5 users even when I use the continuous toxicity measure.

Figure E.14 unpacks the effect on the number of toxic posts shared, based on the empirical decomposition in equation (1) for different user types. This figure shows that for Q5 users, the increase in the ratio of toxic shares to toxic views checks the reduction in the number of toxic posts shared. If this effect had been zero, the total effect on toxic shares would have been more negative for these users. Therefore, the behavioral response in the ratio of toxic posts shared to toxic posts viewed dampens the societal benefits of the intervention which are measured by the decrease in the total number of toxic posts shared.

# 5 Model

In this section, I introduce a model to rationalize opposing effects of the intervention on the absolute and relative quantities of toxic shares. I evaluate counterfactual policies that target toxic content on social media as such policies cannot be implemented on the field.

## 5.1 Setup

I model a strategic interaction between the platform's algorithm and each user. The players' objective functions and strategies are described below.

### 5.1.1 Platform and Algorithm

For each user $i$, the platform maximizes $N_i$, the total number of posts $i$ views, assumed to be a continuous variable.[17] The algorithm assigns a set of posts to each user, $q_i^t$ proportion of which are toxic, and $(1 - q_i^t)$ proportion are non-toxic, with the objective of maximizing overall platform usage and engagement.[18] $N_i$ is endogenously determined by the user for the given proportion of the content-feed that is toxic, or the assignment probabilities chosen by the algorithm $q_i^t$, as shown next in the users' optimization problem.

### 5.1.2 User

Consider a social media user $i$, who picks the utility maximizing number of posts to view, $N_i = N_i^t + N_i^n$, for given assignment probability $q_i^t$ determined by the algorithm because $N_i^t$ (toxic) is determined as $N_i q_i^t$, and $N_i^n$ (non-toxic) as $N_i(1 - q_i^t)$. She also chooses $S_i^t$ and $S_i^n$ posts to share out of the total $N_i^t$ and $N_i^n$ posts viewed, in order to maximize consumption value from both viewing and sharing posts. This determines the continuous interval of posts shared $[0, S_i]$ from $[0, N_i]$ interval of posts viewed, where $S_i = S_i^t + S_i^n$. Effectively, the user chooses the proportion of posts shared that are toxic as $s_i^t = S_i^t / S_i$.

Following Akerlof and Kranton (2000), I introduce the set of categories $C$ based on some observable group characteristics or social identities, such as caste, religion, race or ethnicity. One motivation for this model is that users identify with social group $c_i \in C$, and have social-image concerns that motivate them to conform to the group's norms of engaging with toxic content. These norms presumably stem from immutable preferences of users in that group (Atkin et al., 2021). Then, when user preferences for toxic content are denoted by $p_i^t \in [0, 1]$, each user $i$ is assumed to belong to some category $c_i = c(p_i^t)$. That is, although social categories such as caste or religion are fixed, the underlying preferences for toxic content can also characterize these identities. Therefore, $c_i$ is a mapping that lumps together user preferences into the set of discrete and observable social groups $C$.

As in Butera et al. (2022), users derive public recognition utility from actions that reflect conformity with their perception of social norms. I assume that users learn about their group's tastes for toxic content from the content feeds, consistent with studies in media psychology that show that users rely on algorithmic curation to learn about appropriate discourse in their categories (Masur et al., 2021). Therefore, from a user's point of view, the algorithm's recommendation $q_i^t$ rely on categories $c_i$. Then, the user's objective is to maximize the utility she derives from viewing and sharing posts of different varieties,

$$\max_{s_i^t, S_i, N_i} u(s_i^t, S_i, N_i; q_i^t, c_i) = \underbrace{\beta N_i - \alpha(N_i - S_i)^2 - \eta S_i^2}_{\text{consumption utility}}$$
$$- \delta S \quad \underbrace{(1-\theta)\left(\log\left(\frac{s_i^t}{p_i^t}\right)\right)^2}_{\text{disutility of deviating from own tastes}}$$
$$- \delta S \quad \underbrace{\theta\left(\log\left(\frac{s_i^t}{q_i^t(c_i)}\right)\right)^2}_{\text{disutility of deviating from category's tastes}}$$

where, $\beta$ is the weight assigned to consumption utility received from viewing posts, $N_i$. A user is assumed to incur some disutility if an additional post she views is not shareable according to $\alpha(N_i - S_i)^2$ because while users get positive consumption utility from sharing posts, they also incur a cost if their feeds are too dissimilar from the posts they would like to share. $\eta > 0$ gives the convex cost of sharing. Users have self-image concerns which are modeled as disutility from sharing toxic content in a way that deviates from users' own preferences, $p_i^t$. This disutility is parameterized with $(1 - \theta)$, and $\delta S$ normalizes the utility function in terms of number of posts. I use the log form in the model because it is shown to fit the experimental data on toxic shares well. However, all model predictions are shown to be robust to functional form choices in Appendix B. Further, $\theta \in [0, 1]$ is the weight users put on their perception of their social group's tastes for toxic content. Users are assumed to learn group norms from the algorithm's recommendations $q_i^t(c_i)$ according to the final term which shows that users incur disutility if they deviate from these norms at some rate $\theta$.

[17] I follow Becker (1991) in adopting the continuity assumption.

[18] The platform's problem is a simplification of the actual problem faced by social media platforms, where the platform also optimizes the number of likes, shares, comments, number of ads shown to each user, and the price of advertising.

Thus, $\theta$ measures the "influence" on account of exposure to the algorithmically generated feed. To summarize, the disutility from sharing toxic content depends on how users' sharing behavior differs from a reference level which is given by a combination of what users' think that others do, $q_i^t(c_i)$, as well as their own tastes, $p_i^t$ (DellaVigna et al., 2012).

## 5.2 Dynamics and Equilibrium

The timing of the strategic interaction is as follows. The algorithm chooses assignment probabilities $q_i^t(c_i)$ to maximize engagement. Next, a user decides the total number of posts to view $N_i$, and the total number of posts to share, $S_i$. Finally, she chooses the fraction of shared posts that are toxic, $s_i^t = S_i^t / S_i$, for given exposure and sharing decision.

Consider two time periods, so that $\tau = 0$ and $\tau = 1$ represent the baseline and the intervention periods, respectively. In each time period, the game unfolds according to the timing described above, and I solve for the subgame perfect equilibrium. The algorithm and users are in equilibrium at baseline $\tau = 0$, irrespective of treatment status. Using backward induction, the first result helps to build the estimation strategy for the main parameter of interest, $\theta$. This follows from the first order condition of the consumer's utility maximization problem with respect to the proportion of posts shared that are toxic, $s^t_{i,\tau}$.

**Lemma 1.** *For a utility maximizing agent $i$,*

$$s^t_{i,\tau} = \left(q^t_{i,\tau}(c_i)\right)^{\theta} \left(p^t_i\right)^{1-\theta} \tag{2}$$

*That is, users place a weight of $\theta$ on perceived norms in the equilibrium sharing function.*

Then, the platform's engagement-maximization problem is given by the user's viewing decision, which is derived from her first order condition with respect to $N_{i,\tau}$.

**Lemma 2.** *For a utility maximizing agent $i$,*

$$N_{i,\tau} = \frac{1}{2\alpha\eta}\left[\beta(\alpha+\eta) - \delta\alpha\theta(1-\theta)\left(\log \frac{q^t_{i,\tau}(c_i)}{p^t_i}\right)^2\right] \tag{3}$$

*That is, users view a smaller number of posts when there is a mismatch between their preferences and the algorithmically generated preferences, $q^t_{i,\tau}(c_i) \neq p^t_i$.*

Maximizing $N_{i,\tau}$ with respect to $q^t_{i,\tau}(c_i)$, the recommendation algorithm's first order condition implies that the proportion of toxic posts assigned to user $i$ is exactly equal to her taste for toxic content in equilibrium.

**Lemma 3.** *The platform's engagement-maximization problem implies*

$$q^t_{i,\tau}(c_i) = p^t_i \tag{4}$$

*That is, the algorithm assigns toxic posts with probability equal to user's taste for toxic content at $\tau = 1$ for control users, and $\tau = 0$ for all users irrespective of treatment status.*

This result enables the characterization of user preferences as $q^t_{i,0}(c_i) = p^t_i$ for all $i$ due to the baseline equilibrium ($\tau = 0$). The algorithm also maps individual preferences to their respective social groups $c_i$ as the categories are further characterized in terms of preferences $p^t_i$. Indeed, the content feed informs the user's perception of her category's tastes for toxic content as users know that the content feeds are tailored according to preferences of all users in the same social group $c_i$ because $q^t_{i,\tau}(c_i) = p^t_i$ in equilibrium. However, the treatment exogenously changes $q^t_{i,\tau}(c_i)$ rendering the following comparative statics.

## 5.3 Model Predictions

The control algorithm assigns toxic posts to each user $i$ at time $\tau$ with probability $q_{i,\tau}^t$. However, in the treatment group, the probability of being assigned toxic content is picked uniformly at random, each day during the intervention period, from the set of all possible assignment probabilities in the control group, that is $\bar{q}^t = \mathrm{E}[q_{i,\tau}^t|$ users $i$ in control group$]$. The model enables an analysis of comparative statics for the exogenous variation in assignment probabilities under treatment.

I consider a targeted policy where users with higher proclivity to toxic content, that is $p_i^t > \bar{q}^t$, are treated with diversified or randomized feeds. This is because the empirical analysis demonstrated that the treatment increased the number of toxic posts viewed by users with the lowest proclivity to toxic content. Such an effect is not desirable, either from the platform's perspective (Q1 users do not see what they like), or that of the planner (Q1 users see more toxic content). Diversifying content feeds for users with $p_i^t > \bar{q}^t$ (Q4 and Q5 users) allows a more direct approach to reducing toxic exposure for the most toxic users.[19]

**Proposition 1.** *Treatment effect on exposure to toxic content is negative for user i with higher proclivity towards toxic content. Further, this effect is smaller for larger* $p_i^t$.

Figure 1 confirms the model's predictions to show that the treatment effect on the number of toxic posts viewed is negative for Q4 and Q5 users. This figure also shows that the treatment intensity is larger (in absolute terms) for Q5 users, compared to Q4 users. Next, the model predicts that the users with higher baseline exposure to toxic content view fewer posts overall, and this hypothesis is tested in the data.

**Proposition 2.** *For user* $i$ *with* $\alpha, \beta, \delta, \eta > 0$, $\theta \in [0, 1]$ *and* $p_i^t > \bar{q}^t > 0$,

$$\frac{\partial^2 N_{i,\tau}}{\partial p_i^t \partial \bar{q}^t} \geq 0 \tag{5}$$

*That is, for marginal increases in the average probability of being assigned toxic content* $\bar{q}^t$, *users with higher proclivity to toxic content view more posts.*

Intuitively, this is because the treatment exogenously lowers the probability of being assigned toxic content to the control mean when $p_i^t > \bar{q}^t$. Therefore, marginal increases from $\bar{q}^t$ bring this probability closer to the user's true taste for toxic content, $p^t$. In other words, the treatment effect on the total number of posts viewed is expected to be negative and smaller, and so is predicted to have a bigger impact (in absolute terms) on more toxic users.

[19] Such a targeted policy could not be implemented in the field experiment because the platform does not want to target users or posts by toxicity so as to maintain political neutrality.

In Figure E.15, I simulate the model's predictions on the total number of posts viewed, using arbitrary parameter values that are systematically calibrated later. The control users continue viewing the optimal number of posts in equilibrium but treated users face a higher treatment intensity, and so choose to view fewer posts in total. Given that the treatment intensity is higher for users with more extreme preferences, the model also predicts that users with higher exposure to toxic content at baseline spend even less time on the platform. Figure 2 shows that I cannot reject this hypothesis because Q4 and Q5 users view fewer posts overall, and the reduction in number of posts viewed is larger for Q5 users.

**Proposition 3.** *For user $i$ with $\alpha, \beta, \delta, \eta > 0$, $\theta \in [0,1]$ and $p_i^t > \bar{q}^t > 0$,*

$$\frac{\partial^2 s_{i,\tau}^t}{\partial p_i^t \partial \bar{q}^t} \geq 0 \tag{6}$$

*That is, marginal increases in the average probability of assigning toxic content leads to larger increases in the proportion of shares that are toxic for users who prefer such content.*

For the given rate of influence $\theta$, the model predicts that the decrease in the proportion of toxic posts shared is larger for more toxic users. Bringing this prediction to the data, Figure E.8 shows that the treatment effect on proportion of shares that are toxic is negative for users in Q5. Finally, the model predicts that the probability of sharing toxic posts conditional on viewing them is higher among the treated.

**Proposition 4.** *For user $i$ with $\alpha, \beta, \delta, \eta > 0$, $\theta \in [0,1]$ and $p_i^t > \bar{q}^t > 0$,*

$$\frac{\partial^2 (s_{i,\tau}^t / \bar{q}^t)}{\partial p_i^t \partial \bar{q}^t} \leq 0 \tag{7}$$

*That is, increasing the average probability of assigning toxic content leads to smaller changes in the proportion of toxic shares out of toxic views for users who prefer such content.*

Figure E.15 predicts that the ratio of toxic shares to toxic views is larger for treated users with higher exposure to toxic content at baseline. Figure 5 finds that the average effect on this ratio is driven by Q4 and Q5 users who share toxic content at a higher rate.

## 5.4 Estimation

### 5.4.1 Measurement

Algorithmic content assignment is assumed to be equilibrium at baseline with $q_{i,0}^t = p_i^t$, as shown above. This means that user behavior can be determined by substituting for $q_{i,0}^t$ into the optimal sharing function (2) as $s^t_{i,0} = p^t_i$. Therefore, preferences are measured using baseline sharing behavior during the intervention period to estimate the main parameter of interest $\theta$. Further, I proxy user's probability of being assigned toxic content by the algorithm $q^t_{i,\tau} \equiv v^t_{i,\tau}$, where $v^t_{i,\tau}$ is the proportion of posts viewed that are toxic.[20] As a result, Lemma 1 provides that sharing in equilibrium is a function of sharing at baseline, and the type of posts viewed.[21] I introduce preference shocks into the sharing behavior to get,

$$\log s^t_{i,1} = \theta \log v^t_{i,1} + (1-\theta) \log s^t_{i,0} + \mu w^t_{i,1} \tag{8}$$

where, $w^t_{i,\tau}$ represents *iid* preference shocks or unobserved heterogeneity in sharing behavior. The behavioral parameter $\theta$ is interpreted as the influence of exposure. This is in line with the idea that users expand their view of socially acceptable things to say in public discourse, by observing the content that is recommended to them by the algorithm. $\theta$ cannot be directly estimated through equation (8) as exposure to toxic content is asymptotically constant for all users $i$ in the treatment group, as content is randomly drawn from the same distribution of posts every day, $v^t_{i,1} = \bar{q}^t$. Further, the preference shocks are interpreted as user "moods" correlated with sharing behavior and exposure in each time period, rendering the randomization insufficient for identification. A steady state condition identifies $\theta$.

### 5.4.2 Steady State

This system is in steady state when the probability of viewing and sharing toxic posts $(v^t_{i,\tau}, s^t_{i,\tau})$, are stable over time $\tau$. The steady state condition is also the identifying condition because in the absence of any exogenous changes to assignment probabilities, user behavior should be the same in each time period. That is, $\theta$ is identified when the following assumption is satisfied,

$$\begin{aligned} \log s^t_{i,0} &= \log s^t_{i,1} \\ &= \theta \log v^t_{i,1} + (1-\theta) \log s^t_{i,0} + \mu w^t_{i,1} \end{aligned}$$

I test the validity of this assumption by showing that sharing behavior is identical in each time period in the sample of control users (Appendix C).

**Proposition 5.** *Assume,*

*(A1) User behavior is in equilibrium at baseline,* $s^t_{i,0} = p^t_i$

[20] $v^t_{i,\tau}$ can also be some fraction of $q^t_{i,\tau}$ because these variables are assumed to be continuous as $N_{i,\tau}$ and $S_{i,\tau}$ are not discrete in this framework.

[21] I arrive at equation (8) by substituting $p^t_i$ and $q^t_{i,1}$ in (2) during the intervention period $s^t_{i,1} = \theta \log q^t_{i,1} + (1-\theta) \log p^t_i$ with $s^t_{i,0}$ and $v^t_{i,1}$ respectively.

*(A2) The system is in steady state,* $\log s^t_{i,0} = \log s^t_{i,1}$

*Let* $D_i$ *indicate treatment status. Then, for some updating parameter* $\theta$*,*

$$\mathrm{E}\left[\log\left(\frac{s^t_{i,1}}{s^t_{i,0}}\right)\middle| D_i = 1\right] - \mathrm{E}\left[\log\left(\frac{s^t_{i,1}}{s^t_{i,0}}\right)\middle| D_i = 0\right] = \kappa - \theta\ \mathrm{E}[\log v^t_{i,0} | D_i = 1]$$

*where,* $\kappa$ *is some constant.*

Therefore, $\theta$ is identified using the relationship between the differences in toxic shares (from baseline $\tau = 0$ to intervention period $\tau = 1$) and the level of toxic views at baseline ($\tau = 0$) in the treated sample. The differences in sharing behavior between treatment and control groups account for unobserved heterogeneity.

# 6 Estimates

I use linear approximations of the log-specification above due to the presence of a large number of zeroes in the data on toxic views and shares (Chen and Roth, 2023). Figure E.16a shows that the structural relationship in Proposition 5 is approximately linear, and the slope is negative.

## 6.1 OLS Estimates

Table F.4 shows that a 1% decrease in the proportion of toxic posts viewed during the intervention period decreases the proportion of toxic posts shared by $\hat{\theta}\% = 0.1\%$ only. This demonstrates stickiness in user behavior, as the elasticity in sharing behavior with respect to baseline exposure $(1 - \hat{\theta})$ is 0.9. These estimates support the claim that user behavior is not malleable, and is largely determined by user preferences at baseline. However, the OLS estimates are likely to be attenuated due to measurement error. This is because the treated users sample the toxic posts they view from a Binomial distribution because the full list of assigned posts is generated randomly, and each post can be toxic or not. I correct this using an IV strategy outlined in Appendix C.1.

## 6.2 IV Estimates

I instrument exposure to toxic content in the first half of the posts viewed at baseline, with the average toxicity in the second half of the listed views to correct for measurement error. The setup requires an exclusion restriction which is satisfied because the sampling error in toxic posts viewed in the first half of posts viewed is uncorrelated to the sampling error in the second half of the posts viewed by construction. Then, the IV estimates in Column (2) of Table 3 indicate that the measurement error indeed attenuated the OLS estimates. Column (1) in Table 3 shows the strength of the first stage in the IV specification. The corrected estimate in Column (2) shows that a 1% reduction in exposure to toxic content reduces engagement with toxic content by 0.16%. Therefore, the IV estimates indicate that the elasticity of sharing toxic content with respect to exposure at baseline is close to 0.84, and that user behavior significantly depends on pre-existing behaviors or preferences. I perform validation checks in Appendix C.2.

## 6.3 Model Based Counterfactuals

I calibrate model parameters by matching moments of the empirical distribution of the total number of posts viewed and shared, with the number of toxic posts shared in Appendix C.3.

### 6.3.1 Alternative Behavioral Assumptions

I simulate the effects of diversification targeting more toxic users under different assumptions on user behavior. I describe the treatment effects, when users share the toxic content appearing on their feed mechanically ($\theta = 0$), and when users fully update their behavior in line with new information they are exposed to ($\theta = 1$).

Figure E.17 shows that the percentage change in number of toxic posts shared is decreasing in user preferences for toxic content when users are completely malleable. This is because more toxic users are more likely to be influenced when they view fewer toxic posts and when $\theta = 1$. Therefore, the decrease in number of toxic posts shared is larger in absolute terms when users are fully malleable, than when the users behave according to the observed degree of malleability, $\theta = 0.16$ as shown in Figure E.18a. Finally, the model predicts that the number of toxic posts shared or the constituent parts of this measure, will not change for mechanical users with $\theta = 0$ as shown in Figure E.18b. Similarly, Figure E.18c shows that the treatment effect on toxic sharing is decreasing in user toxicity when $\theta = 1$, only because more toxic users saw fewer toxic posts, thus influencing the sharing probabilities.

### 6.3.2 Model Based Decomposition

I decompose the treatment effect on the total number of toxic posts shared into two channels: (1) Engagement: change in platform usage, or the number of posts of any kind that were viewed and shared by treated users, and (2) Behavior: change in the probability of sharing toxic content for given exposure to toxic content. The main outcome variable, number of toxic posts shared, can be written as,

$$S_{i,1}^t = N_{i,1} \cdot \frac{S_{i,1}}{N_{i,1}} \cdot s_{i,1}^t$$

$$\implies \text{\% change in } S_{i,1}^t = \underbrace{\text{\% change in } N_{i,1}^t + \text{\% change in } \frac{S_{i,1}}{N_{i,1}}}_{\text{change in engagement}} + \underbrace{\text{\% change in } s_{i,1}^t}_{\text{change in behavior}}$$

Figure E.18a shows that the disengagement effect, or the decrease in overall platform usage, drives the decrease in the number of toxic posts shared when $\hat{\theta} = 0.16$. However, when users are fully malleable with $\theta = 1$, as in Figure E.18c, the treatment effect on number of toxic posts shared is driven by the change in proportion of shares that are toxic, $s^t$, or the behavioral changes due to the influence of exposure to diverse content. I also find that the (dis)engagement effect contributes to approximately 55% of the total treatment effect, which is in line with the estimates from the empirical decomposition in Section 3.5.

### 6.3.3 Counterfactual Policies

Social media platforms frequently diversify user feeds by randomizing a portion of the posts that users see. This is because platforms typically want to be at some point on the exploration-exploitation frontier where they are able to retain users by showing them content they like, while continuously learning their preferences (Kleinberg et al., 2022). This paper shows that introducing diversity into feeds may also be beneficial from a societal viewpoint as it may persuade users to share less toxic content.

I simulate the main policy-relavant outcome, number of toxic posts shared, and its component parts in the model-based decomposition, under different mixes of algorithmic and random feeds in Figure 6. This shows that even when 60% of the feed is randomized, the effect on toxic sharing for toxic users is driven by behavior changes represented by $s^t$.

On the other hand, if at least 80% of the feed is randomized, the effect on toxic sharing for toxic users is large and is driven by the engagement effect. This shows that a planner can optimally choose the degree of user feeds to diversify to balance the trade-off between user engagement with social media platforms, and the dissemination of toxic content.

# 7 Conclusion

In February 2019, a TikTok video of a 28-year old man threatening to “butcher” villagers from an oppressed caste group surfaced on TikTok (Christopher, 2019). P Saravanan, the deputy inspector at Tiruttani police station in Tamil Nadu which dealt with this case said, “If you leave a gun on a table, it is partially your (TikTok) responsibility...What we have now is leaving a gun chest open." However, P Madhava Soma Sundaram, Professor of Criminology and Fellow of the Indian Society of Criminology pins more accountability on the demand for harmful content, "I will not blame Tiktok. This is a reflection of our society."

This paper studies the role of user preferences and personalization algorithms in driving engagement with extreme content. Using an individually randomized intervention with 8 million social media users in India, I examine whether the content presented by algorithms substantially impacts user choices, or if users seek out content consistent with their existing behavioral patterns. I show that the while the intervention significantly reduced user exposure to toxic content, there was an increase in the probability of sharing toxic posts conditional on viewing such content. A behavioral model rationalizes these results, and estimates show that the algorithm's influence on user behavior is limited. This leaves little room for policy instruments to alter sharing behavior through reduced exposure to toxic content.

I examine the mechanisms using observable user attributes and baseline behavior. Figure E.19 shows that, irrespective of treatment status, users with a higher number of toxic shares were more likely to **(1)** have adopted the platform early, **(2)** be less active at baseline but more active during working hours, **(3)** be older, **(4)** be more engaged with "politics," but less engaged with "greetings" at baseline. Recall that, Q5 users disengage with the platform upon being treated, whereas Q1 users seek out content on TipTop. This is likely because Q1 users log on to TipTop to consume content that is not found on other platforms (good morning messages, or other "greetings"). On the contrary, Q5 users are more likely to consume political content at baseline, which is substitutable on other platforms because TipTop offers a unique opportunity for users to share posts directly to WhatsApp, making the platform a one-stop shop for posts that users want to share. Moreover, there are no competing platforms that offer this type of 'WhatsApp-able' content in India as most content generation platforms (like Facebook, Instagram, or YouTube) encourage users to stay on their respective apps.[22] In Figure E.20, users with the highest affinity to greetings at baseline did not disengage with the platform. This may explain the inelasticity in Q1.

Figure 3 presents survey evidence to show that treated users with higher affinity to toxic content spent more time on other platforms. On the other hand, no such trend was observed for users with lower affinity to toxic content. This shows that users with higher proclivity to toxic content at baseline are more likely to find such content on other platforms, when the intervention reduced their exposure on TipTop. Therefore, cross-platform regulation is necessary to effectively reduce the spread of toxic content.

[22]In fact, this is the primary objective of the algorithm on these other platforms: to increase the time a user spends on the platform. See this Marketing guide `https://www.socialpilot.co/youtube-marketing/youtube-algorithm`, for instance.

This paper has the following limitations. First, it considers the effects of a specific algorithm on a specific platform. These results are generalizable to other platforms to the extent that they use similar algorithms to personalize content recommendations. Second, this analysis is restricted to the effects of the "random algorithm" for one month only. Future work will focus on understanding the long-term effects of the intervention, using administrative data for later months, and survey data for a random subset of users in the experimental sample. Finally, the broader implications of this intervention on mental health and digital addiction are moot but were outside the scope of this paper. I aim to contribute to these strands of knowledge in future research work using the survey data on mental health outcomes that I collected at the end of the 11-month intervention period.

# Tables and Figures

Figure 1: Treatment intensity by user type

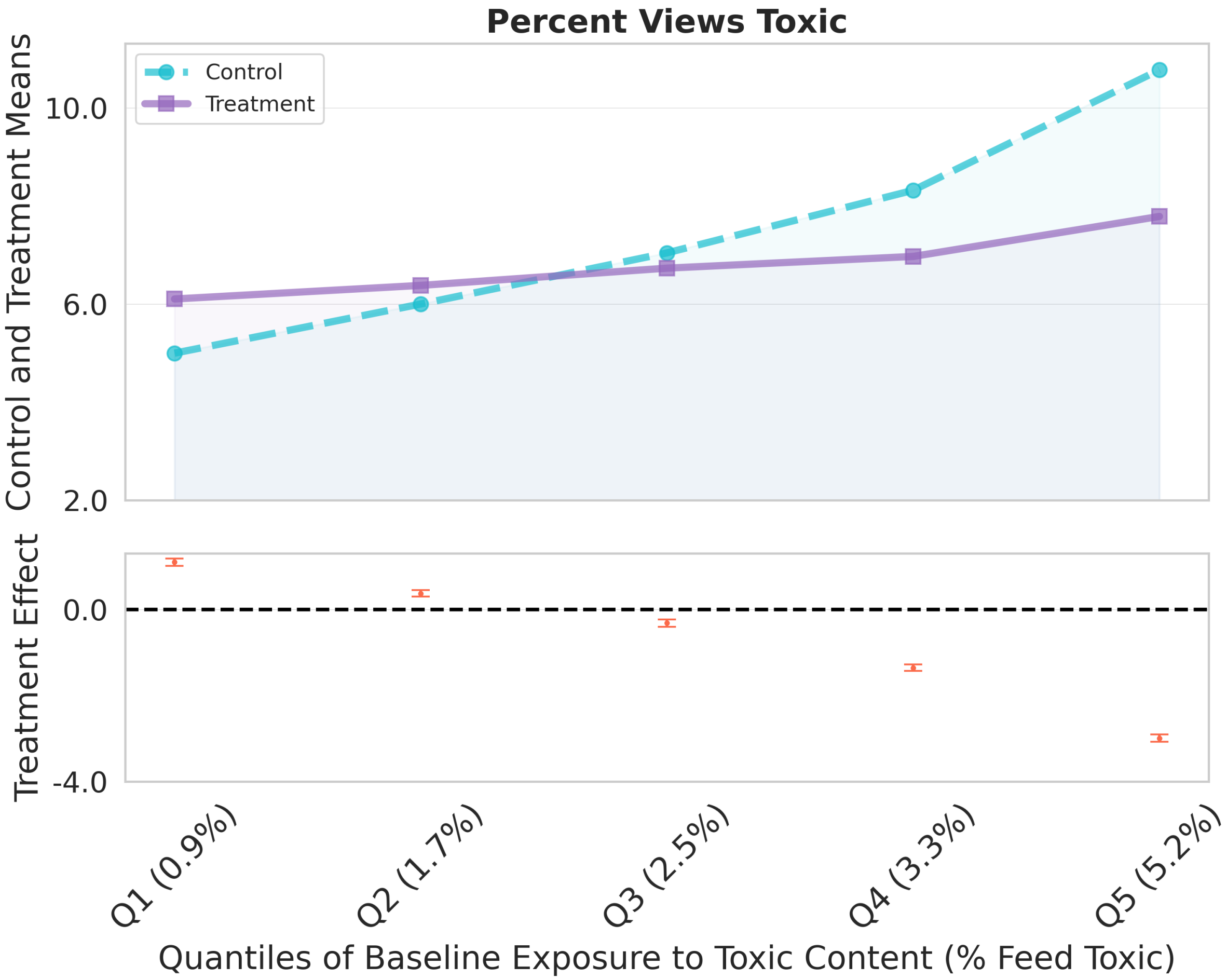

Notes: This figure shows that the treatment intensity, treatment effect on the percent toxic posts viewed, is decreasing with exposure to toxic posts at baseline. This effect is positive for Q1 users and negative for Q5 users, where each quantile is defined using the proportion of toxic posts viewed at baseline. The axis corresponding to the bottom plot shows the magnitude of the treatment effects (as coefficient plots), while the top panel is scaled according to the control mean of the outcome, percentage of posts viewed that are toxic, for each quantile. All regressions are run at the user level with robust standard errors.

Figure 2: Treatment effects on viewing behavior, by user type

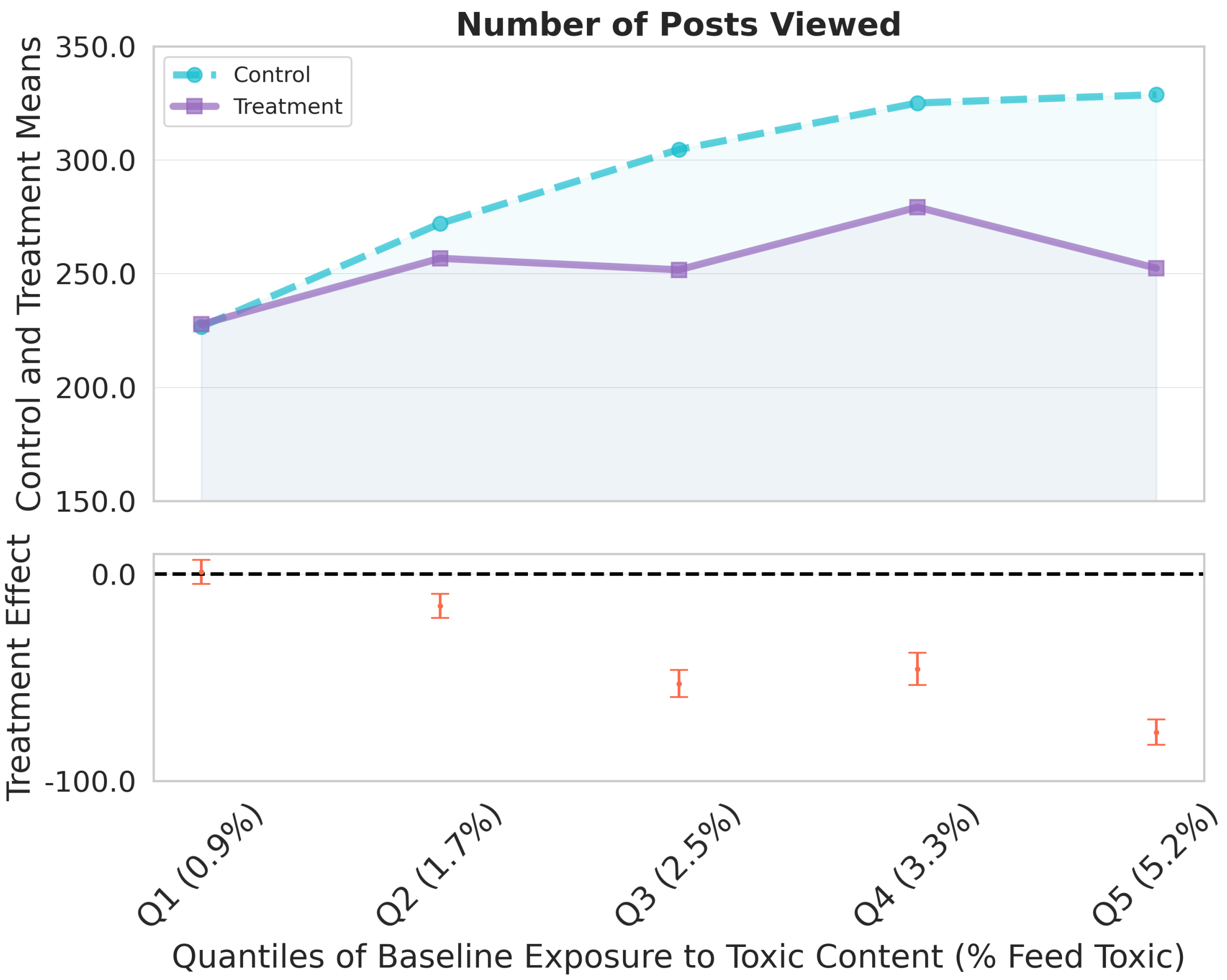

Notes: This figure shows that the total number of posts viewed, or overall engagement with the platform, also changes by treatment status and user type. In fact, the treatment effect on the total number of posts viewed is larger (in absolute terms) for users with higher exposure to toxic content at baseline. The axis corresponding to the bottom plots shows the magnitude of the treatment effects (as coefficient plots), while the top panel is scaled according to the control mean of the outcomes for each quantile. All regressions are run at the user level with robust standard errors.

Figure 3: Substitution with other platforms, by user type

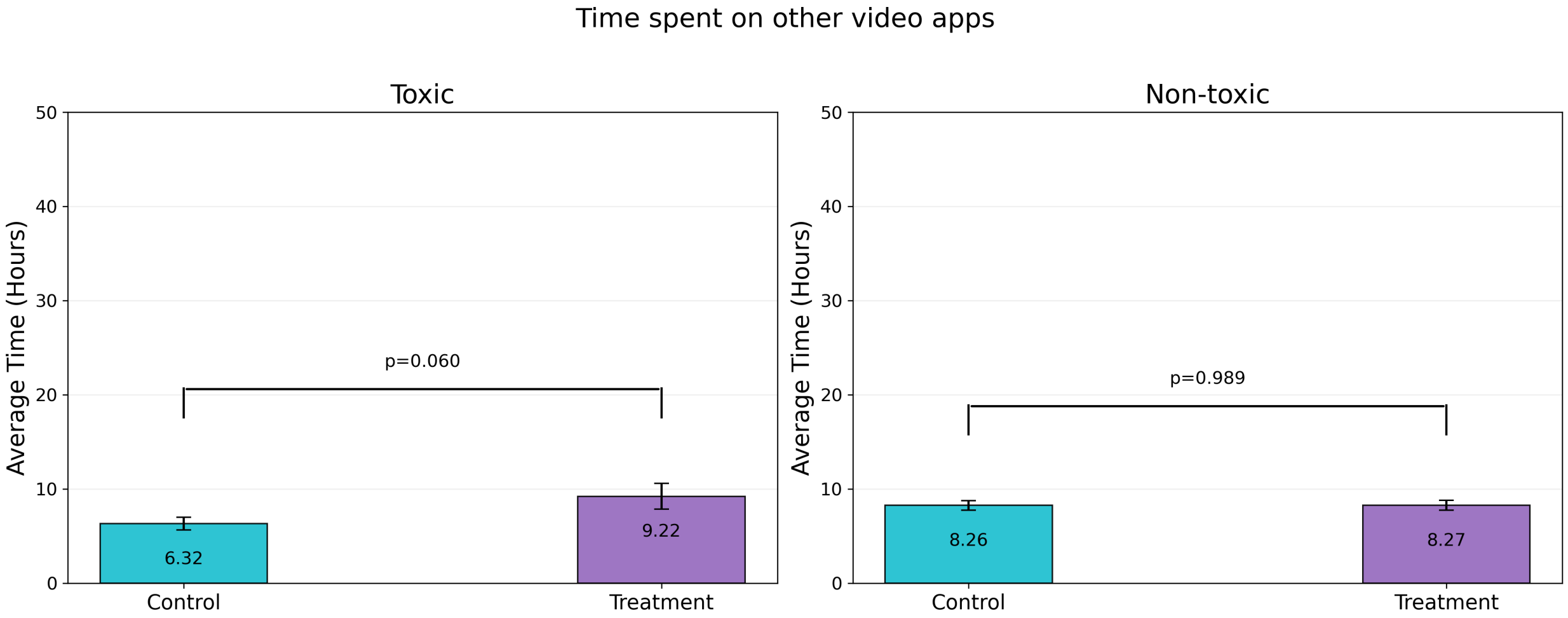

Notes: This Figure shows that users with higher proclivity to toxic content at baseline were more likely to spend more time on other platforms upon being treated (with a p-value of 0.06). A subset of users in the experimental sample were randomly selected for a follow-up survey (N = 8, 387), and asked how much time they spent on a range of other social media platforms, and the time they spent on the TV, or telephone conversations, or in-person interactions. The survey was conducted at the end of the intervention period, with 4, 236 users randomly sampled from the treatment group, and the remaining 4, 151 users sampled from the control group.

Figure 4: Treatment effects on sharing behavior, by user type

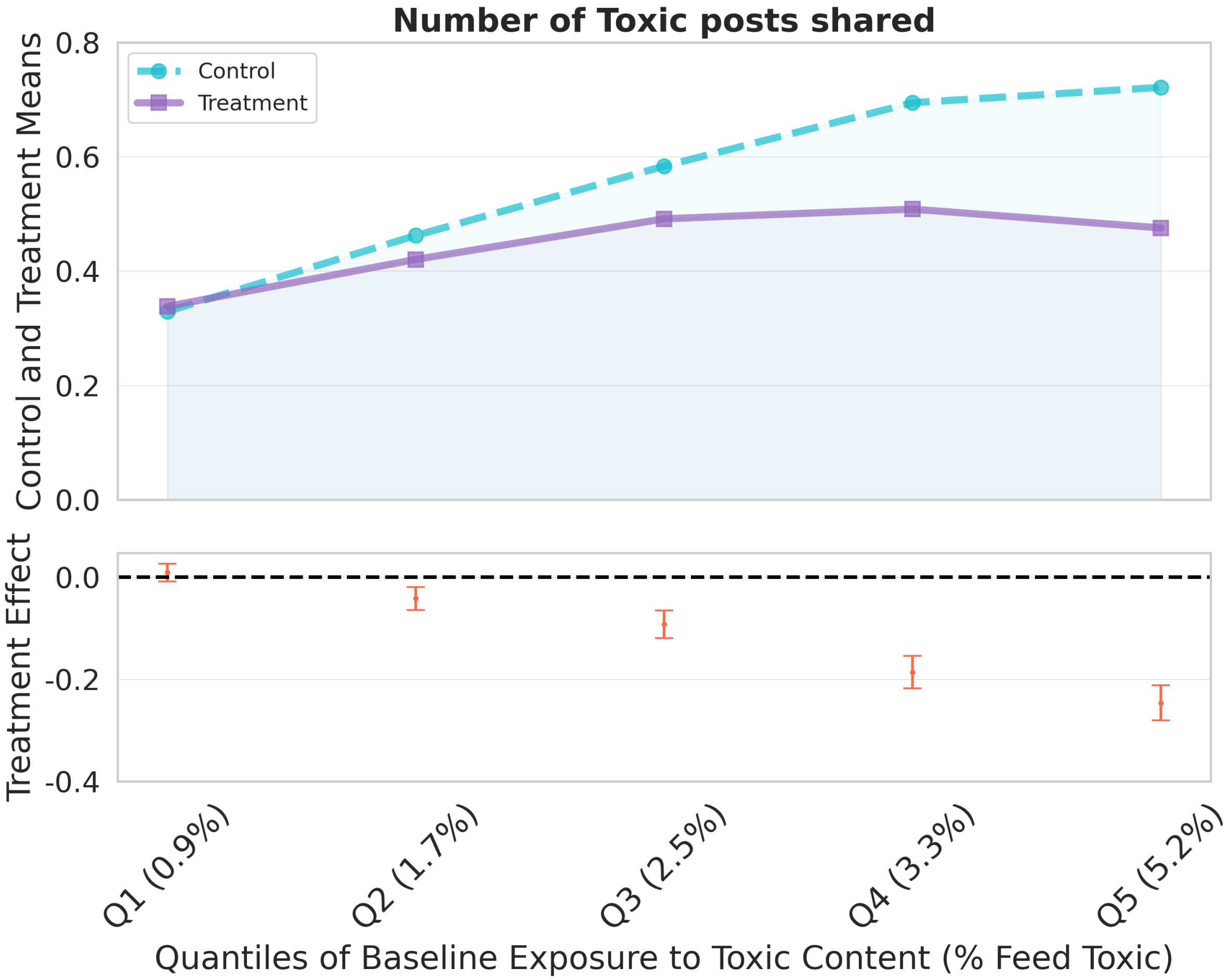

Notes: This figure shows that the treatment effect on the number of toxic posts shared is negative for toxic users (Q3–Q5) but is not statistically significant for Q1 users. While the number of toxic posts shared by toxic users has decreased, these users disengage from the platform by sharing fewer posts overall. From this figure, it remains unclear whether toxic users share fewer toxic posts because they are exposed to less toxic content they can share, are influenced by the non-toxic content they are exposed to, or because they are disengaging with the platform. The axis corresponding to the bottom plots shows the magnitude of the treatment effects (as coefficient plots), while the top panel is scaled according to the control mean of the outcomes for each quantile. All regressions are run at the user level, and inference about the treatment effects is based on robust standard errors.

Figure 5: Evidence on inelasticity in toxic sharing and seeking out behavior, by user type

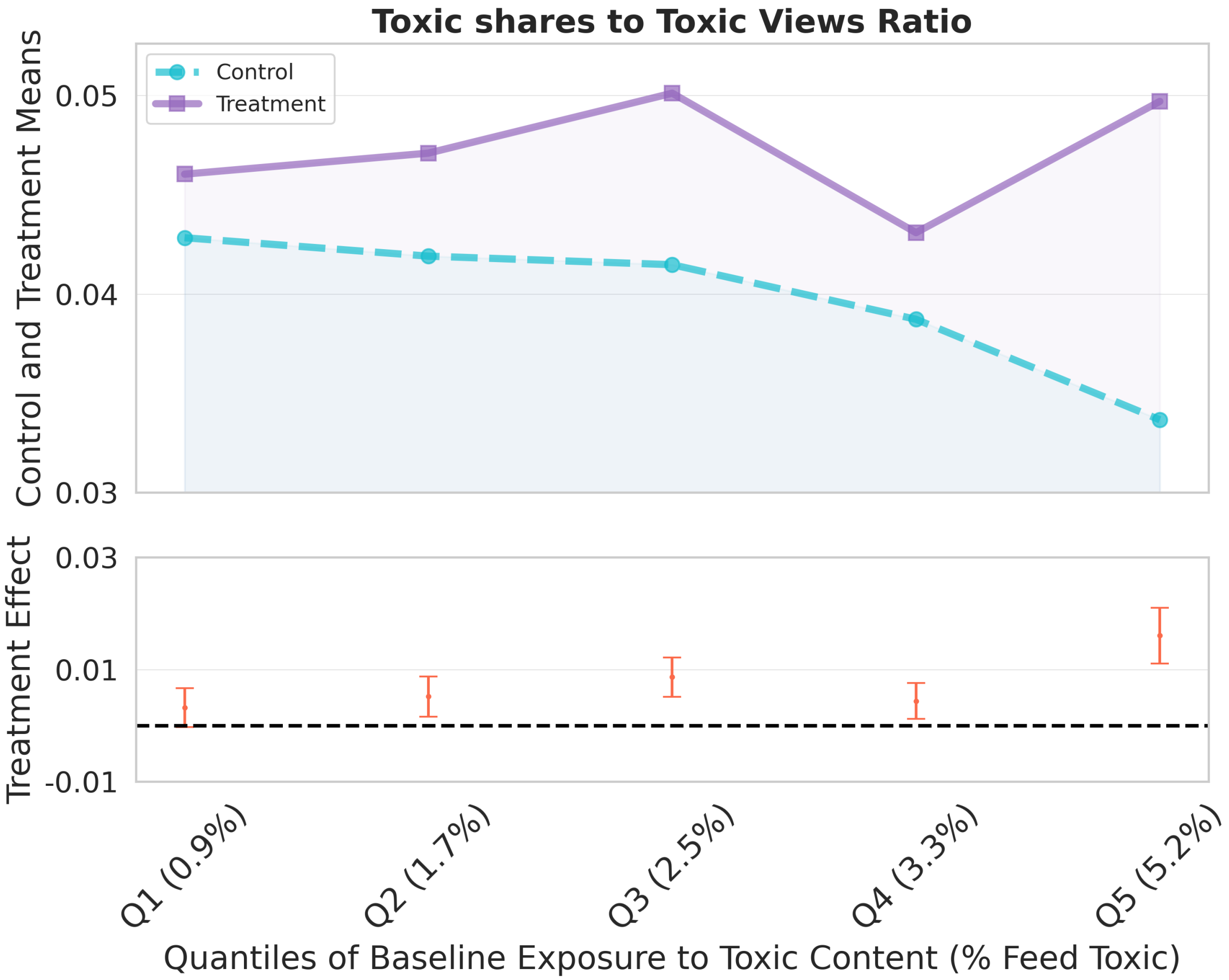

Notes: This figure shows that user behavior immalleable or "sticky". Treated users shared a higher proportion of the toxic posts they viewed during the intervention period, and the treatment effect on this measure is the largest for Q5 users. This suggests that the societal benefit of the policy is blunted as the reduction in the total number of toxic posts shared would have been larger if users had not shared a higher proportion of the toxic posts they viewed. The axis corresponding to the bottom plots shows the magnitude of the treatment effects (as coefficient plots), while the top panel is scaled according to the control mean of the outcomes for each quantile. All regressions are run at the user level, and inference about the treatment effects is based on robust standard errors.

Figure 6: Counterfactuals for different levels of randomization in content feeds

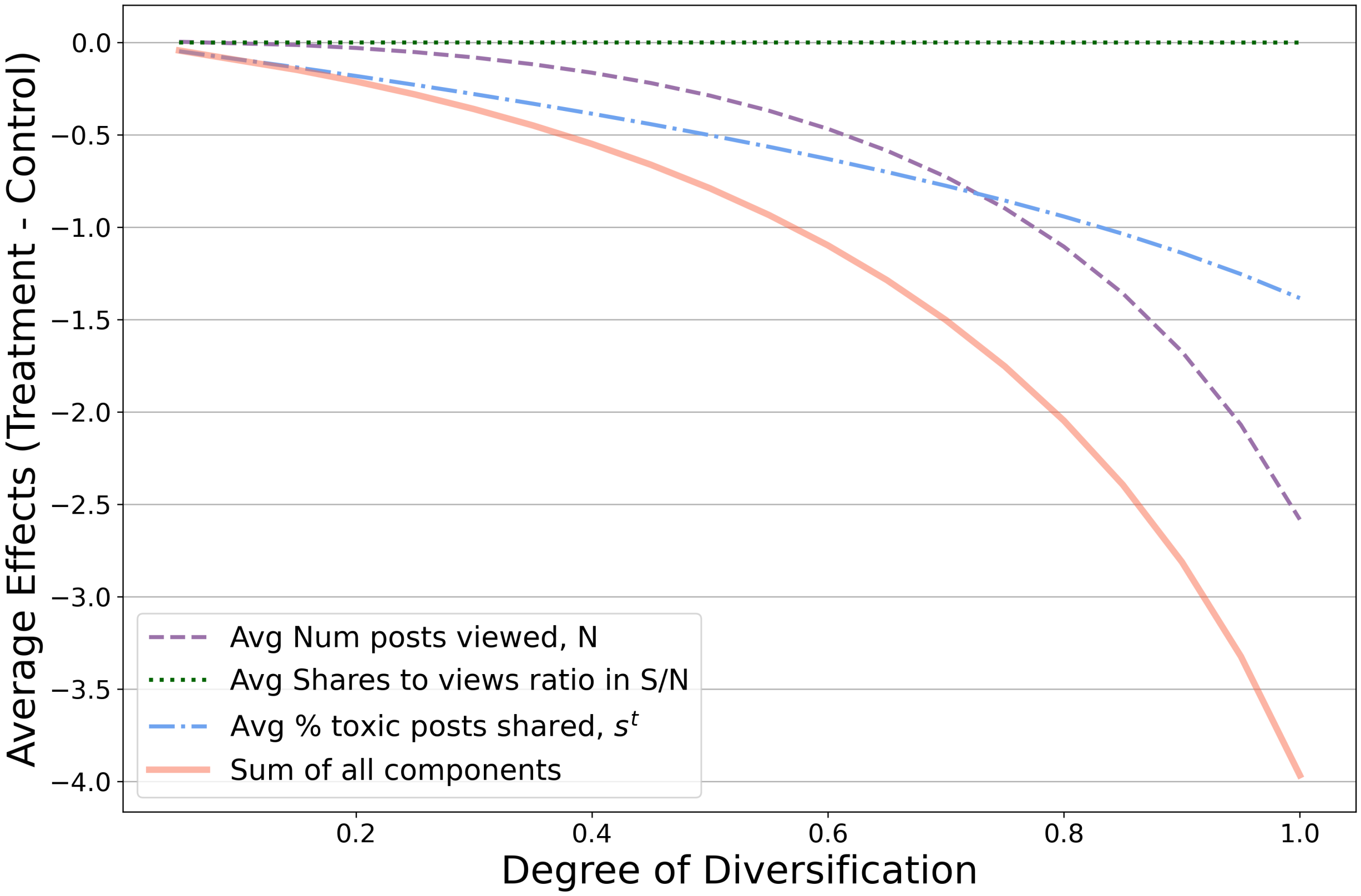

Notes: This figure simulates the counterfactual policy predictions for different levels of randomization in content feeds. The different degrees of randomization are achieved by considering linear combinations of the probabilities of being assigned toxic content in the control and treatment groups. That is, the counterfactual probabilities of being assigned toxic content under different policy regimes is given by $q_i^{t,a} = a \cdot \bar{q}^t + (1-a) \cdot q_i^t$. This shows that when $a = 60\%$, the decrease in the number of toxic posts is driven by the reduction in the probability of toxic users being assigned toxic content. This is ideal for a policymaker who wants to reduce the number of toxic posts viewed and shared without affecting the platform's overall engagement. However, as the degree of randomization increases to 80%, the reduction in toxic user engagement contributes more to the decrease in the number of toxic posts shared. Therefore, the policymaker can choose the degree of randomization, $a$, to balance this trade-off between reducing toxic engagement and maintaining overall engagement with the platform.

Table 1: Balance in treatment assignment across user characteristics and baseline behavior

| Variable | Control Mean | Difference (T - C) | Std.Err. |
|---|---|---|---|
| **Observable User Characteristics** | | | |
| State: gujarat | 0.021 | -0.019 | 0.014 |
| State: uttar pradesh | 0.105 | -0.012 | 0.012 |
| City: aligarh | 0.002 | 0.019 | 0.027 |
| City: bareilly | 0.002 | -0.010 | 0.024 |
| City: dehradun | 0.001 | 0.012 | 0.028 |
| City: faizabad | 0.002 | -0.038 | 0.026 |
| City: hardoi | 0.002 | -0.020 | 0.025 |
| City: jaunpur | 0.003 | -0.028 | 0.022 |
| City: khandwa | 0.001 | -0.007 | 0.037 |
| City: latur | 0.001 | -0.068 | 0.033 |
| City: north east delhi | 0.001 | -0.054 | 0.034 |
| City: pratapgarh | 0.002 | 0.031 | 0.024 |
| City: raipur | 0.004 | -0.005 | 0.023 |
| City: sitapur | 0.002 | -0.017 | 0.026 |
| Gender: Male | 0.699 | -0.002 | 0.003 |
| Age: 19-30 | 0.006 | 0.000 | 0.016 |
| Week: 2016-28 | 0.000 | -0.662 | 10.698 |
| Week: 2022-38 | 0.012 | -0.748 | 10.696 |
| **Baseline Behavior** | | | |
| Num Posts Viewed | 777.126 | 0.000 | 0.000 |
| Num Posts Shared | 22.045 | -0.000 | 0.000 |
| Num Logins | 9.250 | -0.000 | 0.000 |
| Time Spent (in hours) | 16.341 | -0.000 | 0.000 |
| Prop Activity during Daytime | 0.097 | -0.001 | 0.004 |
| Prop Activity during Weekends | 0.346 | -0.007 | 0.005 |
| Num Searched per Post Viewed | 0.175 | 0.001 | 0.002 |
| Prop Views in Humor Genre | 0.051 | -0.009 | 0.030 |
| Prop Views in News Genre | 0.058 | -0.008 | 0.030 |
| Prop Shares in Bollywood Genre | 0.010 | -0.037 | 0.012 |
| Prop Shares in News Genre | 0.009 | -0.010 | 0.014 |
| Prop of Views Toxic (%) | 2.681 | 0.007 | 0.007 |
| Prop of Shares Toxic (%) | 2.241 | -0.029 | 0.042 |
| Tox Share to Tox View Ratio | 1.023 | -0.000 | 0.000 |
| F-statistic: | 0.984 | p-value: | 0.506 |
| N | | 231814 | |

Notes: This table shows balance in treatment assignment across all observable characteristics, using a single regression run at the user level: $D_i = \beta_0 + \sum_c \beta_c \mathbb{1}_i(\text{user characteristic} = c) + \varepsilon_i$, where $D_i$ is a binary variable equal to 1 when user $i$ was assigned to the treatment group. The table shows coefficients corresponding to a randomly selected set of user characteristics, with weeks representing the date on which a user created their account. Additionally, none of the observable characteristics are correlated with treatment assignment. I cannot reject the null hypothesis of joint insignificance, with an F-statistic of 0.984 and a p-value of 0.506. The regression is estimated at the user level, and robust standard errors are in parentheses.

Table 2: Experimental Effects on Post Views and Shares

| | Num Posts Viewed | Num Posts Shared |
|---|---|---|
| Treatment Effect | -35.497** | -6.367** |
| | (2.208) | (0.206) |
| Control Mean | 246.654** | 18.396** |
| | (1.361) | (0.131) |
| | Num Toxic Posts Viewed | Num Toxic Posts Shared |
| Treatment Effect | -5.024** | -0.093** |
| | (0.172) | (0.010) |
| Control Mean | 18.806** | 0.474** |
| | (0.129) | (0.006) |
| | % Toxic Posts Viewed | Toxic Shares to Views Ratio |
| Treatment Effect | -0.641** | 0.007** |
| | (0.033) | (0.001) |
| Control Mean | 7.416** | 0.040** |
| | (0.018) | (0.001) |
| N | 231814 | |

Notes: This table shows that the treatment led to overall disengagement with the platform, decrease in the total number of toxic posts viewed and shared, but an increase in the rate at which users shared toxic content they viewed. I find that the treatment effect on the number of posts viewed and shared, and the number of toxic posts viewed and shared, in one month is negative and statistically significant. This suggests that the treatment effect on engagement with toxic content would have been larger if users had not shared a higher proportion of the toxic posts they viewed. Each cell reports estimates of the regression coefficient from a linear regression of the outcome aggregated at the user level over all days in the first month of the intervention period (February 10 to March 10, 2023). Robust standard errors are in parentheses. $p < 0.05^{*}, p < 0.01^{**}, p < 0.001^{***}$.

Table 3: Structural estimation of influence parameter $\theta$, with measurement error correction

| | (1) Proportion of Toxic Posts Viewed (Baseline, half-2) | (2) Proportion of Toxic Posts Shared (Intervention - Baseline) | (3) |
|---|---|---|---|
| Proportion of Toxic Posts Viewed (Baseline, half-1) | $0.572^{***}$ | $-0.155^{**}$ | |
| | (0.004) | (0.058) | |
| Proportion of Toxic Posts Viewed (Baseline, half-2) | | | $-0.183^{**}$ |
| | | | (0.065) |
| $N$ | 63041 | 63041 | 63041 |

Notes: This table provides estimates for the structural parameter $\theta$ in the model of sharing behavior, where $\theta$ captures the rate at which users update their behavior according to perceived social norms. $\theta$ is therefore the influence effect of one month's exposure to non-personalized feeds. Column (1) shows the relevance of the instrument, i.e., the proportion of toxic posts viewed computed using only the first half of posts viewed by a user at baseline, when they were arranged in a random order ($half1$). This instrument is used to correct the measurement error resulting from treated users randomly sampling toxic posts from their feeds. The independent variable in Column (1) is the proportion of toxic posts viewed at baseline, computed using only the second half of posts viewed by a user at baseline, when they were arranged in a random order ($half2$). Column (2) shows the results of an 2SLS regression on the first difference in proportion of toxic posts shared between baseline and the intervention period. Here, the independent variable is $half1$, which is instrumented with $half2$. Column (3) estimates the model with classical measurement error correction in STATA, where the correlation between $half1$ and $half2$ serves as the reliability measure for the proportion of toxic posts viewed. The estimated slope coefficient estimates $\gamma_1$ is always negative and statistically significant. Consequently, the estimated $\theta$ is positive, with a value of 0.16 based on the preferred specification in Column (2). The baseline period is December 2022, and the intervention period data span from February 10, to March 10, 2023. Robust standard errors are in parentheses. $p < 0.05^{*}, p < 0.01^{**}, p < 0.001^{***}$.

# A Matrix Factorization Model

Matrix Factorization algorithms provide some approximation of users' latent features or attributes that predict user retention and engagement. This is done using data on previous engagement with posts on the platform to gauge the type of posts users have shown affinity towards in the past. The algorithm factorizes a matrix of engagement at the user-post level for some abstract set of user and post features as described in the example below.

Figure A.1: An example of SVD decomposition into two-dimensional user embeddings $U$, eigenvalues $\Sigma$, and movie embeddings $V^T$

(a) Matrix representation of vector embeddings

$$R = \begin{pmatrix} 1 & 0 \\ 0 & 2 \\ 3 & 0 \end{pmatrix} = \underbrace{\begin{pmatrix} \frac{1}{\sqrt{10}} & 0 \\ 0 & 1 \\ \frac{3}{\sqrt{10}} & 0 \end{pmatrix}}_{U} \cdot \underbrace{\begin{pmatrix} \sqrt{10} & 0 \\ 0 & 2 \end{pmatrix}}_{\Sigma} \cdot \underbrace{\begin{pmatrix} 1 & 0 \\ 0 & 1 \end{pmatrix}}_{V^T}$$

(b) Graphical representation of vector embeddings

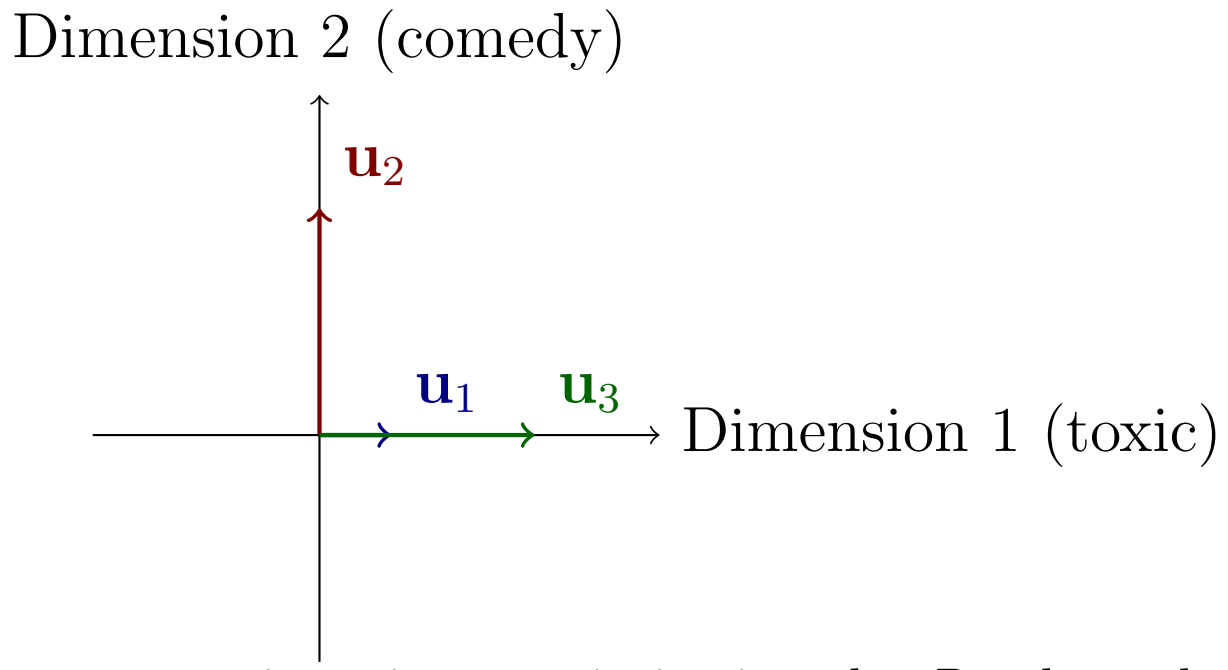

Notes: In this example, a user-movie rating matrix is given by $R$, where three users rate two movies on a scale of 1 to 5. The idea is to learn user tastes in some low-dimensional space of latent features because the dimensionality of the $R$ matrix rises with the number of users and movies. Singular Value Decomposition (SVD) breaks this matrix down as **(1)** $U$ represents the user embeddings (u1 and u2), showing how users relate to the abstract features; **(2)** $\Sigma$ is a diagonal matrix containing singular values ($\sigma_1$ and $\sigma_2$), which scale the importance of each feature; **(3)** $V^T$ represents the movie embeddings (v1 and v2), showing how movies relate to the abstract features. By multiplying $U$, $\Sigma$, and $V^T$ back together, the original matrix $R$ is reconstructed. The embeddings in $U$ and $V$ are plotted in a 2D space to visualize their relationships in Panel (b). These plots show that the first user is more interested in the first movie (or movies of that type), while the second user is more interested in the second movie (or movies of that type). The two dimensions represent abstract features that summarize the original data's structure and relationships. For example, dimension 1 could represent the toxicity while dimension 2 could represent the comedy genre.

## A.1 Illustration: Control Algorithm

Consider an example with three users and two movies in Figure A.1. I use singular value decomposition (SVD) to factorize the engagement matrix into two-dimensional user and post latent features. If we interpret dimension 1 of the factor matrices as movies relating to toxic genre, and dimension 2 as movies relating to comedy genre, then the factorization process generates a vector of weights for each user with respect to these attributes. In this example, the weights (or embeddings) reveal that users 1 and 3 have a higher proclivity for toxic movies, while user 2 is likely to rate comedy movies higher. As a result, these attribute weights enable a platform to serve toxic movies to users 1 and 3, and comedy movies to user 2, in order to maximize user satisfaction.

More generally, this factorization process generates a vector of weights for each user with respect to some post attributes, so that a cross product of weights for user and post latent features gives the predicted engagement matrix. The user features produced are latent representations of user behavior revealed in the past, and are produced by minimizing a known loss function using Stochastic Gradient Descent. These latent features are represented as a multi-dimensional embedding vector, where each element in the vector represents the weight each user is predicted to put on some latent post attributes (Athey et al., 2021). These vector-weights in the space of some latent post/ user features generate ranking of various posts that each user is recommended in the future.

## A.2 Illustration: Treatment Algorithm

In this experiment, the content recommendations for the control group are generated as per the usual personalization algorithm. For the treatment group, the algorithm is modified to replace actual user embeddings with randomly selected user embeddings from the control group distribution. In the example below, user 2 is randomly chosen to be treated, and the embeddings for user 2 are replaced with the average of the embeddings for users 1 and 3.

Figure A.2: Matrix representation of vector embeddings, for treated and control users

$$\underbrace{\begin{pmatrix} \frac{1}{\sqrt{10}} & 0 \\ \color{red}{\rho_{21}} & \color{red}{\rho_{22}} \\ \frac{3}{\sqrt{10}} & 0 \end{pmatrix}}_{U} \cdot \underbrace{\begin{pmatrix} \sqrt{10} & 0 \\ 0 & 2 \end{pmatrix}}_{\Sigma} \cdot \underbrace{\begin{pmatrix} 1 & 0 \\ 0 & 1 \end{pmatrix}}_{V^T}$$

Notes: This figure shows the user embeddings for the control group (in black) and the treatment group (in red). The treatment group embeddings, e.g. user 2, are generated by randomly selecting from the distribution of control group embeddings. Effectively, the treatment embeddings are concentrated around the mean of the control group embeddings as the embedding vectors are uniformly drawn from an epsilon ball centered around the mean control embedding on each day of the intervention period.

The embeddings generated for each treated user are concentrated around the average of the embeddings for the control group users because the treatment embeddings are generated by randomly selecting from the distribution of control group embeddings on each day during the intervention by application of the Law of Large Numbers (LLN). This is depicted in Figure A.3, for the simulated (two-dimensional) recommendation algorithm using 1000 users in the control group, 1000 users in the treatment group, 50 posts, 30 days, and 2 dimensions.

It may be expected that in bringing the treatment group embeddings closer to the mean, the treatment biases content exposure among the treated towards more popular posts. This is because the average user's embeddings are likely to be closer to the preferences of the largest number of users on the platform, potentially making them "viral" (Salganik et al., 2006). However, Table A.1 shows that the treatment group was exposed to less popular posts than the control groups because the random numbers picked to generate preference weights for the treatment group were not representative of actual user preferences on the platform.

Figure A.3: Distribution of simulated two-dimensional embedding vectors

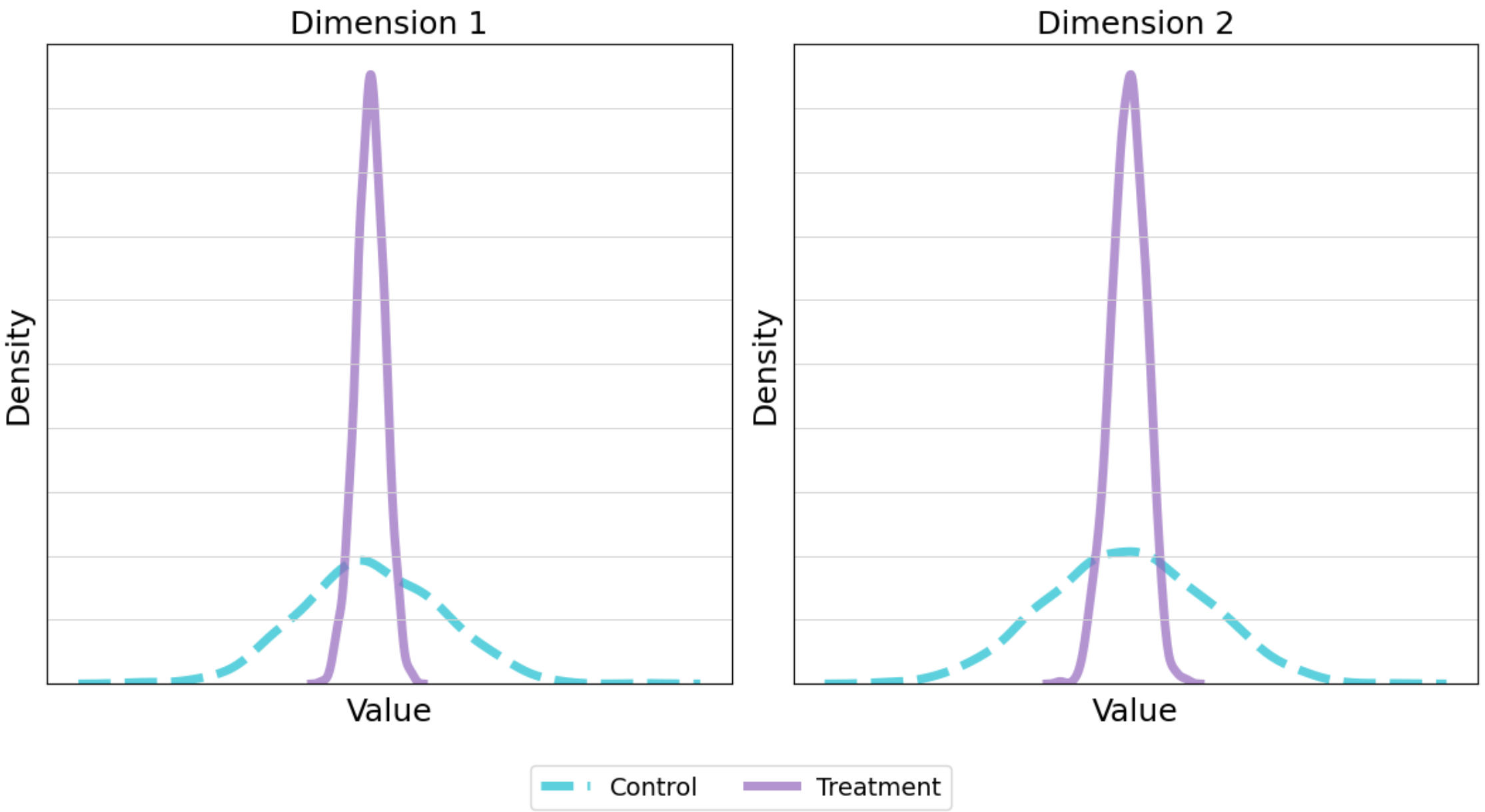

Notes: This figure shows that the two dimensions (components) of the embedding vector follow a Gaussian distribution, where the embeddings were simulated using a simple SVD algorithm and a randomly generated matrix of engagement with 1000 control users, 50 posts, 30 days, and 2 dimensions. Then, the randomly selected embeddings for the treated users are centered around the mean of each embedding dimension, and the spread of control user embeddings is larger than the embeddings generated for treated users as predicted by the LLN. This is because the treatment embeddings are drawn uniformly at random, each day during the intervention. Other properties are shown in Figure A.4.

Table A.1: Popularity of posts viewed by users in the treatment group

| | Views on posts viewed | Likes on posts viewed | Shares on posts viewed |
|---|---|---|---|
| Treatment | -140732.408** | -1549.188** | -3966.425** |
| | (758.901) | (7.527) | (28.271) |
| Constant | 241586.576** | 3093.363** | 5999.583** |
| | (682.964) | (6.491) | (26.112) |
| Obs | | 231814 | |

Notes: This Table shows that, contrary to expectations, the treatment group was exposed to less popular posts than the control group. It is possible that in bringing the treatment group's preference weights closer to those of an average user, the intervention recommended posts that are more appealing to the widest audience. However, this is not observed in the data as the average embedding is not representative of actual users. $p < .0001^{***}$, $p < .01^{**}$, $p < .05^{*}$.

Figure A.4: Properties of content recommendations from a simulated personalization algorithm

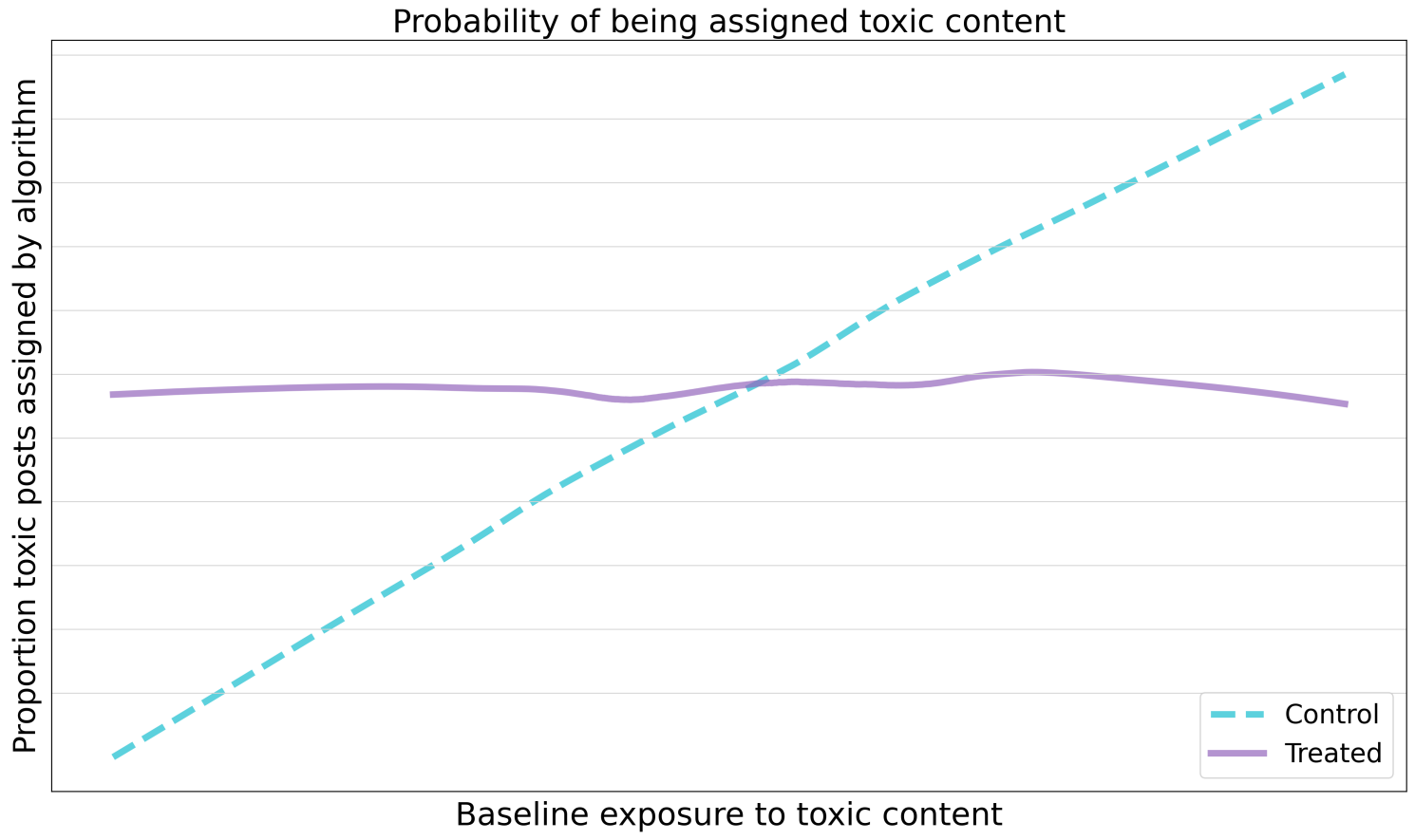

(a) Distribution of the first dimension of the embedding vector across treatment and control

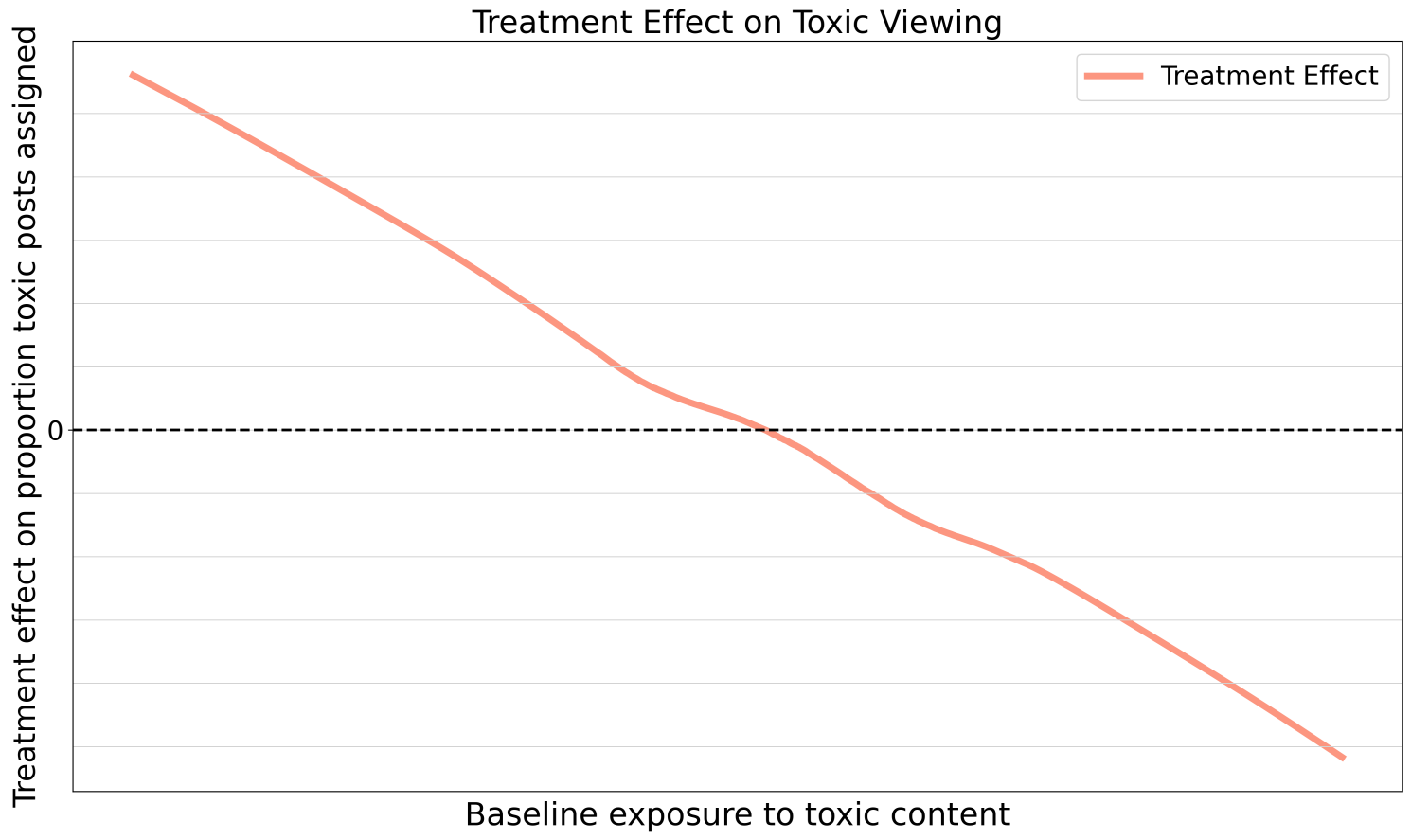

(b) Treatment effect on the embedding values assigned, sorted by baseline embedding value

Notes: This Figure shows that there is a positive correlation between the user preferences (measured using embedding vectors at baseline), with the type of posts recommended by a simple personalization algorithm. The algorithm used to simulate the embeddings for both treatment and control groups uses Singular Value Decomposition to factorize a simulated matrix of engagement with 1000 users in the control group, 1000 users in the treatment group, 50 posts, 30 days, and 2 dimensions. This generates two-dimensional embedding vectors for each user and each post, where each dimension users' preference weights on different post attributes, e.g. tragedy, toxicity, comedy, etc. To fix ideas, this graph shows the first dimension of the embedding vector, which represents the toxicity of the post (as an example). The embedding values for the control users form an upward sloping curve, with respect to user preferences for toxic content, which is the first dimension of the embedding vector. In breaking this correlation between user preferences and the preferred content, treatment is expected to have a smaller effect in absolute terms on users with embeddings closer to the average, at baseline. This is because the treatment algorithm assigns toxic content with the average probability in the control group, as the treated users are simply assigned the average control embedding as shown by the flat curve in Panel (a). On the other hand, users with more extreme preferences experienced bigger absolute effects.

# B Details of Behavioral Model

## B.1 Equilibrium

Users' best response characterizes one of the main outcome variables, i.e. proportion of toxic posts to share, $s^t_{i,\tau} = S^t_{i,\tau}/S_{i,\tau}$ in Lemma 1. The optimal sharing strategy is a combination of user's own tastes and the content she is shown, $q^t_{i,\tau}$, weighted by $(1-\theta)$ and $\theta$, respectively. The distribution of toxic posts on a user's content feed informs her about the type of content that a similar user is engaging with, and is therefore, socially acceptable. The user also decides the number of non-toxic posts she will share, if any, upon viewing posts in their feed.

**Lemma B.1.** *For a utility maximizing agent $i$,*

$$S_{i,\tau} = \frac{1}{2(\eta+\alpha)}\left[2N_{i,\tau}\alpha - \delta\theta(1-\theta)\left(\left(\log p^t_i\right)^2 - 2\log q^t_{i,\tau}\log p^t_i + \left(\log q^t_{i,\tau}\right)^2\right)\right] \tag{9}$$

*That is, total number of posts shared is higher for more engaged users, with higher $N_{i,\tau}$; but is decreasing in the cost of sharing and viewing, $\eta$ and $\alpha$ respectively.*

*Proof.* This follows from the first order condition of the user's utility maximization problem, after substituting optimal $s^t_{i,\tau}$ in the utility function. □

Then, Lemma 2 shows that when users are assigned content randomly, they are likely to spend less time on the platform. This is because the recommendations do not match user preferences, as extreme treated users are recommended the average user's feed. Subsequently, $N^t_{i,\tau}$ in equilibrium helps in determining the total number of posts shared, $S_{i,\tau}$.

For the given timing of the game, I finish characterizing the equilibrium by solving for the algorithm's optimal assignment probabilities. The platform's customization algorithm is trained to maximize the expected number of posts viewed in order to increase eyeballs on advertisement posts that are interspersed on the users' ranked content feed. Therefore, the platform feeds the objective function in (3) to the algorithm, which in turn optimally chooses $q^t_{i,\tau}$ to maximize advertisement revenues as per Lemma 3. The model provides comparative statics, that generate implications tested in the data.

If users were mechanical, they would all have the same behavioral response such that the proportion of toxic posts shared is equal to the proportion of toxic posts viewed, irrespective of treatment status and time period. The comparative statics show that the treatment effects on toxic sharing are unlikely to be mechanical. This means that users put a positive weight on the new information they receive when making sharing decisions. The model predicts mechanical behavior if and only if the influence parameter, $\theta$ equals 0 for mechanical users. I show that this parameter is non-trivial.

**Lemma B.2.** *User $i$ with $N_{i,\tau}, S_{i,\tau} > 0$, is said to behave 'mechanically' when*

$$\theta = \beta = \eta = 0$$

*That is, when $\theta = 0$, the elasticity of the proportion of toxic posts shared with the respect to the proportion of toxic posts viewed is 1.*

*Proof.* If, $\theta = 0$, the utility maximization problem becomes,

$$\max_{s^t,S,N} = \alpha(N-S)^2 - \delta S\left(\log\frac{s^t}{p^t}\right)^2 - \eta S^2 \tag{10}$$

Utility is maximized with respect to $s^t$ when $s^t_{i,\tau} = p^t_i$. Then, by definition,

$$\frac{S^t_{i,\tau}}{S_{i,\tau}} = s^t_{i,\tau} = p^t_i$$

We know that in equilibrium, $q_{i,\tau} = p^t$. Then, assuming users view all the posts they are assigned, we have, $N^t_{i,\tau} = q^t_{i,\tau}N$. Therefore,

$$\frac{S^t_{i,\tau}}{S_{i,\tau}} = \frac{N^t_{i,\tau}}{N_{i,\tau}} = p^t_i = q^t_{i,\tau}$$

Then the treatment implies that,

$$\frac{\partial s^t_{i,\tau}}{\partial \bar{q}^t} = \frac{\partial v^t_{i,\tau}}{\partial \bar{q}^t} = 1$$

where, $v^t_{i,\tau} = \frac{N^t_{i,\tau}}{N_{i,\tau}}$, and, elasticity of toxic sharing with respect to toxic viewing is

$$\frac{\partial s^t_{i,\tau}/\partial \bar{q}^t}{\partial v^t_{i,\tau}/\partial \bar{q}^t} = 1$$

□

When $\theta = 0$, users are considered mechanical as they share a fixed proportion of toxic content they view in each time period. The negation of this implication is also true, and is tested empirically to analyze if user behavior is malleable or sticky. That is, if users do not behave mechanically, then exposure has an influence on user behavior, i.e. $\theta > 0$.

I find that the treatment effect on the proportion of toxic posts shared is distinct from the effect on the proportion of toxic posts viewed. In stating Fact III before, I rejected the hypothesis that the elasticity of toxic sharing with respect to toxic viewing equals 1, with a p-value of 0.002. This shows that there are behavioral responses to diversifying content feeds, even though the influence of exposure is relatively small, as I show with the estimated model parameters.

## B.2 Robustness to Functional Form Assumptions

In Figure B.1, I show that the model's key predictions are unchanged upon relaxing the assumption of the logarithmic utility function. The utility function, similar to Gentzkow and Shapiro (2010), is written as,

$$\max_{s_i^t, S_i, N_i} u(s_i^t, S_i, N_i; q_i^t, c_i) = \beta N_i - \alpha(N_i - S_i)^2 - \eta S_i^2 - \delta S\big[(1-\theta)(s_i^t - p_i^t)^2 + \theta(s_i^t - q_i^t(c_i))^2\big]$$

Figure B.1: Model predictions by user tastes for toxic content, $p^t$

(a) Total number of posts viewed (b) Total number of posts shared

(c) Proportion of shares that are toxic (d) Ratio of toxic shares to toxic views

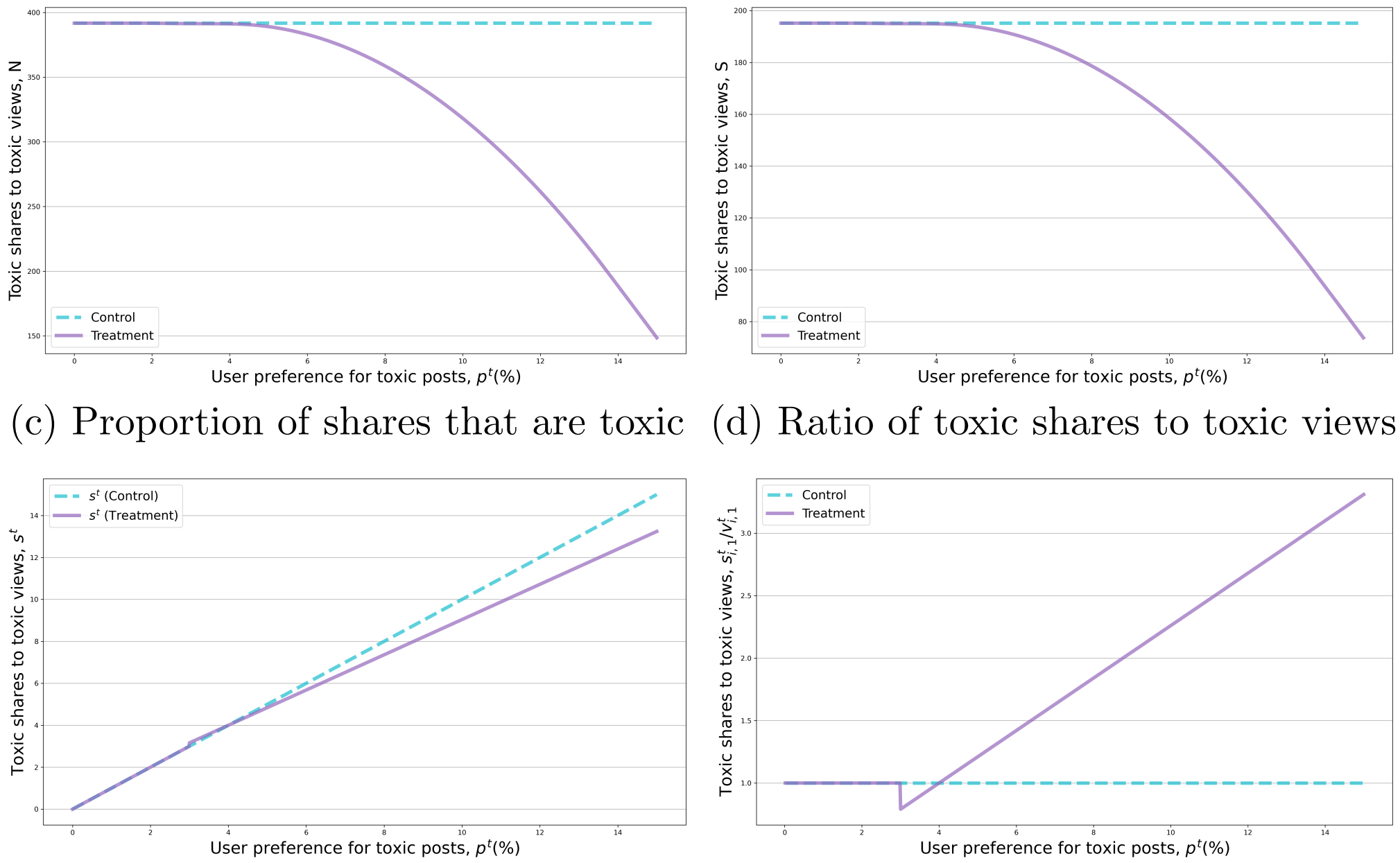

Notes: This figure provides the model's predictions relaxing the functional form assumptions, for key outcomes when the feed is randomized only for users with $p^t \geq \bar{q}^t$, where user type is defined by their tastes for toxic content, $p^t$. Panels (a) and (b) show that more toxic users (toward the right in the $p^t$ distribution) are expected to view and share fewer posts upon being treated. Panel (c) shows that the treatment effect on the proportion of toxic shares is expected to be negative for toxic users. This is due to the larger reduction in total platform usage among toxic users and behavioral changes in the probability of sharing toxic content, both resulting from reduced exposure to such content. Panel (d) predicts that the ratio of toxic shares to toxic views is increasing in $p^t$. These predictions are obtained using calibrated parameters from the structural model by matching moment conditions for heterogeneous users.

# C Details of Structural Estimation

## C.1 Measurement Error

The design of the experiment may induce measurement error in the proportion of toxic content viewed. This is because of *sampling errors*, i.e. users view only a fraction of content in the ranked lists of content (in a set order), that the algorithm generates for them in each time period.

Among *treated users at baseline*, each toxic post *viewed* is assumed to be a Bernoulli trial with probability $q_{i,0}^t$. In each session therefore, the total number of toxic posts viewed is subject to measurement error, on account of the sampling procedure itself. However, since the sampling distribution of toxic views is known, the estimates can be corrected for measurement error using IV approaches (Schennach, 2016).

Consider the following linear classical measurement error set up. Suppose, $v^{t*}_{i,0}(1)$ denote the true proportion of toxic content viewed respectively, that are observed with measurement error in the data.

$$v^t_{i,0}(1) = v^{t*}_{i,0}(1) + ev^t_{i,0}(1)$$

where, $ev^t_{i,0}$ denote the measurement error in the proportion of toxic content viewed. In general, assume that $Cov(v^{t*}_{i,0}(1), ev^t_{i,0}(1)) = 0$. The estimators constructed from the strategy above are therefore, likely to suffer from attenuation bias due to the unobserved measurement error on the right-hand side of the estimating equation. I construct an instrumental variable to address this issue.

Note that $v^t_{i,0}$ is the average of toxic posts viewed over all the posts viewed (of any type) by a user. Consider the proportion of toxic posts viewed out of half of the total posts viewed,

$$v_{i,0}^{\frac{t}{2(-)}}(1) = \frac{\sum_{j=1}^{J/2} t_{ij,0}(1)}{J/2}, \quad v_{i,0}^{\frac{t}{2(+)}}(1) = \frac{\sum_{j=1+J/2}^{J} t_{ij,0}(1)}{J/2}$$

where, $j \in \{1, \ldots, J\}$ indexes each post viewed by user $i$, so that $t_{ij,0}$ is a binary variable indicating whether post $j$ was toxic or not, and $J/2$ indexes the median post. The first expression averages over the first half of posts per user (arranged in a random order) and is henceforth referred to as $half-1$. Similarly, $v_{i,0}^{\frac{t}{2(+)}}$ denotes the fraction of toxic posts out of the second half of the total posts viewed (for brevity, this variable is henceforth referred to as $half2$). However, assuming that the measurement errors pertaining to each half of the posts, are uncorrelated to each other for every user, this fraction computed over the first half of posts can be instrumented by this variable constructed using the second half of the posts. That is to say,

$$Cov(ev_{i,0}^{\frac{t}{2(-)}}(1), ev_{i,0}^{\frac{t}{2(+)}}(1)) = 0 \tag{AME}$$

Under this exclusion restriction, the attenuation bias in a 2SLS estimate of $\gamma_1$ is reduced to zero.

**Proposition C.1.** *Measurement error in average toxic views is corrected by instrumenting the fraction of toxic posts viewed in the first half of posts viewed ($half1$), with the fraction of toxic posts viewed in the second half of posts viewed ($half-2$) by a user in a session.*

*Proof.* The measurement error in these variables constructed using half the viewed posts, is written as

$$v_{i,0}^{\frac{t}{2(-)}}(1) = v_{i,0}^{\frac{t*}{2(-)}}(1) + ev_{i,0}^{\frac{t}{2(-)}}(1)$$
$$v_{i,0}^{\frac{t}{2(+)}}(1) = v_{i,0}^{\frac{t*}{2(+)}}(1) + ev_{i,0}^{\frac{t}{2(+)}}(1)$$

where, as before $Cov(v_{i,0}^{\frac{t*}{2(-)}}(1), ev_{i,0}^{\frac{t}{2(-)}}(1)) = 0$ and $Cov(v_{i,0}^{\frac{t*}{2(+)}}(1), ev_{i,0}^{\frac{t}{2(+)}}(1)) = 0$.

Note the first stage regression using $half - 2$ as the instrumental variable,

$$\begin{aligned} v_{i,0}^{\frac{t}{2(-)}}(1) &= \alpha_0 + \alpha_1 v_{i,0}^{\frac{t}{2(+)}}(1) + \mu_{i,0} \\ &= \alpha_0 + \alpha_1 (v_{i,0}^{\frac{t*}{2}}(+) + ev_{i,0}^{\frac{t}{2(+))}}(1) + \mu_{i,0} \end{aligned}$$

where, $Cov(v_{i,0}^{\frac{t}{2(+)}}(1), \mu_{i,0}) = 0$. Then, any bias in the estimates from the IV specification, due to measurement error in fraction of toxic posts viewed would depend on

$$\begin{aligned} Cov(v_{i,0}^{\frac{t}{2(-)}}(1), v_{i,0}^{\frac{t}{2(+)}}(1)) &= Cov(v_{i,0}^{\frac{t*}{2(-)}}(1) + ev_{i,0}^{\frac{t}{2(-)}}(1), v_{i,0}^{\frac{t*}{2(+)}}(1) + ev_{i,0}^{\frac{t}{2(+)}}(1)) \\ &= Cov(v_{i,0}^{\frac{t*}{2(-)}}(1), v_{i,0}^{\frac{t*}{2(+)}}(1)) \end{aligned}$$

Therefore, the IV approach eliminates measurement error, due to the exclusion restriction stated in (AME). This shows that the IV estimation strategy only depends on the true distribution of the main explanatory variable. □

## C.2 Validation of structural estimates

I validate my estimation procedure that measures the rate at which users update their sharing behavior upon being randomly exposed to more non-toxic content during the intervention period. This model correctly estimates the updating-behavior only for treated users, because for these users, exposure to toxic content in the baseline period is related to the engagement with such content only through the channel of behavioral response.

$\theta$ is not estimable using the sample of control users as these users are always in equilibrium, meaning that viewing and sharing behaviors are perfectly correlated. Therefore, estimates that employ the control sample are expected to be distinct from the main estimates derived using data on treated users only. Figure E.16c shows that this is indeed the case. Additionally, Figure E.16b shows that exposure to toxic content *during the intervention period* has a much smaller effect on the odds of sharing such content. This also validates the main result, because the intervention period exposure is very likely concentrated around the average user's exposure, and is expected to produce different estimates.

## C.3 Calibration

I match moments of the empirical distribution of various outcomes, with the distributions simulated by the model using $\hat{\theta} = 0.16$. This enables calibration of four main parameters of the model: (1) $\beta$, the consumption value of viewing posts, (2) $\alpha$, the disutility from viewing unshareable posts, (3) $\eta$, the cost of sharing an additional post, (4) $\delta$, the utility weight on conformity with societal norms. I use the method of simulated moments to estimate these parameters, using the data $\{s_{i,1}^t, v_{i,1}^t, S_{i,1}, N_{i,1}\}$, which is the proportion of toxic posts shared and viewed respectively, as well as the number of posts shared and viewed, respectively. I compute the empirical mean of each of these outcomes, separately for users with above and below median exposure to toxic content at baseline.

$$\mathrm{E}[X] = \frac{1}{n/2}\sum_{i=1}^{n} x_i$$

Then, the model is defined by the following functions using the equilibrium conditions,

$$s^t(v^t, p^t, \theta) = \left(v^t\right)^{\theta}\left(p^t\right)^{1-\theta} \tag{11}$$

$$N(\delta, \theta, \alpha, \eta, \beta, v^t, p^t) = \frac{-\alpha\delta\theta\left(\log\frac{v^t}{p^t}\right)^2 + \beta(\alpha+\eta)}{2\alpha\eta} \tag{12}$$

$$S(\delta, \theta, \alpha, \eta, N, v^t, p^t) = \frac{\delta\theta(1-\theta)\left[\left(\log p^t\right)^2 - 2\log v^t \log p^t + \left(v^t\right)^2\right] + \frac{N}{\eta}}{2(\alpha+\eta)} \tag{13}$$

where, $v^t$, the proportion of viewed posts that are toxic. The moment conditions for users with lower proclivity to toxic content are given as,

$$\mathrm{E}_1[s^t] = \int_0^m \left(v^t\right)^{\theta}\left(p^t\right)^{1-\theta} \cdot f(v^t)dv^t \tag{14}$$

$$\mathrm{E}_1[N] = \int_0^m N(\delta, \theta, \alpha, \eta, \beta, v^t, p^t) \cdot f(v^t)dv^t \tag{15}$$

$$\mathrm{E}_1[S] = \int_0^m S(\delta, \theta, \alpha, \eta, N, v^t, p^t) \cdot f(v^t)dv^t \tag{16}$$

where, $m$ denotes the median value of the proportion of toxic posts shared, $v^t$, at baseline. Similarly, I write the moment conditions for users with higher proclivity to toxic content as,

$$\mathrm{E}[s^t] = \int_m^\infty \left(v^t\right)^{\theta}\left(p^t\right)^{1-\theta} \cdot f(v^t)dv^t \tag{17}$$

$$\mathrm{E}[N] = \int_m^\infty N(\delta, \theta, \alpha, \eta, \beta, v^t, p^t) \cdot f(v^t)dv^t \tag{18}$$

$$\mathrm{E}[S] = \int_m^\infty S(\delta, \theta, \alpha, \eta, N, v^t, p^t) \cdot f(v^t)dv^t \tag{19}$$

I use numerical integration methods to evaluate these integrals, assuming $v^t \sim EVT1$. Subsequently, the empirical moments are matched with the simulated moments. I use the Nelder-Mead simplex method to estimate the parameters of the model, which converge to the following values in 800 iterations, in this case (Gao and Han, 2012).

## C.4 Discussion

I use a structural model to formalize the analysis for the following reasons. First, the model's equilibrium characterization of user types allows an analysis of the treatment effect on toxic sharing. This is because user preferences are not observed, but are inferred from the assignment probabilities of toxic posts at baseline, when the system is assumed to be in equilibrium. Second, the model provides micro-foundations for user engagement with harmful content. In the model, users update their view of socially acceptable content in order to conform with other users of similar type (Fang and Loury, 2005). Treated users were served an average user's feed, and thought to update their opinion of what other users of the same type might be viewing.

Third, the model decomposes the treatment effect into two channels with various counterfactual policies that cannot be implemented in the data. This includes **(1)** the endogenous response in the total number of posts viewed and shared, and **(2)** the influence of exposure to diverse content on the proportion of toxic posts shared. Treated users endogenously responded to diversity in content assignment by viewing fewer posts, or spending less time on the platform. However, the model shows that this effect was more pronounced for users who were previously engaged with more extreme toxic content (henceforth, toxic users). Finally, the structural model estimates the malleability of user behavior by identifying $\theta$.

### C.4.1 Simplifying Assumptions

Figure C.1: Testing simplifying assumption in action-signalling model

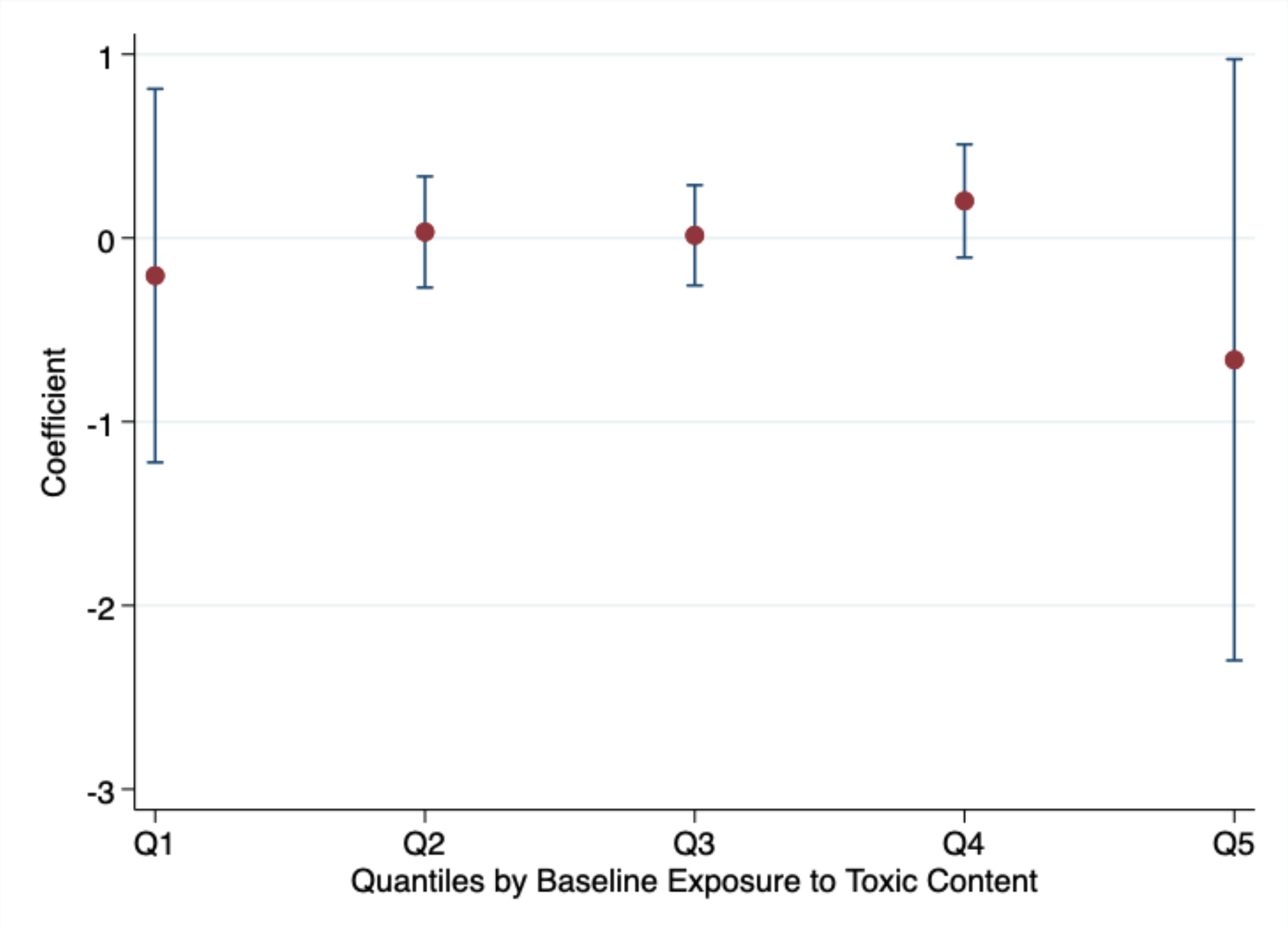

Notes: This Figure shows that I cannot reject the hypothesis that heterogeneous users are influenced by exposure at equal rates. The plot was obtained by estimating the main structural equation from the model, in different sub-samples of users, based on their baseline toxic exposure.

In writing the utility for structural estimation, I made four simplifying assumptions: **(1)** consumption as well as signalling utilities are additively separable for each content type, **(2)** user behavior in the action-signalling model is updated at some constant rate $\theta$ across all users, **(3)** deviating from the reference point of own and society's tastes generates disutility which is quadratic in nature.

The first assumption rules out strategic complementarities between different kinds of posts. This is tenable due to the fact that users scrolling through social media are assumed to be viewing posts one at a time, and do not know if the next post they will view is going to be toxic or not.

I test the second assumption, that is, the constant effects with respect to the rate of updating user behavior in the action-signalling model. Figure C.1 supports this assumption, as the estimates of $\theta$ obtained from samples of different types of users are indistinguishable from each other. Finally, I have assumed the costs of using social media to be quadratic for ease of computation. The model does not stray far from the literature on strategic interactions in the presence of social signalling, especially when such models are estimated using structural methods, for instance, in Butera et al. (2022).

### C.4.2 Identifying Assumptions

The main identifying assumption is that the probability of sharing toxic content is equal in steady state equilibrium. Since, the control users remain in steady state during the intervention and were chosen randomly, I test this assumption in the sample of control users,

$$s^t_{i,0} = s^t_{i,1} \tag{IA}$$

and estimate parameters of the following regression using normalized proportion of toxic content shared in each time period,

$$s^t_{i,1}(D_i = 0) = \delta_1 s^t_{i,0}(D_i = 0) + \varepsilon_{i,1}$$

Under this identifying assumption (IA) I expect $\delta_1 = 1$ in the sample of control users. Table C.1 shows that I cannot reject the hypothesis $\delta_1 = 1$ in the measurement error corrected case.

Table C.1: Testing identifying assumption in structural model using control sample

| | (1) | (2) | (3) |
|---|---|---|---|
| | Probability of sharing toxic post during intervention period | | |
| Proportion of toxic posts shared at baseline | 0.112*** (0.012) | | 0.820*** (0.091) |
| Proportion of toxic posts among first half of posts shared at baseline | | 0.290*** (0.057) | |
| $N$ | 52663 | 52663 | 52663 |

Notes: This Table tests the identifying assumption, derived from the steady state condition $s^t_{i,0} = s^t_{i,1}$. That is, all else equal, the probability of sharing toxic content for each user is expected to be equal in each time period. Column (3) shows that the measurement error corrected estimates of the slope coefficient is close to 1. The sample includes control users who shared at least one post at baseline. Robust standard errors in parentheses. $p < .0001^{***}$, $p < .01^{**}$, $p < .05^{*}$

# D Proofs for Theoretical Framework

## Proof of Lemma 1

For a utility maximizing agent $i$,

$$s_{i,\tau}^t = \left(q_{i,\tau}^t\right)^\theta \left(p_i^t\right)^{1-\theta}$$

That is, users place a weight of $\theta$ social norms, as perceived by the user through her feed, while choosing the proportion of posts shared that are toxic.

*Proof.* The claims follow from users' first order condition (with respect to $s_{i,\tau}^r$) from the utility maximization problem. □

## Proof of Lemma 2

For a utility maximizing agent $i$,

$$N_{i,\tau} = \frac{1}{2\alpha\eta}\left[\beta(\alpha+\eta) - \delta\alpha\theta(1-\theta)\left(\log\frac{q_{i,\tau}^t(c_i)}{p_i^t}\right)^2\right]$$

That is, users view a smaller number of posts when there is a mismatch between their preferences and the algorithmically generated preferences, $q_{i,\tau}^t \neq p_i^t$.

*Proof.* I begin by substituting the optimal sharing behavior (from Lemmas 1 and B.1) into the utility function. User's first order condition, with respect to the total number of posts viewed generates the required expression. This shows that $N_{i,\tau}$ is decreasing in $\left(\log\frac{q_{i,\tau}^t}{p_i^t}\right)^2 > 0$. Therefore, $N_{i,\tau}$ is maximized when $q_{i,\tau}^t = p_i^t$. □

## Proof of Lemma 3

$$q_{i,\tau}^t = p_i^t \tag{20}$$

That is, the algorithm assigns toxic posts with probability equal to user's intrinsic tastes for toxic content.

*Proof.* This follows directly from the first order conditions of an algorithm that is set to maximize $N_{i,\tau}$ in (3), by choosing $q_{i,\tau}^t$ optimally. The same result follows if the algorithm's objective is defined more broadly, choosing $q_{i,\tau}^t$ to maximize $N_{i,\tau}^t$, or $S_{i,\tau}^t$, or some linear combination of the two. This is because the number of posts viewed and shared is decreasing in $\left(\log\frac{q_{i,\tau}^t}{p_i^t}\right)^2 \geq 0$, which equals zero when $q_{i,\tau}^t = p_i^t$. □

## Proof of Proposition 1

Let $D_i$ be a binary variable indicating treatment status, so that $D_i = 1$ for treated users. Then, for users with $p_i^t > \bar{q}^t > 0$,

$$v_{i,1}^t(D_i = 1) - v_{i,1}^t(D_i = 0) = \bar{q}^t - q_{i,\tau}^t < 0$$

where, $v_{i,1}^t$ is the proportion of posts viewed that are toxic for user $i$ at $\tau = 1$.

*Proof.* $\bar{q}^t - q_{i,\tau}^t < 0$, for users with higher baseline exposure to toxic content with $p_i^t > \bar{q}^t$ as $\bar{q}^t$ is the average user's probability of being assigned toxic content. Then, assuming users view everything they are assigned, $v_{i,1}^t(D_i = 1) = \bar{q}^t$. The fact that $v_{i,1}^t(D_i = 0) = q_{i,1}^t = p_i^t$ under the equilibrium condition for the control group completes the proof. □

## Proof of Proposition 2

For user $i$ with $\alpha, \eta, N_{i,\tau} > 0$, and $p_i^t > \bar{q}^t$,

$$\frac{\partial^2 N_{i,\tau}}{\partial p_i^t \partial \bar{q}^t} \geq 0$$

That is, the reduction in the total number of posts viewed, on account of the treatment, is larger for users with higher proclivity to toxic content.

*Proof.* Lemma 2 implies

$$N_{i,\tau} = \frac{1}{2\alpha\eta}\left[\beta(\eta+\alpha) - \delta\alpha\theta(1-\theta)\left(\log\frac{q_{i,\tau}^t}{p_i^t}\right)^2\right]$$

With random content assignment during the intervention period ($\bar{q}^t$),

$$\frac{\partial N_{i,\tau}}{\partial \bar{q}^t} = \frac{-1}{2\alpha\eta}\left[\frac{2}{\bar{q}^t}\delta\alpha\theta(1-\theta)\log\frac{\bar{q}^t}{p_i^t}\right]$$

Note that, $p_i^t > \bar{q}^t$ is both necessary and sufficient for the derivative to be positive. That is, for users with higher proclivity to toxic content, randomly increasing the probability of assigning such content increases the number of posts viewed. Consider, the cross derivative with respect user tastes, $p_i^t$ gives,

$$\frac{\partial^2 N_{i,\tau}}{\partial p_i^t \partial \bar{q}^t} = \frac{1}{2\alpha\eta}\left[\frac{2}{\bar{q}^t p_i^t}\delta\alpha\theta(1-\theta)\right] \geq 0$$

because $\theta \in [0,1]$, $\bar{q}^t, p_i^t \in (0,1)$, and $\alpha, \eta, \beta, \delta > 0$.

Then, for $p_i^t > \bar{q}^t$, random increases in probability of assigning toxic content increases the number of posts viewed, and the increase is larger for more toxic users. Conversely, when exogenous reductions in $\bar{q}^t$ decrease the number of posts viewed for toxic users, the reduction is larger for more toxic users. □

## Proof of Proposition 3

For user $i$ with $\eta, N_{i,\tau}, S_{i,\tau} > 0$,

$$\frac{\partial^2 s_{i,\tau}^t}{\partial p_i^t \partial \bar{q}^t} \geq 0$$

That is, the treatment effect on the proportion of toxic posts shared is negative and smaller for users with higher proclivity to toxic content.

*Proof.* From Lemma 1 shows that,

$$s_{i,\tau}^t = \left(q_{i,\tau}^t\right)^{\theta} \left(p_i^t\right)^{1-\theta}$$

Then, we can see that

$$\frac{\partial^2 s_{i,\tau}^t}{\partial p_i^t \partial \bar{q}^t} = \theta(1-\theta)\left(q_{i,\tau}^t\right)^{\theta-1} \left(p_i^t\right)^{-\theta} \geq 0$$

for $\theta \in [0, 1]$, and $q_{i,\tau}^t, p_i^t \in (0, 1)$. □

## Proof of Proposition 4

For user $i$ with $\alpha, \eta > 0$, $\theta \in [0, 1]$ and $p_i^t > \bar{q}^t > 0$,

$$\frac{\partial^2 (s_{i,\tau}^t / \bar{q}^t)}{\partial p_i^t \partial \bar{q}^t} \leq 0 \tag{21}$$

That is, marginal increases in the average probability of assigning toxic content leads to smaller changes in the proportion of toxic shares out of toxic views for users who prefer such content.

*Proof.* As before,

$$\frac{\partial^2 (s_{i,\tau}^t / \bar{q}^t)}{\partial p_i^t \partial \bar{q}^t} = -(1-\theta)^2 q^{\theta-2} p^{-\theta} \leq 0$$

□

## Proof of Proposition 5

Assume,

(A1) User behavior is in equilibrium at baseline, $s_{i,0}^t = p_i^t$

(A2) The system is in steady state, $\log s_{i,0}^t = \log s_{i,1}^t$

Then, for some updating parameter $\theta$ for all treated users $i$,

$$\mathrm{E}\left[\log\left(\frac{s_{i,1}^t}{s_{i,0}^t}\right)\middle| D_i = 1\right] - \mathrm{E}\left[\log\left(\frac{s_{i,1}^t}{s_{i,0}^t}\right)\middle| D_i = 0\right] = \kappa - \theta\mathrm{E}[\log v_{i,0}^t | D_i = 1]$$

where, $\kappa$ is some constant, and $D_i$ indicates treatment status.

*Proof.* Lemma 1 and equation (8) gives the optimal sharing function,

$$s_{i,1}^t = \left(v_{i,1}^t(q_{i,1}^t)\right)^{\theta}\left(s_{i,0}^t\right)^{1-\theta} e^{\mu w_{i,1}^t} \tag{22}$$

where, $w_{i,1}^t$ is some *iid* taste-based shock. Then, the steady state condition gives

$$s_{i,0}^t = s_{i,1}^t\left(v_{i,1}^t(q_{i,1}^t), s_{i,0}^t, w_{i,1}^t\right) \tag{23}$$

$$\implies s_{i,0}^t = v_{i,1}^t e^{w_{i,1}^t \frac{\mu}{\theta}} \tag{24}$$

Substituting $s_{i,0}^t$ into (22) for control users,

$$s_{i,1}^t = v_{i,1}^t e^{w_{i,1}^t \frac{\mu}{\theta}} \text{ and } s_{i,0}^t = v_{i,0}^t e^{w_{i,0}^t \frac{\mu}{\theta}} \tag{25}$$

On the other hand, for treated users, substituting the steady state condition (23) into the optimal sharing function (22) gives

$$s_{i,1}^t = \left(v_{i,1}^t\right)^{\theta}\left(\bar{q}^t\right)^{1-\theta} e^{w_{i,1}^t \frac{\mu}{\theta}} \tag{26}$$

Moving subscripts to previous period, we get

$$s_{i,0}^t = \left(v_{i,0}^t\right)^{\theta}\left(\bar{q}^t\right)^{1-\theta} e^{w_{i,0}^t \frac{\mu}{\theta}} \tag{27}$$

Let, $D_i$ indicate treatment status of user $i$. Then, plugging in values from equations (23), (26), and (27), into the difference in the difference in sharing behavior across $\tau = 1$ and $\tau = 0$ across treatment and control groups,

$$\mathrm{E}\left[\log\left(\frac{s_{i,1}^t}{s_{i,0}^t}\right)\middle| D_i = 0\right] - \mathrm{E}\left[\log\left(\frac{s_{i,1}^t}{s_{i,0}^t}\right)\middle| D_i = 1\right] = \theta\Big(\underbrace{\log \bar{q}^t}_{=\text{constant}} - E[\log v_{i,0}^t | D_i = 1]\Big)$$

This follows from the equilibrium condition for control users where $v_{i,1}^t = v_{i,0}^t = s_{i,0}^t$. First differences between treatment and control on the left-hand side account for unobserved heterogeneity in sharing behavior. □

# E Supplementary Figures

Figure E.1: Landing page and trending tab on TipTop

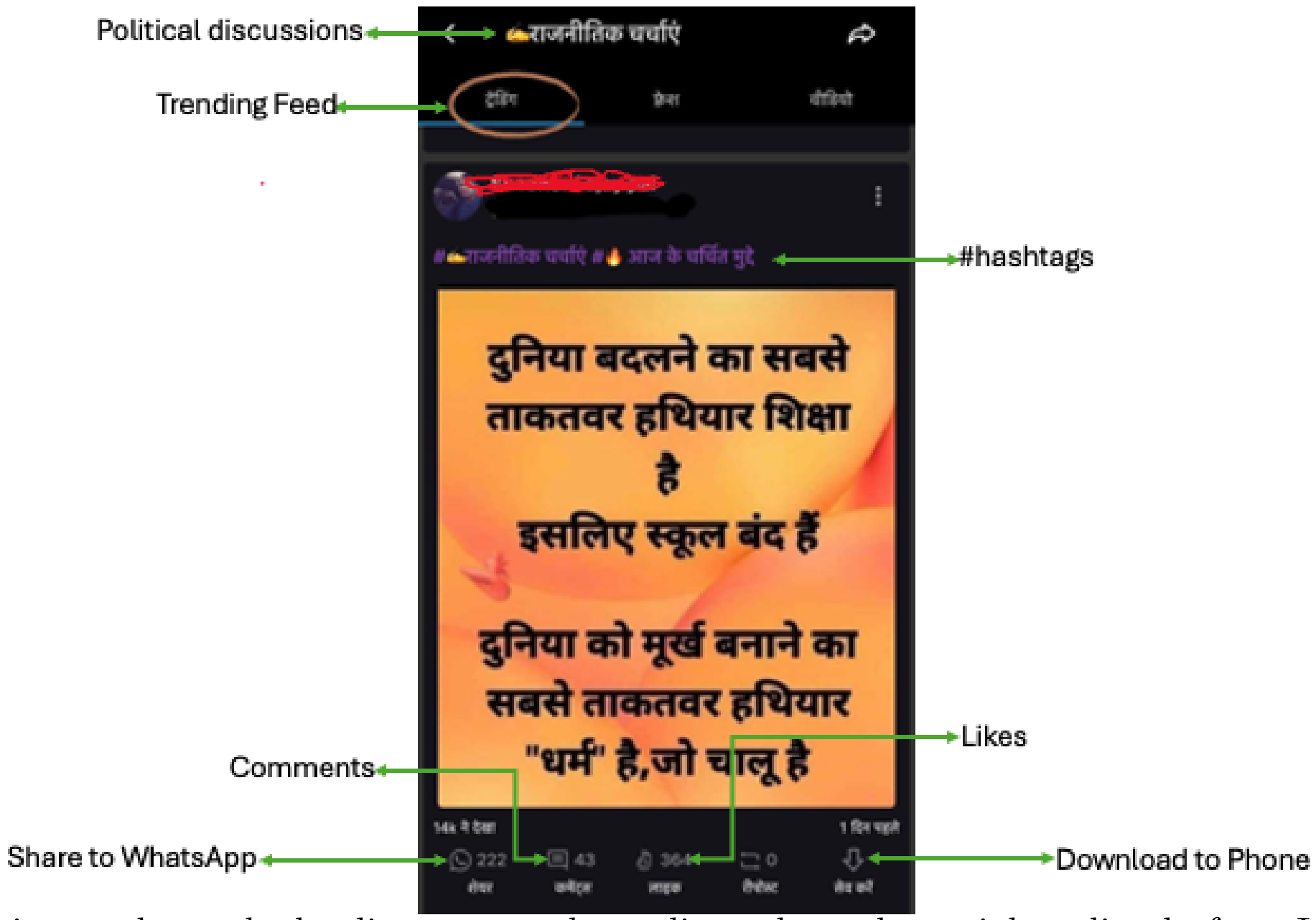

Notes: This image shows the landing page and trending tab on the social media platform I collaborate with, TipTop. Users see a feed of image posts and the creator generated hashtags on the landing page, much like Instgram. Users can share, comment, like, or download the post to their phones. Sharing refers to sharing on WhatsApp, and not on the platform itself, for instance on user's own profile. This makes TipTop's interface very different from other platforms like X (formerly, Twitter), where users can share posts with their followers, through their profile on the platform. A user can see other users who liked and commented on a post, but not the users who shared the post. TipTop posts are classified into broad categories or genres like 'politics' (in this image), 'devotional,' 'romance,' 'Bollywood,' 'greetings,' and 'educational.'

Figure E.2: Network structure and content consumption on TipTop

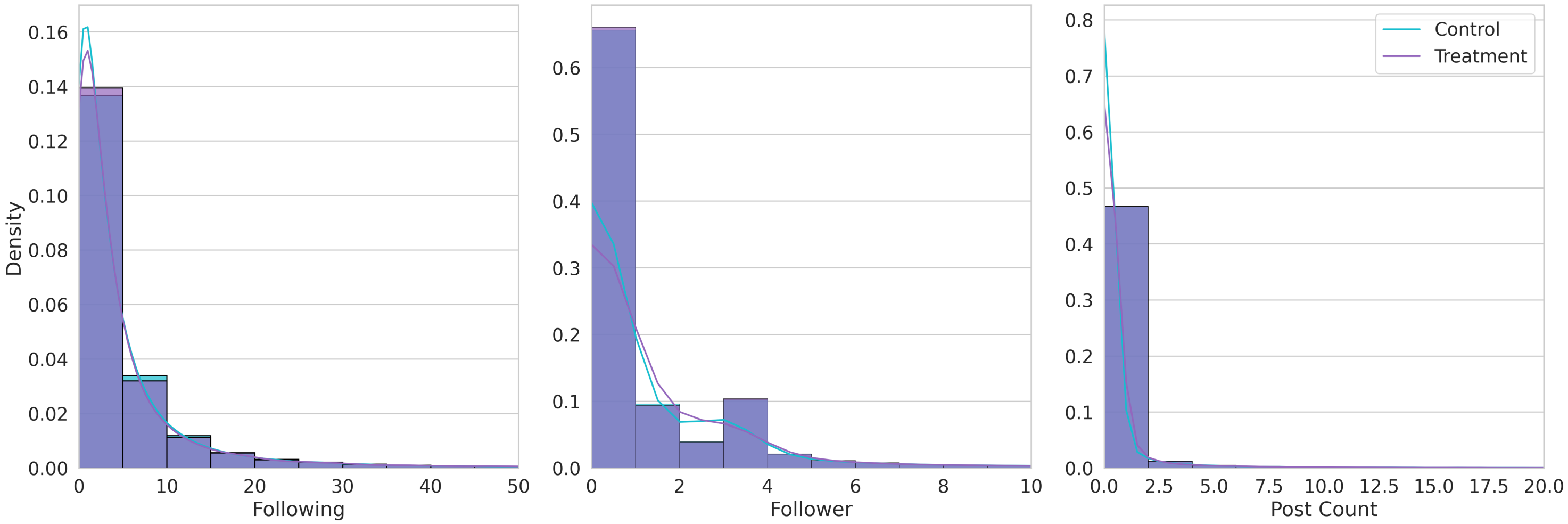

Notes: This figure shows that TipTop's user interface and social networks are distinct from other social media platforms. In particular, despite having an option to follow other users, the platform is content-based, and users interact with content rather than with other users. This is in contrast to platforms like X (formerly, Twitter), where users can engage with users they 'follow.' Additionally, TipTop is distinct from other platforms because users consume content produced by influencers (and not by friends and family), and very rarely produce their own content, as seen in the number of posts produced by users in the experimental sample, in the panel on the right-corner.

Figure E.3: Confusion matrices for different cut-offs in toxicity scores

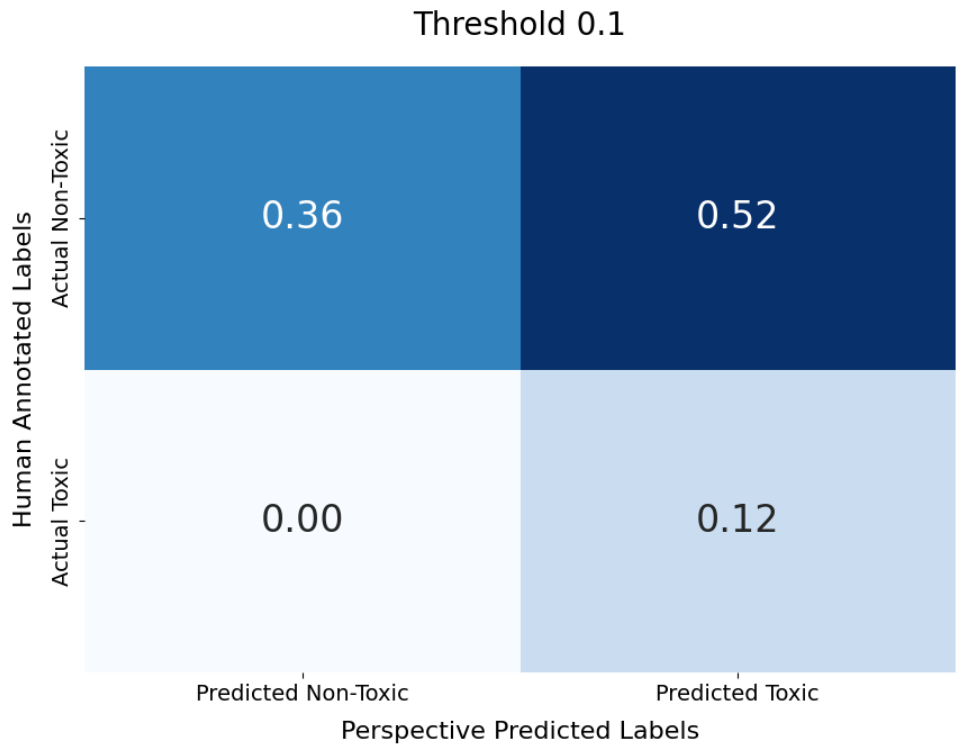

(a) Toxicity score cut-off = 0.1. F1-score = 0.321

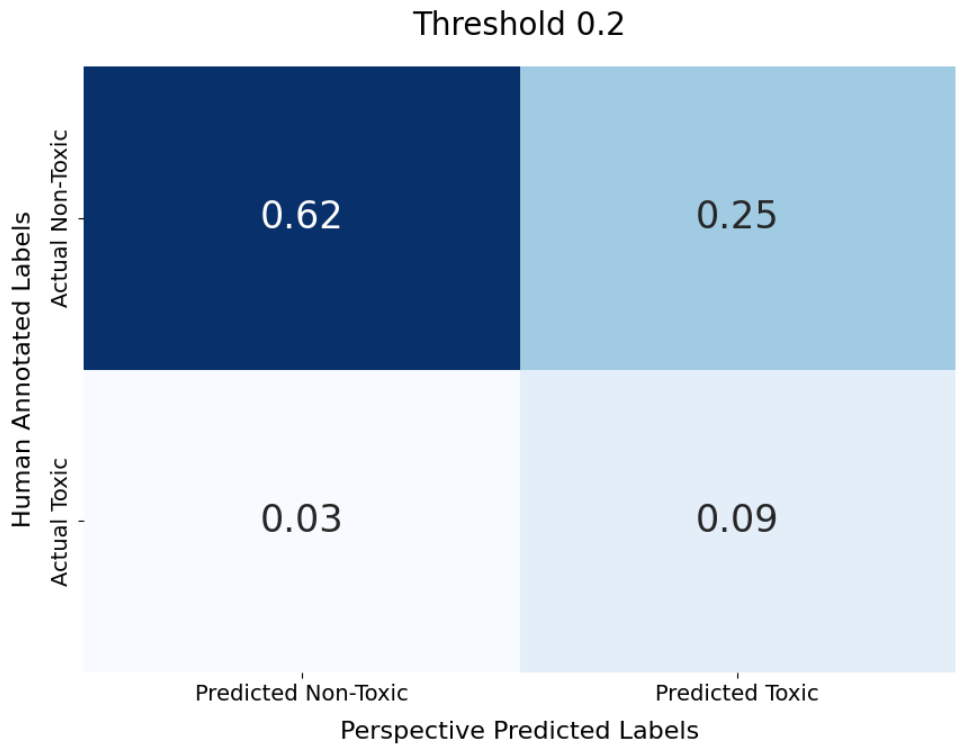

(b) Toxicity score cut-off = 0.2. F1-score = 0.391

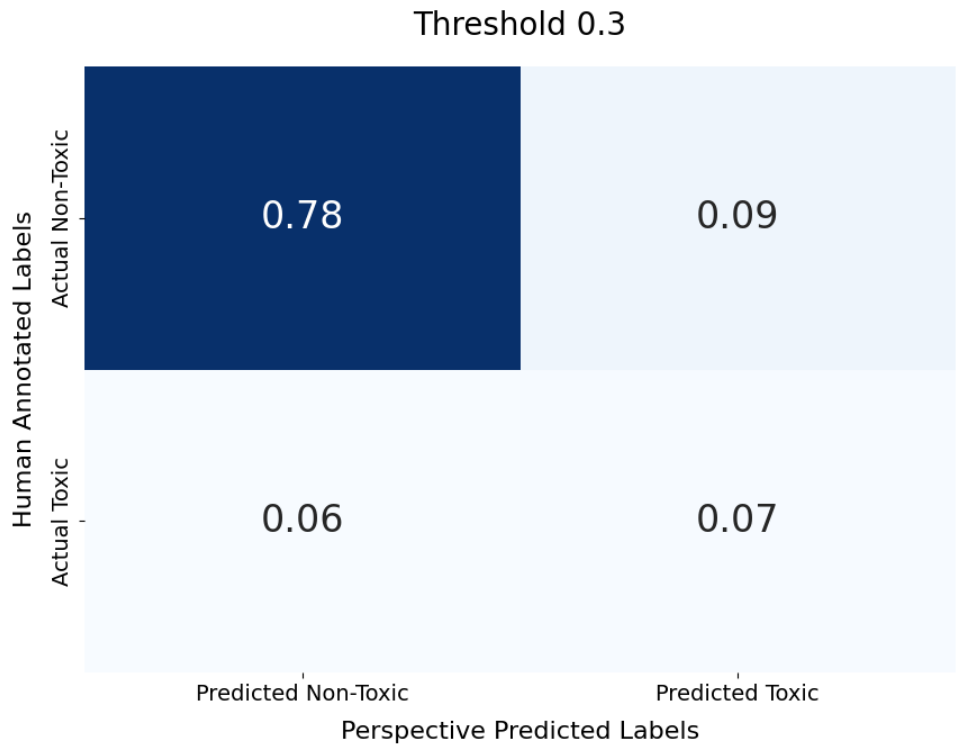

(c) Toxicity score cut-off = 0.3. F1-score = 0.459

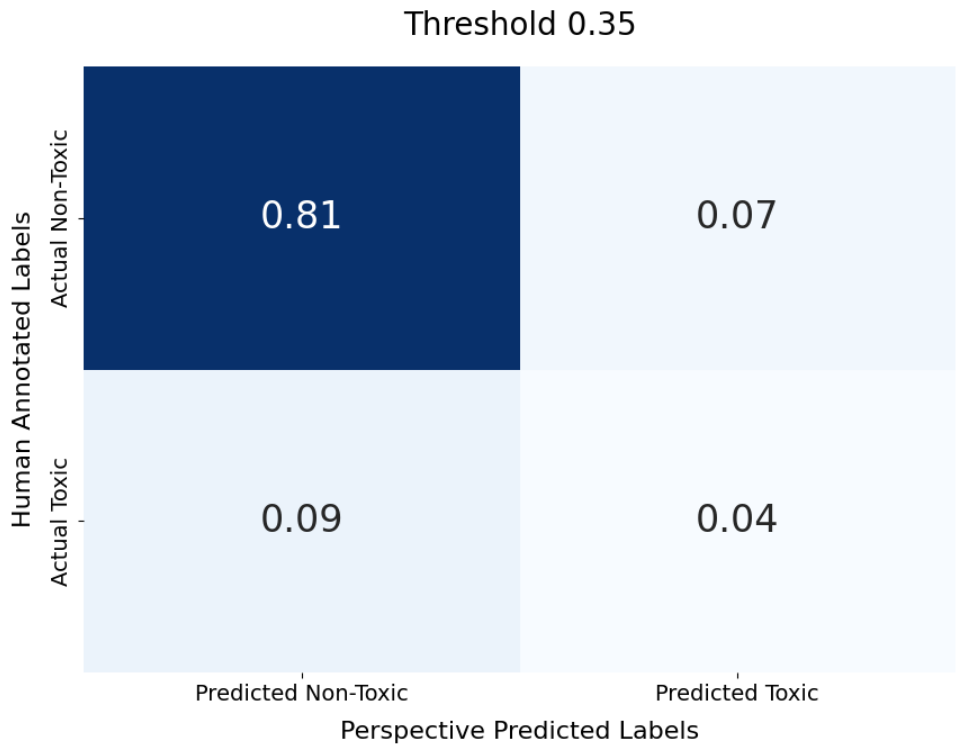

(d) Toxicity score cut-off = 0.35. F1-score = 0.341

Notes: These confusion matrices show Type I and Type II errors for four thresholds for classifying a post as toxic, namely 0.1, 0.2, 0.3, 0.35. User posts were assigned continuous toxicity scores using the Perspective API, and then classified as being toxic or not using different thresholds. These scores were compared with posts annotated as hateful by two human annotators hired at Brown University. The trade-off involved in choosing a threshold is between correctly identifying toxic posts and minimizing false positives, as a lower cut-off will increase the number of false positives, and a higher cut-off will increase the number of false negatives. The threshold of 0.2 was chosen because toxic posts are correctly identified at this threshold with high accuracy. This is the most important criterion for the classification task, because toxic posts are a rare occurrence in the data. I also validate this cut-off using the F1 score, which is highest for the 0.2 threshold among thresholds that maximize classification of truly toxic posts (0.1 and 0.2). Results are robust to different thresholds.

Figure E.4: Change in distribution of exposure to different topics

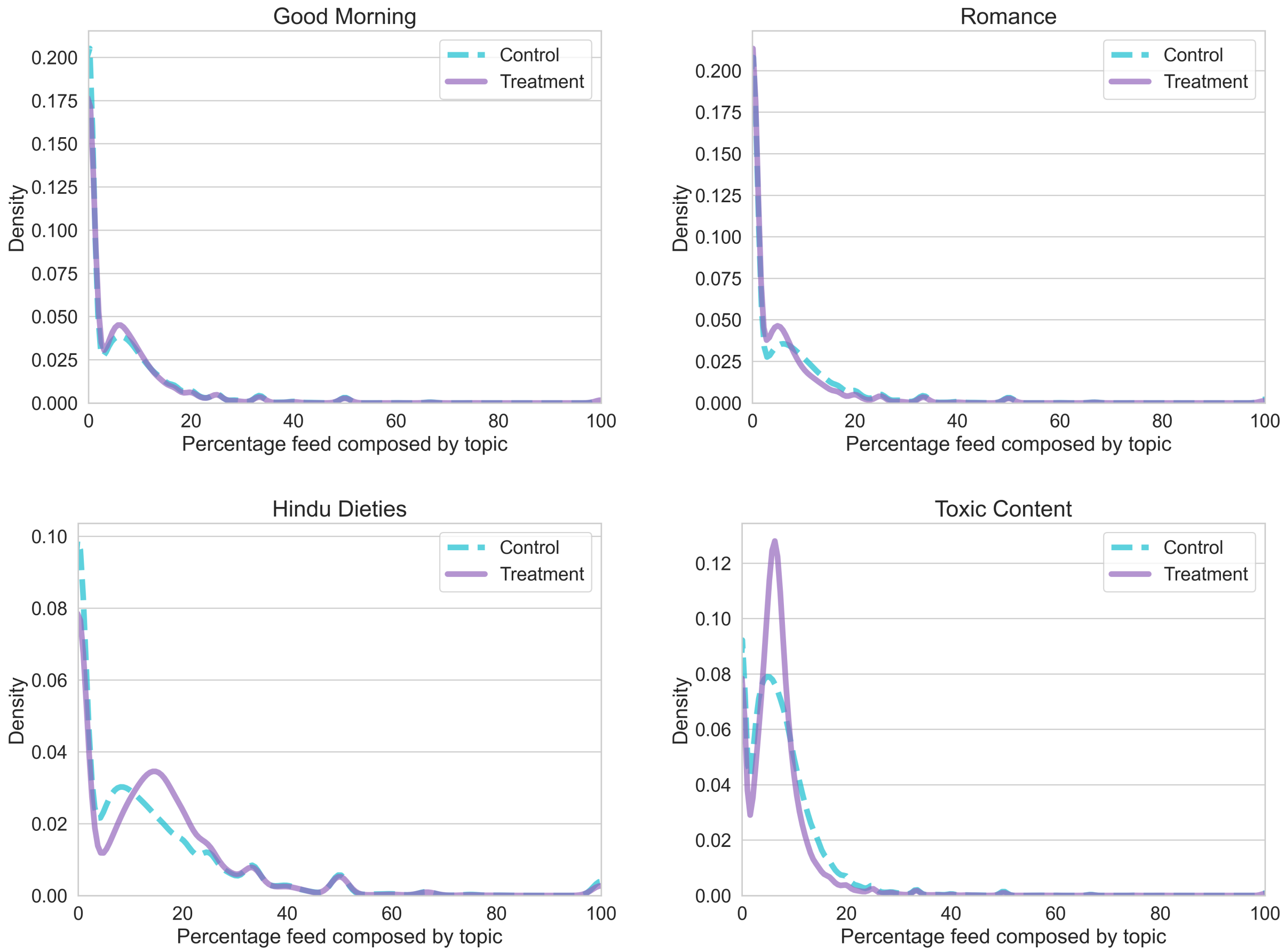

Notes: This Figure shows the change in the distribution of exposure to different types of content for treated users. The top two panels show minimal change in the distribution of exposure to good morning posts and romantic content. The bottom two panels show the change in the distribution of exposure to toxic content, as well as religious posts on various Hindu deities (one for each day of the week). The distribution of toxic and religious content is more concentrated and shifted to the left for toxic content, and shifted to the right for religious content. The topics were modeled using an LDA topic model, and the dominant topic in each post was computed. Classification of posts into genres was done by TipTop engineers using an undisclosed algorithm. Post topic and toxicity are not mutually exclusive.

Figure E.5: Change in distribution of exposure to different types of content

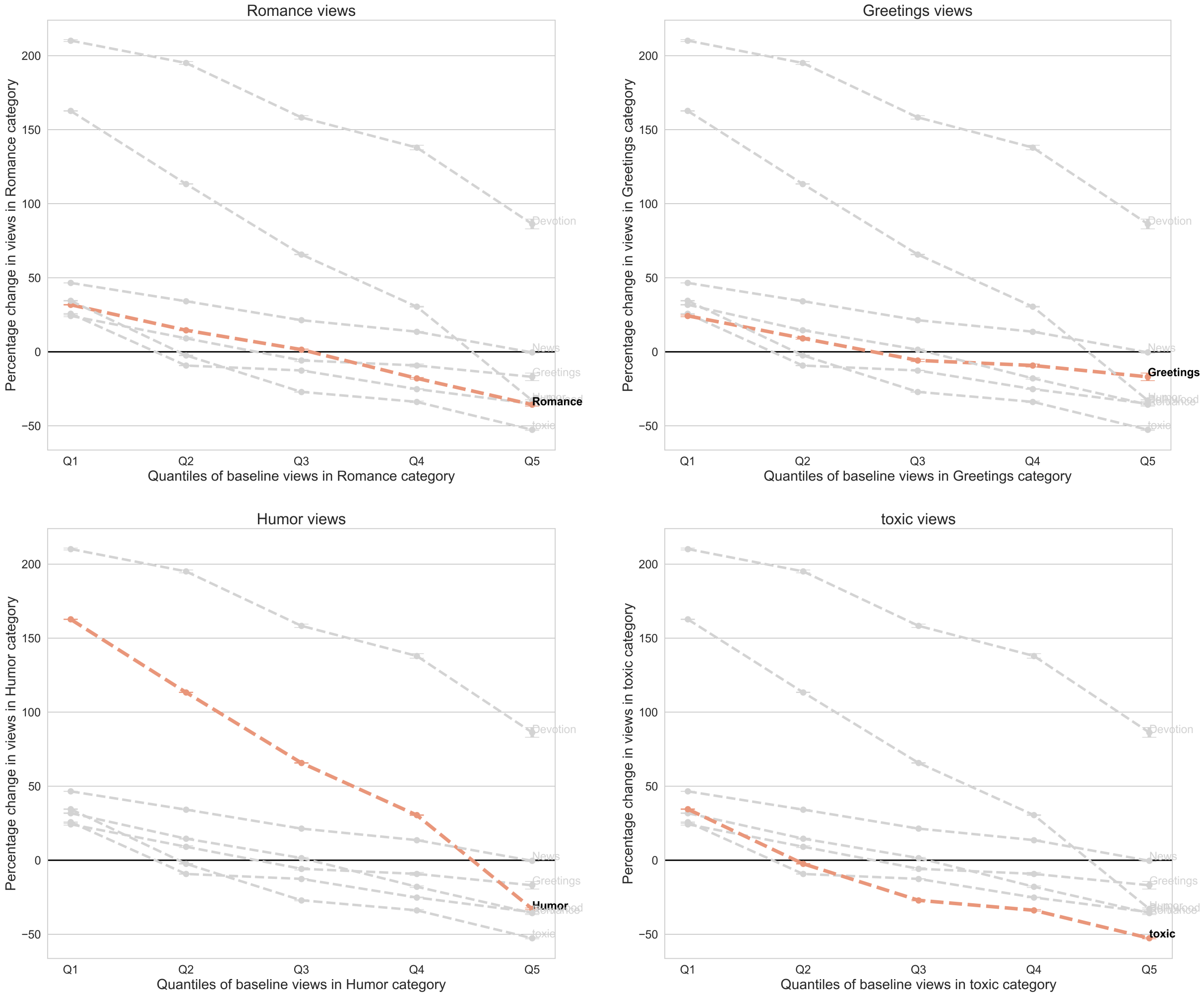

Notes: This Figure shows that the percent reduction in exposure to toxic content is largest for users with the highest exposure to such content at baseline, compared with change in exposure for top users in other broad content categories. This provides an experiment to measure the effect of exogenously reducing exposure to harmful content on user behavior. In each plot, the quantiles represent the percentage of feed that is constituted by a given content type at baseline. For example, Q5 of the top left panel consists of users who saw the most romantic content at baseline, and similarly Q5 users in the top right panel had the highest exposure to greetings content at baseline. Post toxicity and content categories are not mutually exclusive.

Figure E.6: Distribution of exposure to and engagement with toxic content during intervention period

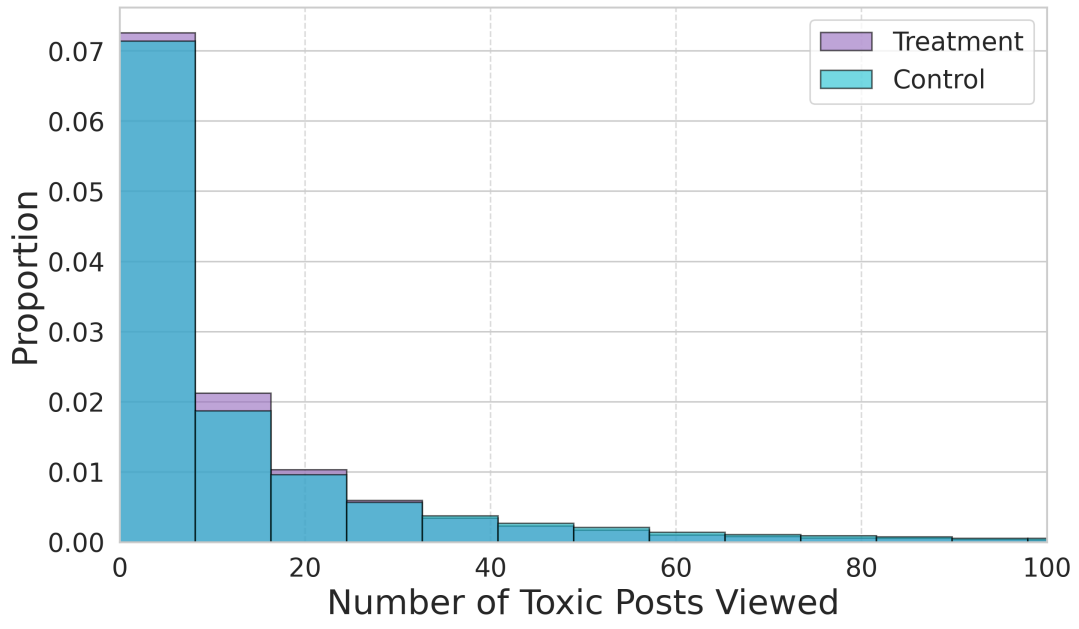

(a) Number of toxic posts viewed

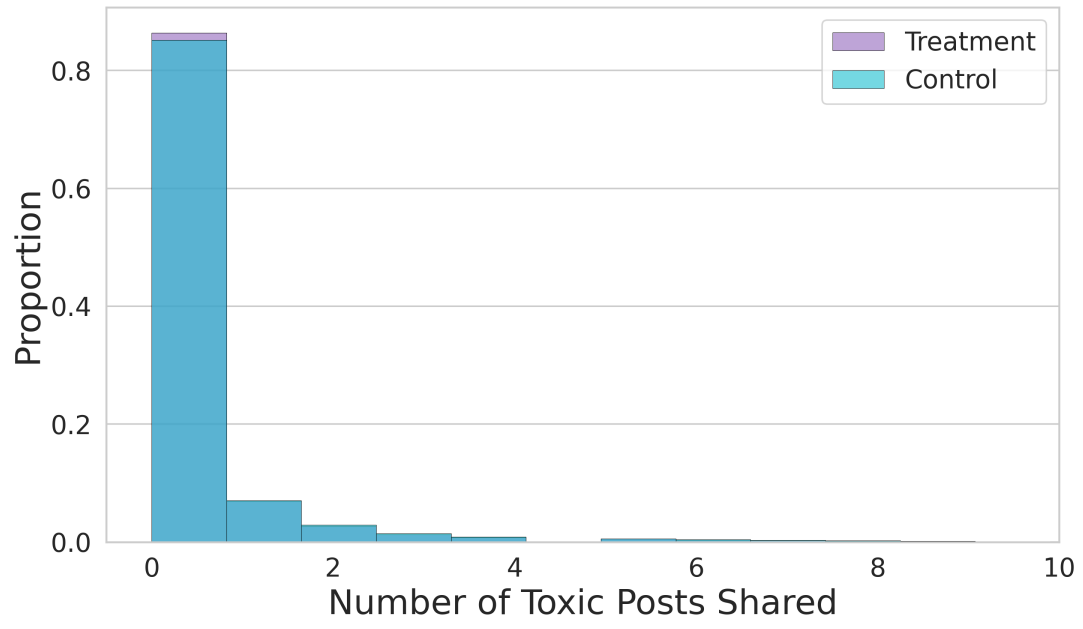

(b) Number of toxic posts shared

Notes: This Figure plots the raw data on the number of toxic posts viewed by users during the intervention period. Panel (a) shows that treated users were less likely to view toxic posts when using the dichotomous toxicity measure constructed using the 0.2 threshold. Panel (b) plots the raw data on the number of toxic posts shared and also finds that treated users shared a lower number of toxic posts on average. However, these plots do not show which users were most affected by the treatment.

Figure E.7: Preferences over redistribution, by user type

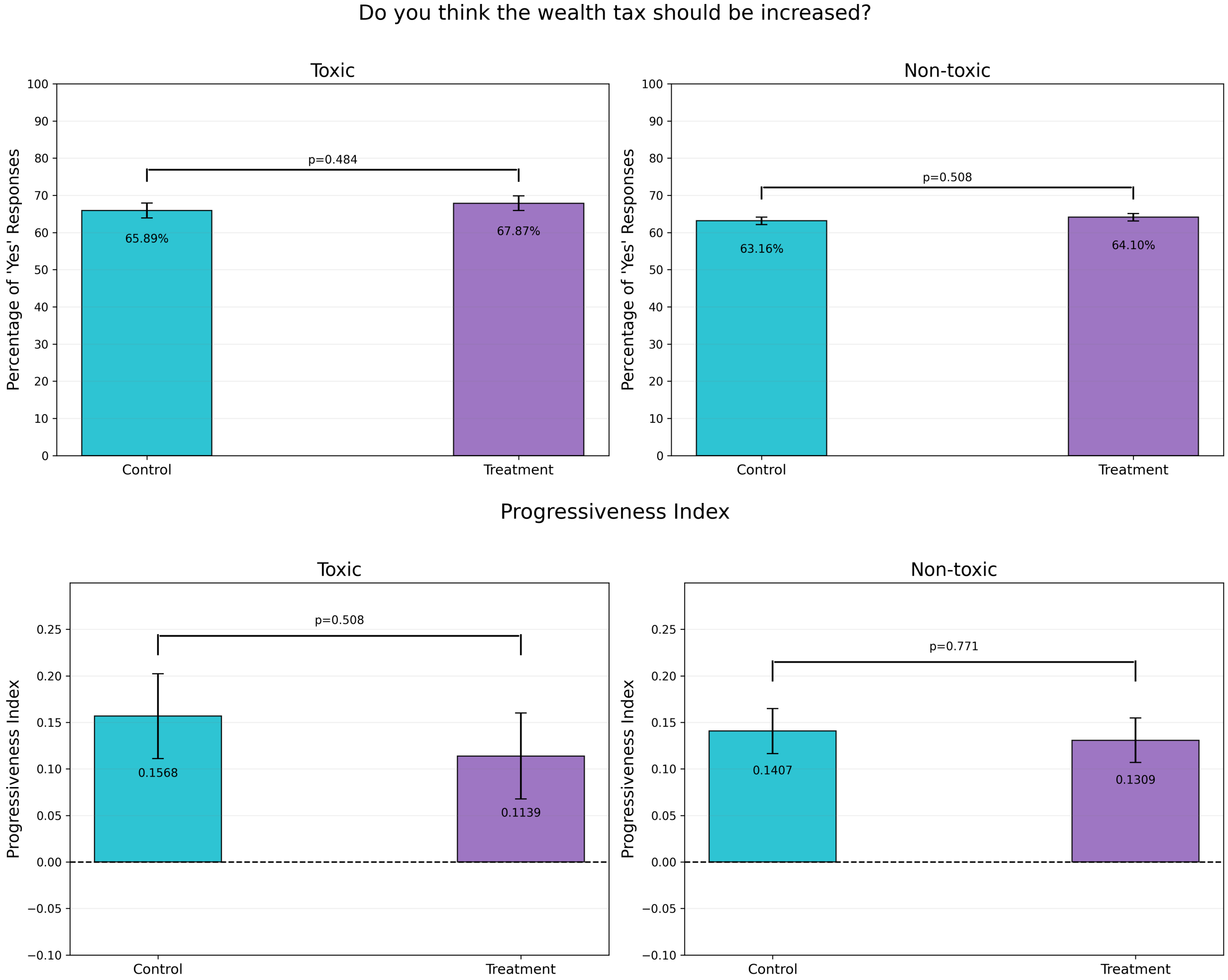

Notes: This Figure shows that the treatment did not affect users' preferences over redistribution, as reflected in the survey data (N = 8,387). This is consistent with the main results that the intervention led to very limited behavioral changes. Users in the random sample survey were asked if they thought that wealth should be redistributed, and the surveyor explained what a wealth tax would mean, in the telephonic surveys. Respondents could say 'Yes,' 'No,' or 'Don't know.' The uncertain responses were dropped before computing these percentages, standard errors, and p-values. Based on these responses, I also created a progressiveness index, from respondents' answers to different questions relating to affirmative action and wealth redistribution for out-group members. Respondents were further divided into toxic and non-toxic groups, based on their exposure to toxic content at baseline in the admin data. If a user's exposure to toxic content was above the median level at baseline, they were classified as a toxic user (left panel). Baseline toxicity in views does not predict treatment status on account of the random assignment and sample selection.

Figure E.8: Treatment effects on toxic behavior as percentage of total engagement (views and shares)

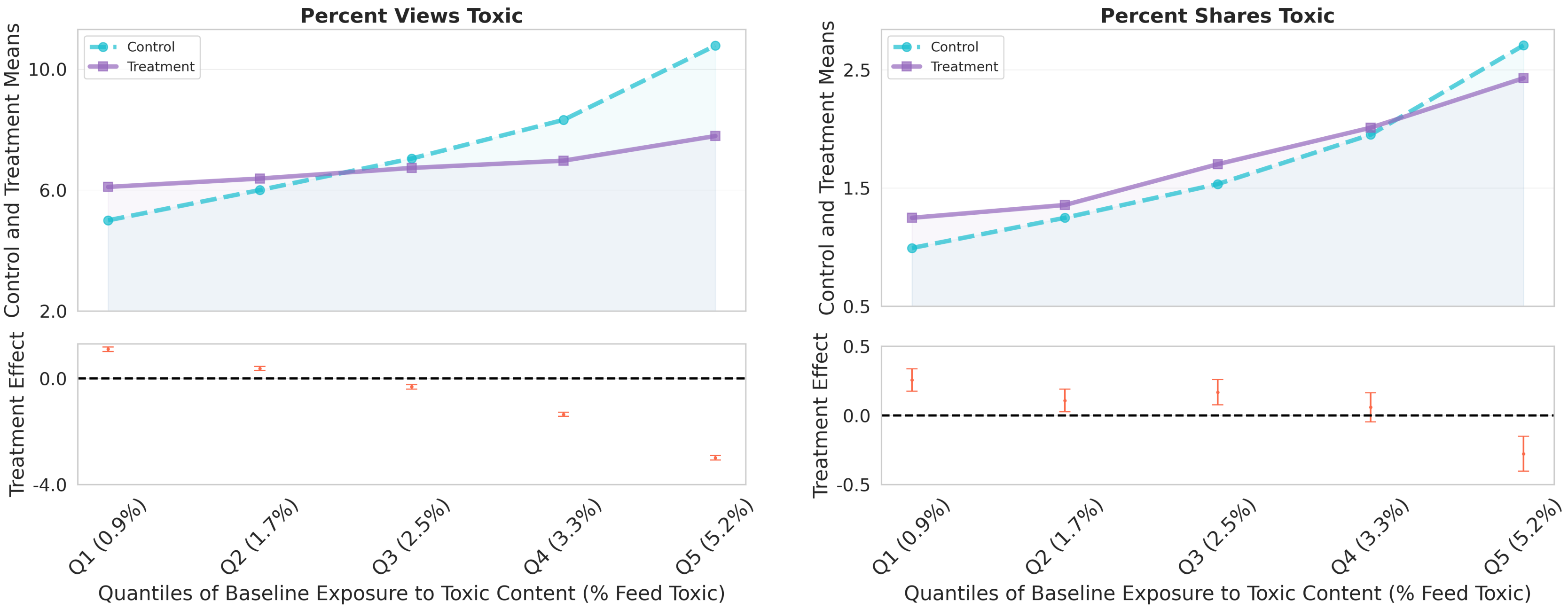

Notes: This figure shows that the treatment effect on the proportion of posts shared that are toxic is non-negative for all users except those in Q5 (with the highest exposure to toxic content at baseline). This is true in the cases of Q3 to Q4, where user type is toxic enough that the treatment effect on the proportion of views that are toxic is negative (left panel). The axis corresponding to the bottom plots show the magnitude of treatment effects (as coefficient plots), while the top panel is scaled according to the control mean of the outcomes for each quantile. All regressions were run at the user level with robust standard errors.

Figure E.9: Experimental effects on overall platform activity

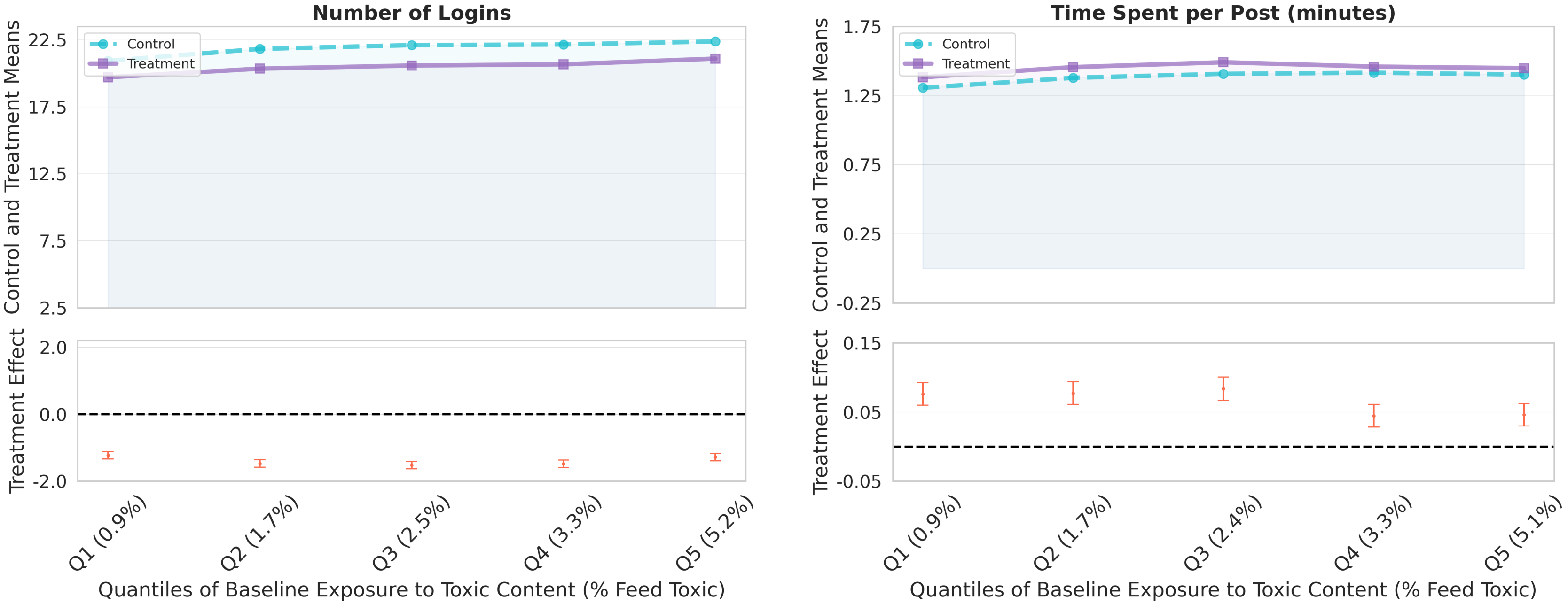

Notes: This figure shows that the treatment effect, on the overall activity of users on the platform is negative for all users. There is a decrease in the number of times a user logged in (left panel), but the effects on logins are indistinguishable across user types. There is an increase in time spent per post (right panel), but the increase is much smaller for more toxic users (Q4 and Q5). This means such users are more likely to scroll through posts, and spend less time on each post. The axis corresponding to the bottom plots show the magnitude of treatment effects (as coefficient plots), while the top panel is scaled according to the control mean of the outcomes for each quantile. All regressions were run at the user level with robust standard errors.

Figure E.10: Experimental effects on toxic behavior with respect to political posts only

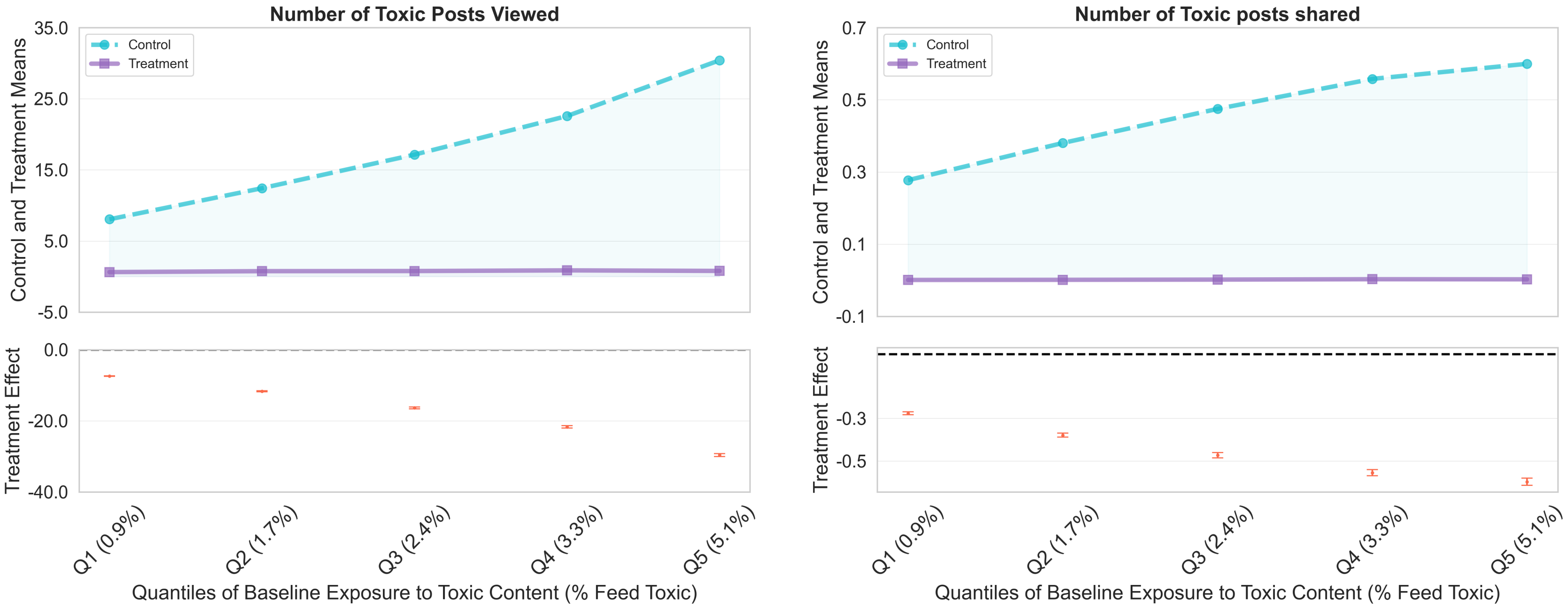

Notes: This figure shows that the treatment effect on the number of toxic posts viewed and shared is negative in the sample of political posts only. This is done to understand the effect of the intervention on user behavior with respect to anti-minority content. Further, the effect is consistently largest for Q5 users. It is unclear why Q1 users do not witness increases in exposure to toxic content under the political category but could possibly be driven by churn due to changing definition of political content. The algorithm that classifies posts into tag genres like politics is undisclosed. The axis corresponding to the bottom plots show the magnitude of treatment effects (as coefficient plots), while the top panel is scaled according to the control mean of the outcomes for each quantile. All regressions were run at the user level with robust standard errors.

Figure E.11: Experimental effects on liking behavior, by user type

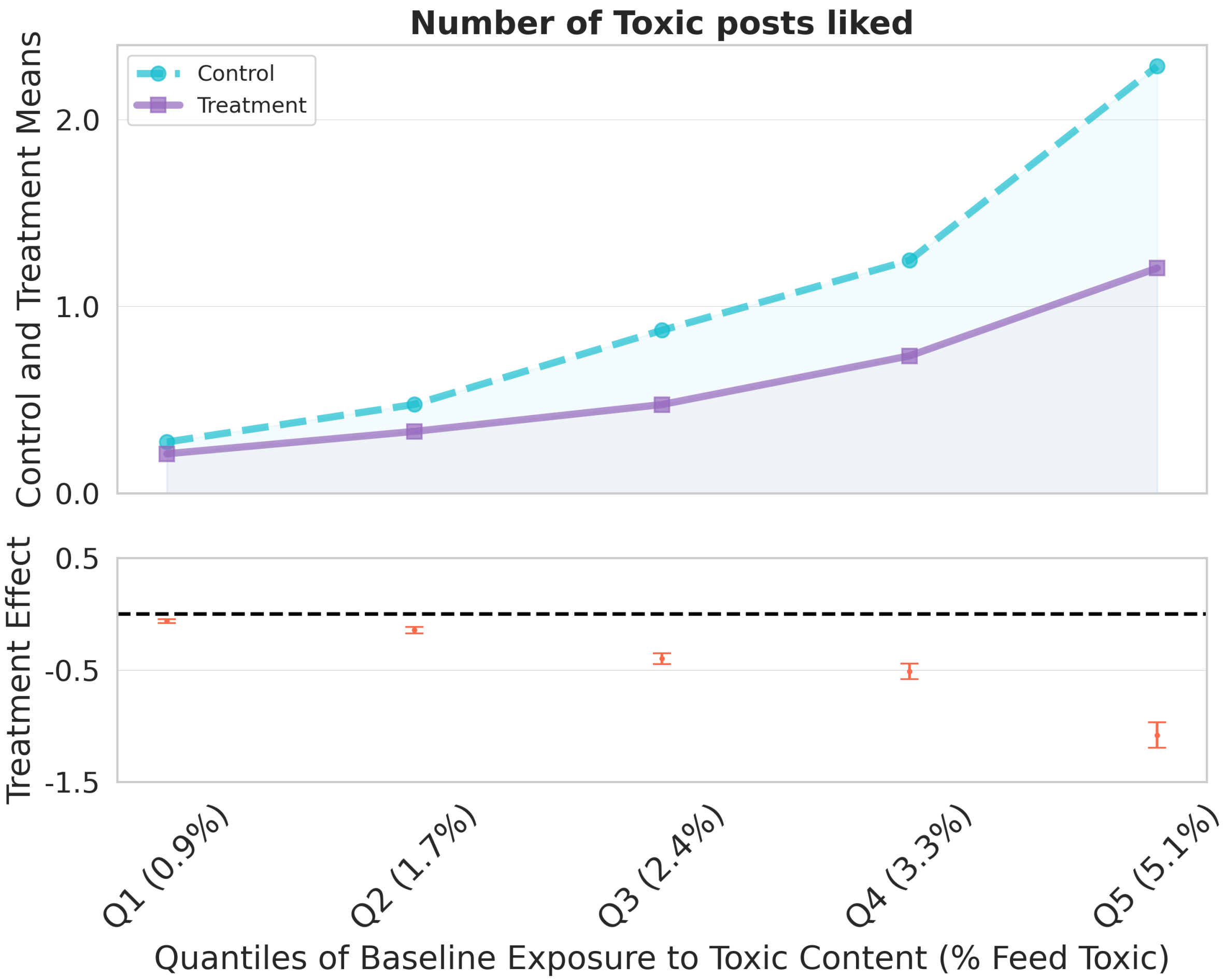

Notes: This figure shows that the treatment effect on the number of toxic posts liked is negative and the smallest for Q5 users. The figure illustrates that sharing behavior indeed captures user preferences because the magnitudes of change with respect to both liking and sharing of toxic content are in the same direction in the respective quantile of users. The axis corresponding to the bottom plots show the magnitude of treatment effects (as coefficient plots), while the top panel is scaled according to the control mean of the outcomes for each quantile. All regressions were run at the user level, and inference about the treatment effects is based on robust standard errors.

Figure E.12: Auxiliary evidence of users seeking out preferred content

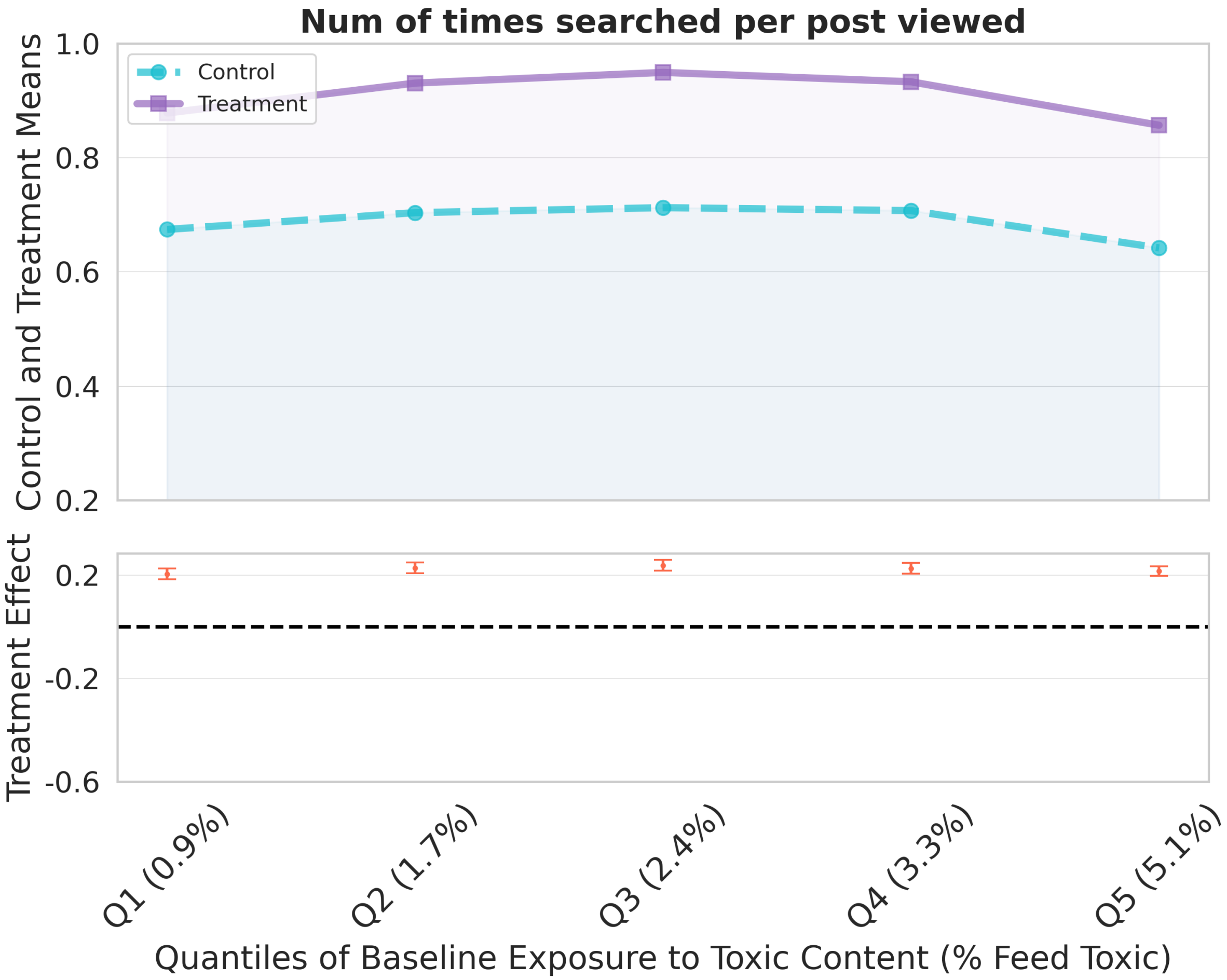

Notes: This figure shows that treated users employed the text-search feature on the platform to seek out preferred content. Further, Q5 users are more likely than Q1 users to search content for every post they view on the landing page, although this difference is not significant. The bottom panel shows the magnitude of treatment effects (coefficient plots), and the top panel shows the means of the outcome variable in each quantile of treatment and the control group. All regressions were run at the user level, and robust standard errors were computed.

Figure E.13: Change in sharing relative to toxic views for continuous toxicity scores

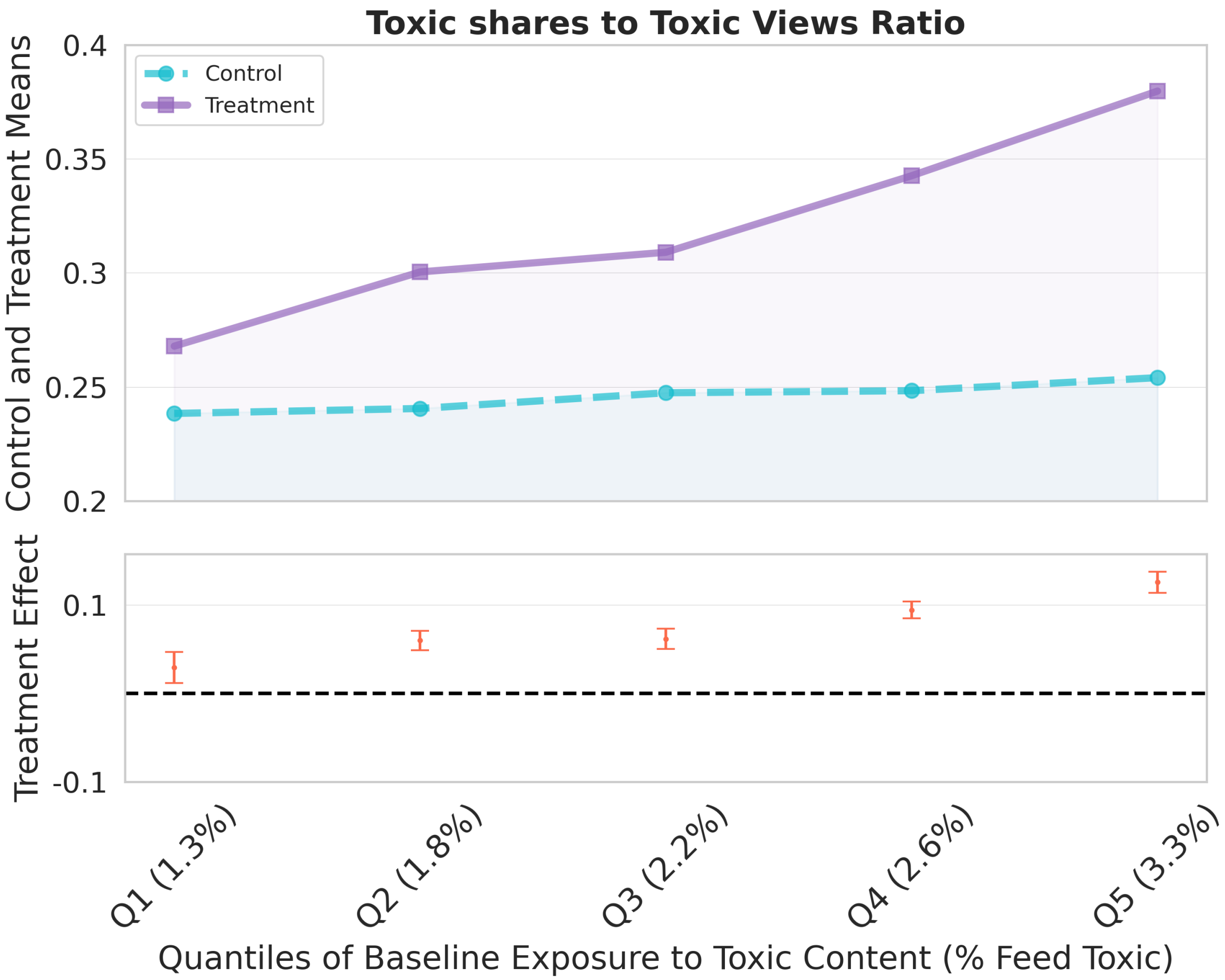

Notes: This Figure replicates the main result that the ratio of toxic shares to toxic views is increasing in users' baseline toxicity, when toxicity is measured by averaging over the continuous toxicity scores of posts viewed or shared. On average, the ratio of toxic shares to toxic views is higher among the treated, and this is driven by users with higher proclivity to toxic content. The axis corresponding to the bottom plots show the magnitude of treatment effects (as coefficient plots), while the top panel is scaled according to the control mean of the outcomes for each quantile. All regressions were run at the user level with robust standard errors.

Figure E.14: Empirical decomposition of treatment effects on toxic shares

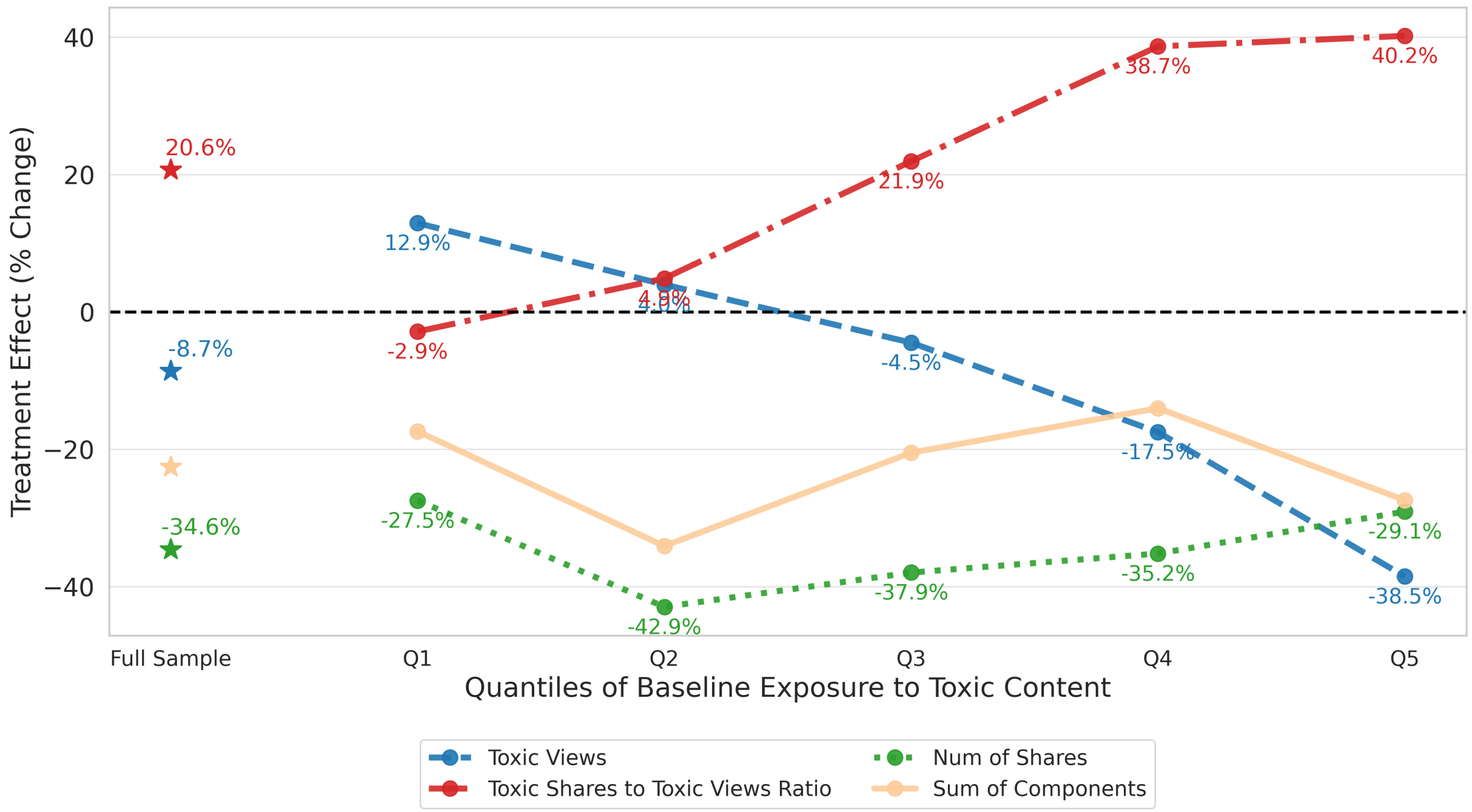

Notes: This Figure shows the empirical decomposition of the treatment effect on toxic shares and finds that the treatment effect for Q5 users is driven by the effect on exposure to toxic content and overall disengagement with the platform. The treatment effect is decomposed into the disengagement effect due to the intervention as well as behavioral change induced by exposure to diverse information. The proportion of views that are toxic represents the effect of exposure, the ratio of toxic shares to toxic views represents the behavioral change effect, and the change in total shares shows the disengagement effect. The exposure effect dominates the effect on toxic shares on average, and for Q5 users, and the behavior change goes in the opposite direction of disengagement effect in both these samples. Finally, the absolute effect of disengagement is larger than that of behavior change for Q5 users.

Figure E.15: Model predictions by user tastes for toxic content, $p^t$

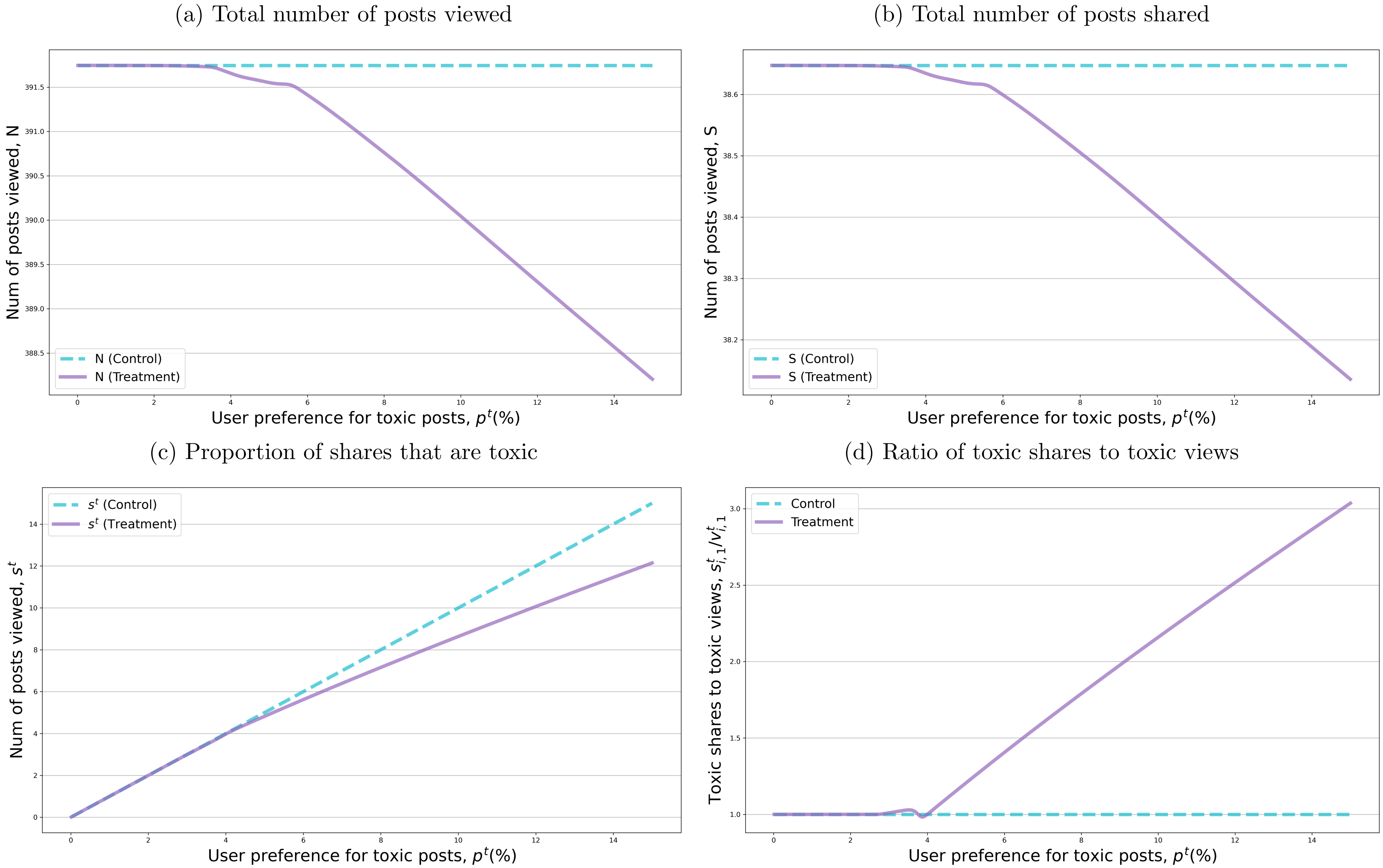

Notes: This figure provides the model's predictions for key outcomes when the feed is randomized only for users with $p^t \geq \bar{q}^t$, where user type is defined by their tastes for toxic content, $p^t$. Panels (a) and (b) show that more toxic users (toward the right in the $p^t$ distribution) are expected to view and share fewer posts upon being treated. Panel (c) shows that the treatment effect on the proportion of toxic shares is expected to be negative for toxic users. This is due to the larger reduction in total platform usage among toxic users and behavioral changes in the probability of sharing toxic content, both resulting from reduced exposure to such content. Panel (d) predicts that the ratio of toxic shares to toxic views is increasing in $p^t$. These predictions are obtained using calibrated parameters from the structural model by matching moment conditions for heterogeneous users.

Figure E.16: Structural Estimates and Validation

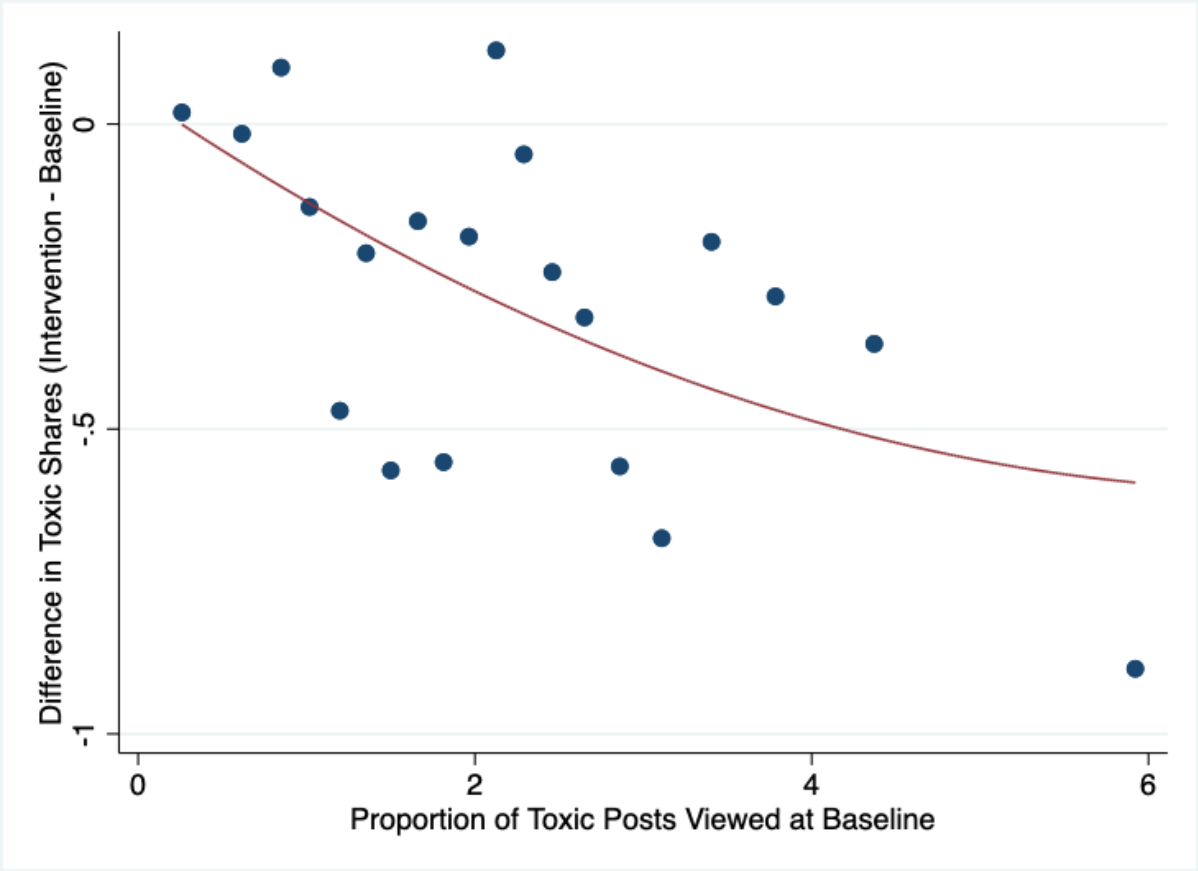

(a) Baseline views and intervention period shares in the treatment group

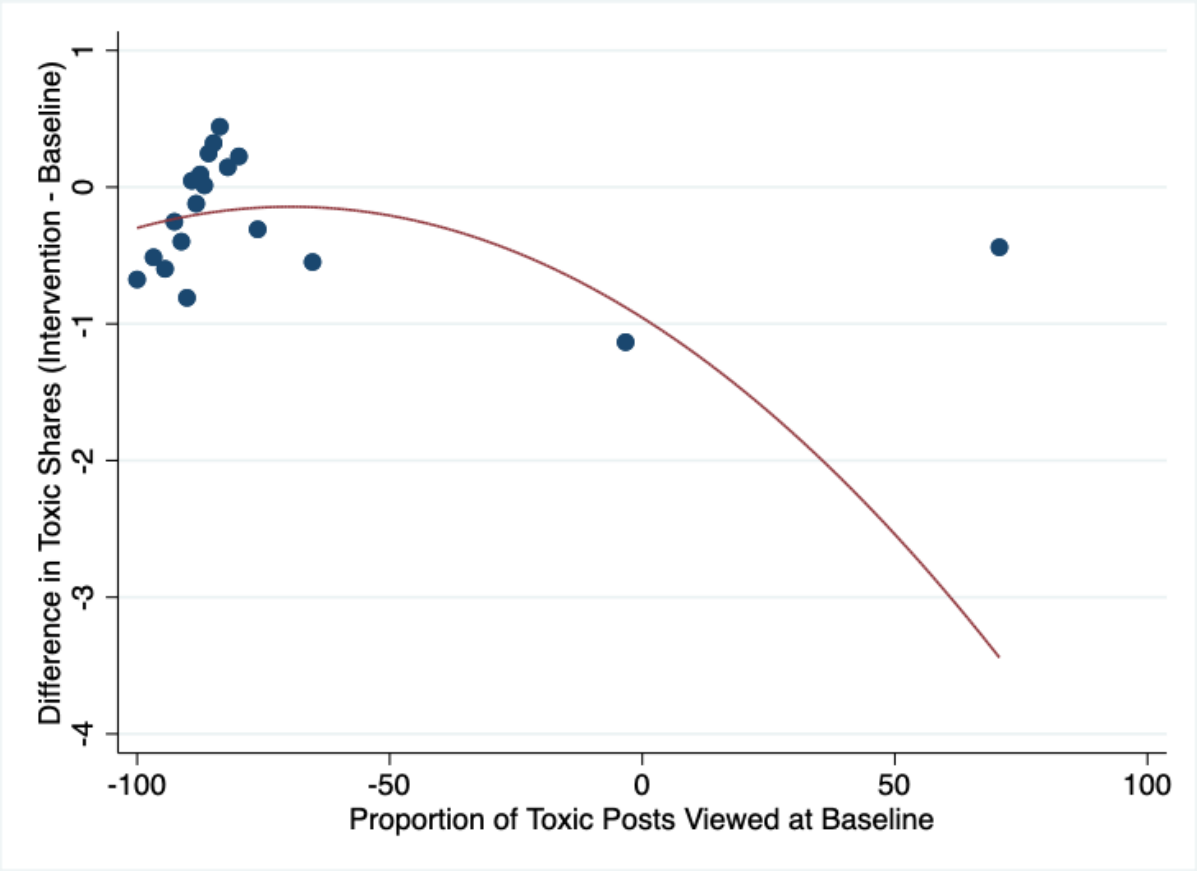

(b) Intervention period views and shares in the treatment group

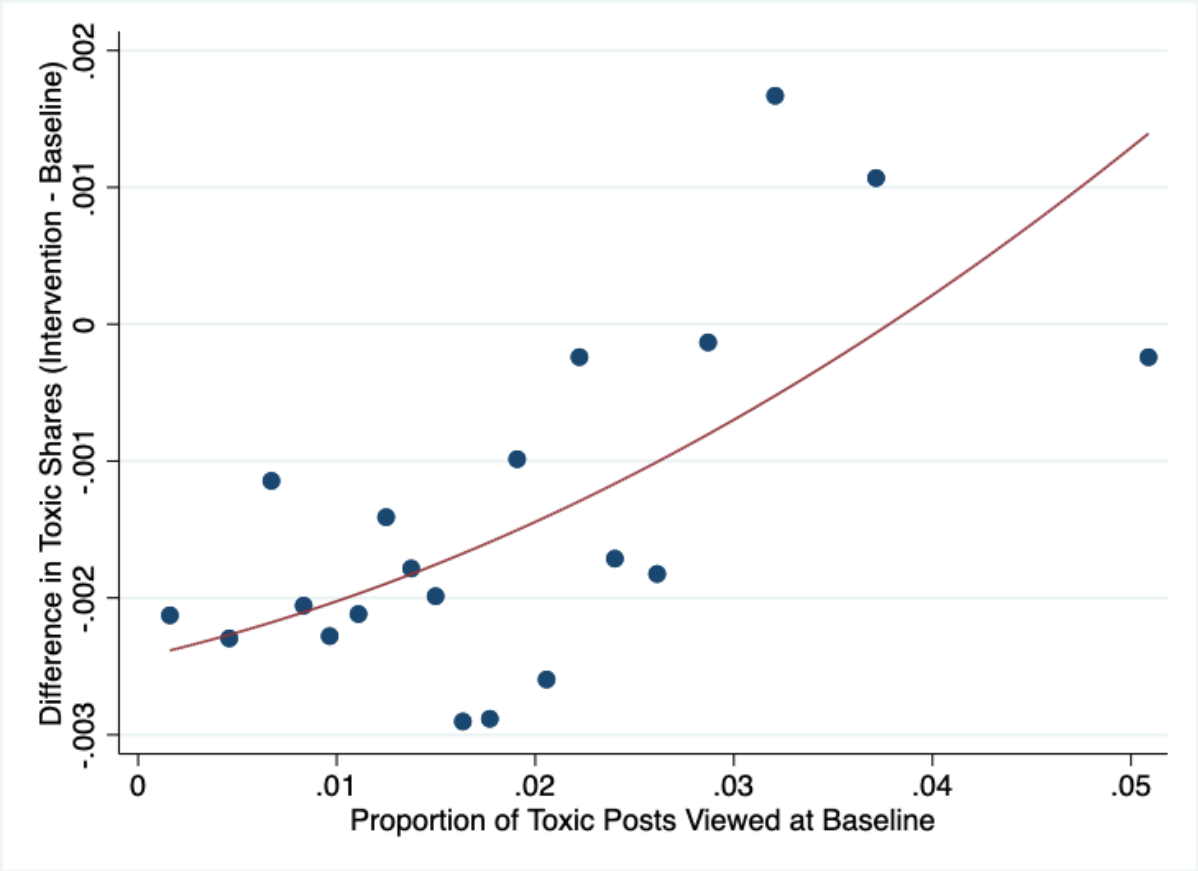

(c) Baseline views and intervention period shares in the control group

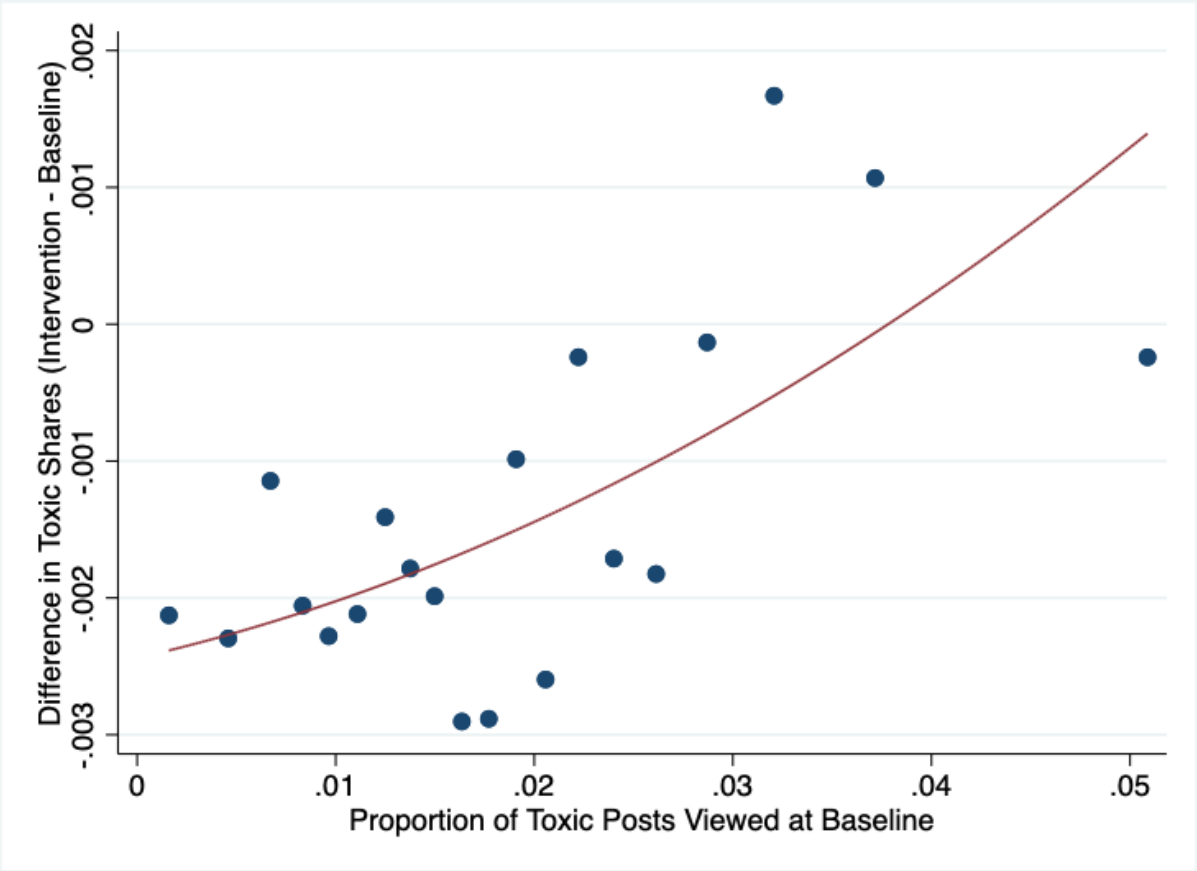

(d) Intervention period views and shares in the control group

Notes: Panel (a) shows that $\theta = -\gamma_1$ is positive, and the relationship between differences in toxic shares and toxic views at baseline approximates a linear one as predicted by the structural model. Panel (b) shows that the relationship between differences in toxic shares (from baseline to intervention period) and in the toxic views during the intervention, produces a relationship that can be positive, as well as distinct from $-\theta$. The intervention period variation in toxic views is concentrated around the mean, by design of the intervention. As a result, this variation is not informative about the rate at which users update their behavior according to the perceived behavior of others or the perceived social norms. Panels (c) and (d) reiterate that the relationship between toxic views and differences in toxic shares, in the control group, do not convey any meaningful information because control users are always in steady state. This means that the said relationship is not estimable in the control group. The binscatter plots constructed using the control group data are distinct from the main plot in panel (a).

Figure E.17: Treatment effects on total number of toxic posts shared for different influence factors, $\theta$

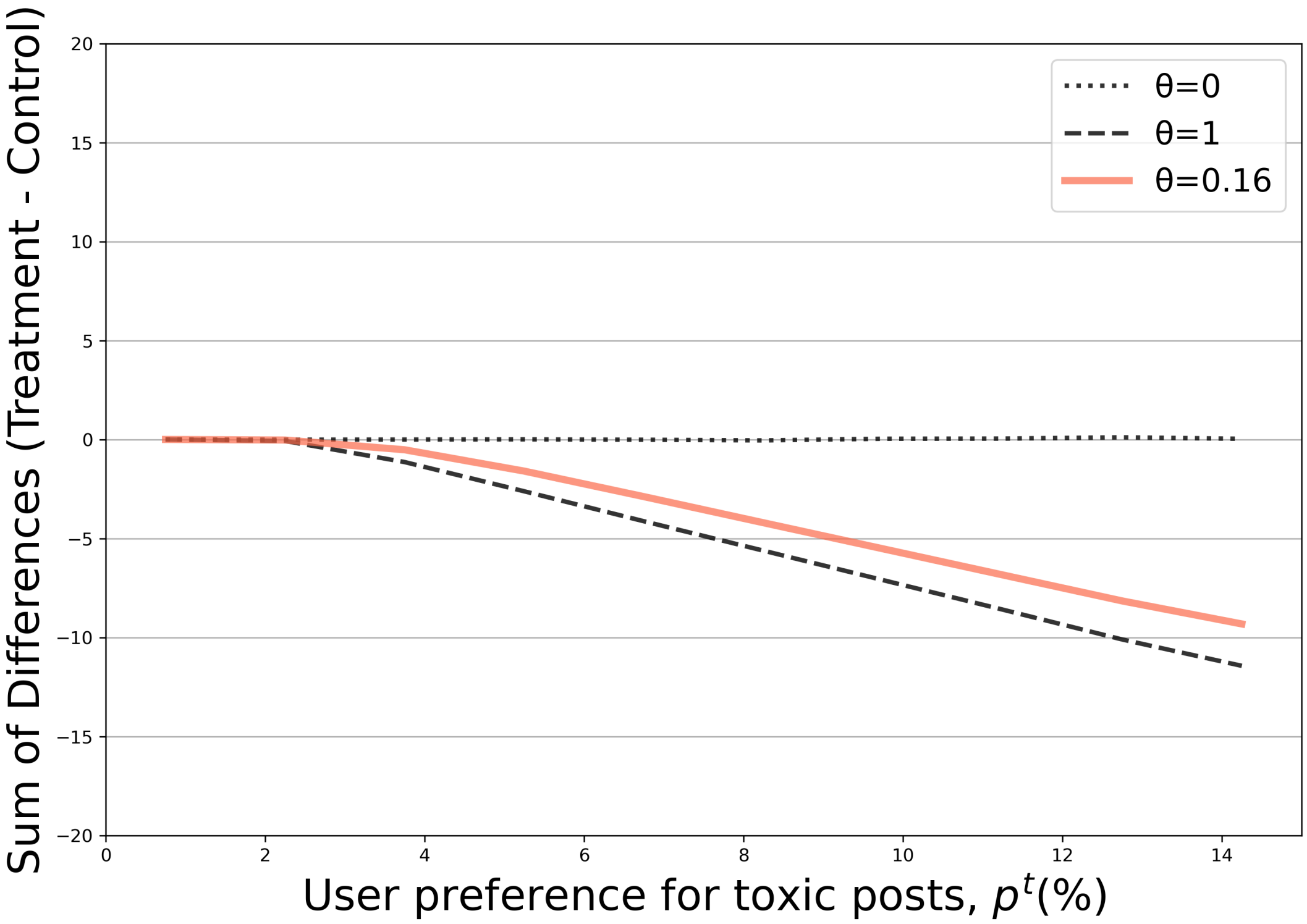

Notes: This figure shows that the simulated treatment effects on number of toxic posts shared is negative for more toxic users, when the rate at which exposure influences behavior is estimated at $\theta = 0.16$. This shows that for the parameter values calibrated using the method of matching moments (See Appendix C.3 for details), the structural model correctly predicts that the treatment effect on the number of toxic posts shared is negative for toxic users. The simulations are based on counterfactual interventions that Q3-Q5 users only. The counterfactual effects are simulated for different influence regimes: $\theta = 0$, when users share content *mechanically*, and $\theta = 1$, when users are thought to be fully *malleable*. The treatment effect on the number of toxic posts shared is constant at zero, in the case of mechanical users (i.e. $\theta = 0$). However, when $\theta = 1$, users with higher proclivity to toxic content share the highest number of toxic posts because they are fully influenced by the content they are exposed to.

Figure E.18: Decomposition of treatment effects, in different updating regimes

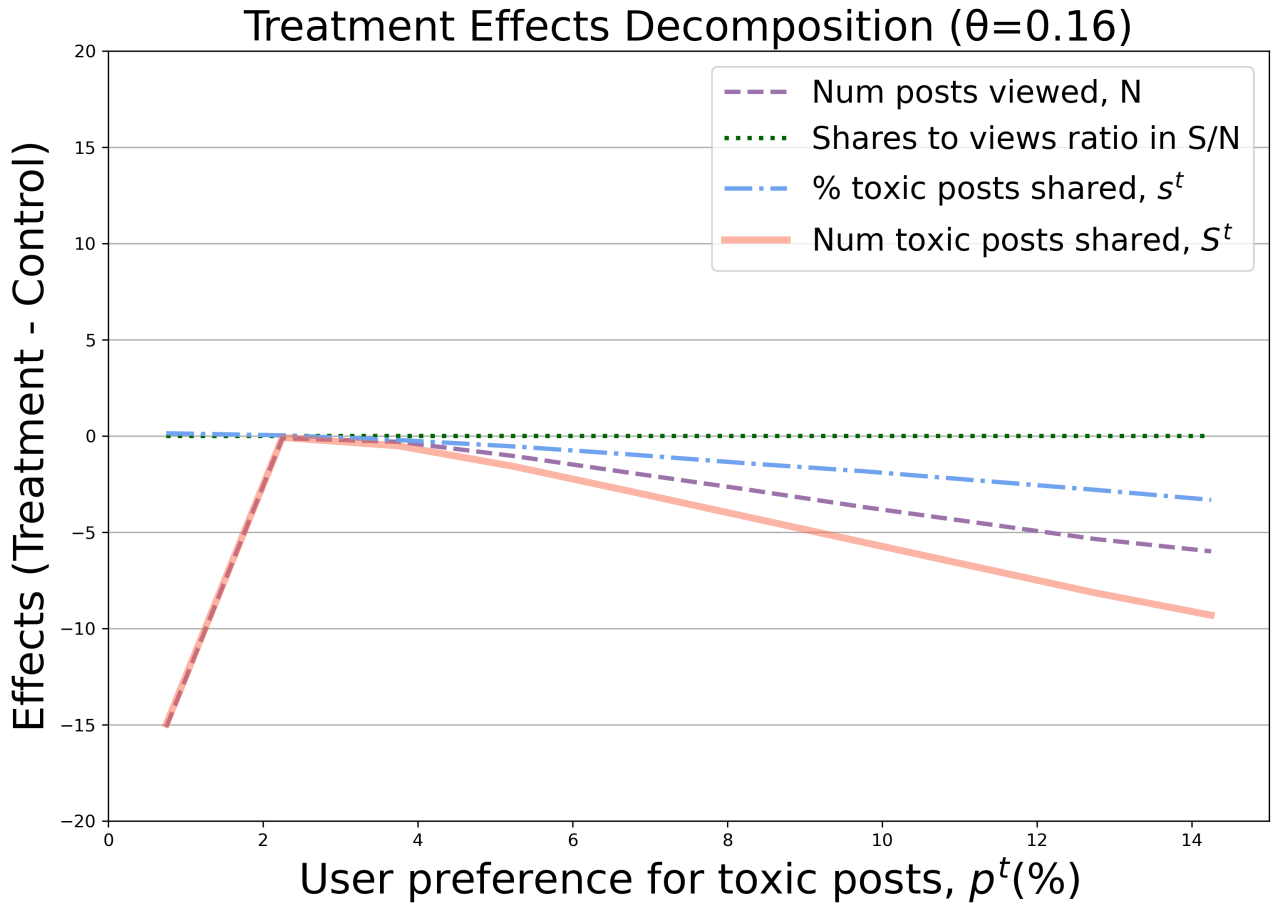

(a) Estimated $\theta = 0.16$

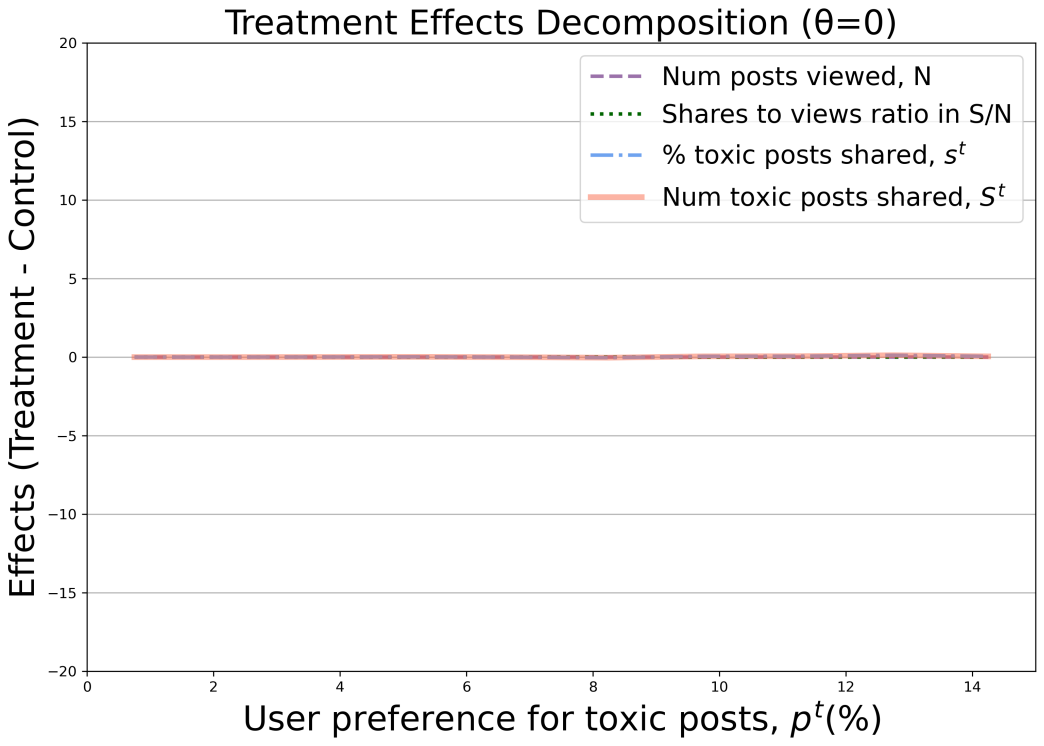

(b) Malleable users, $\theta = 0$

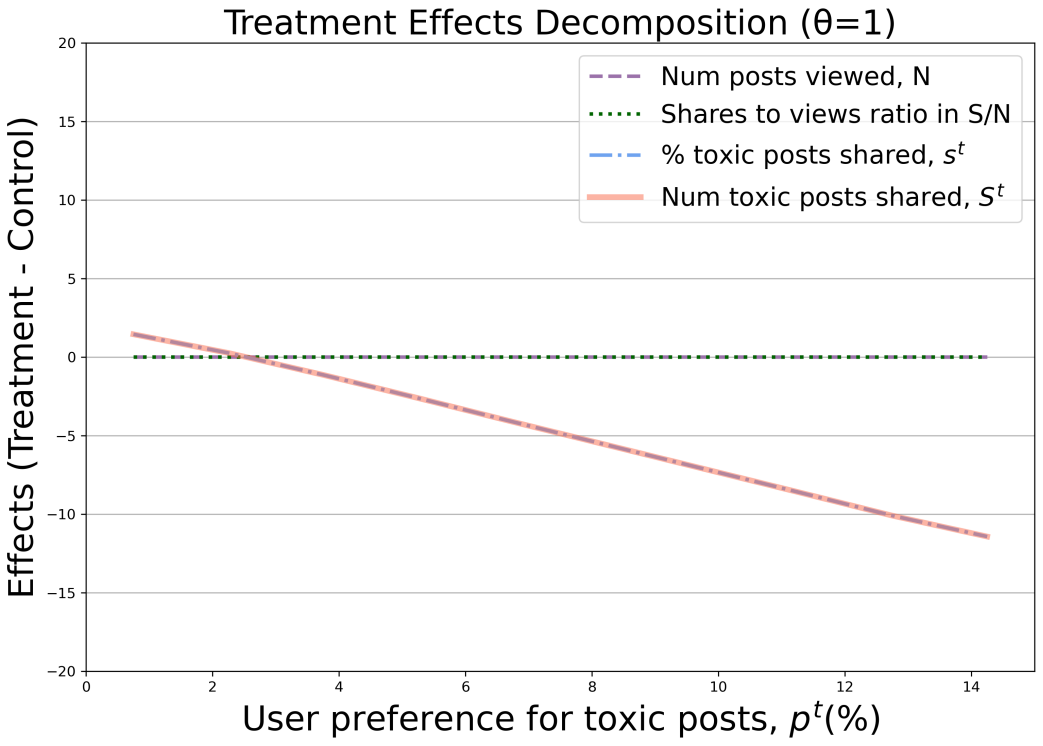

(c) Mechanical users, $\theta = 1$

Notes: This Figure shows that if users were updating their behavior at the same rate $\theta$, the decrease in the number of toxic posts shared is largely driven by the disengagement effect, especially for more toxic users (on the right extreme of the $p^t$ distribution). It decomposes the treatment effect into its two constituent parts, namely, the engagement effect, on number of posts viewed $N$, as well as the shares to views ratio $S/N$, and the influence effect, on the probability of sharing toxic content $s^t$. Panel (a) shows that the reduction in total views (or the disengagement effect) has a higher contribution to the reduction in toxic shares, than the reduction in the probability of sharing toxic content. Panel (b) shows that there is no change in behavior if users were completely mechanical ($\theta = 0$). Panel (c) shows that counterfactual effect is entirely driven by the influence effect if users were completely malleable, and that the number of toxic posts would increase for non-toxic users if $\theta = 1$. The model generated simulated outcomes that are consistent with the data, on the right side of the $p^t$ distribution.

Figure E.19: Suggested mechanisms driving engagement with toxic content

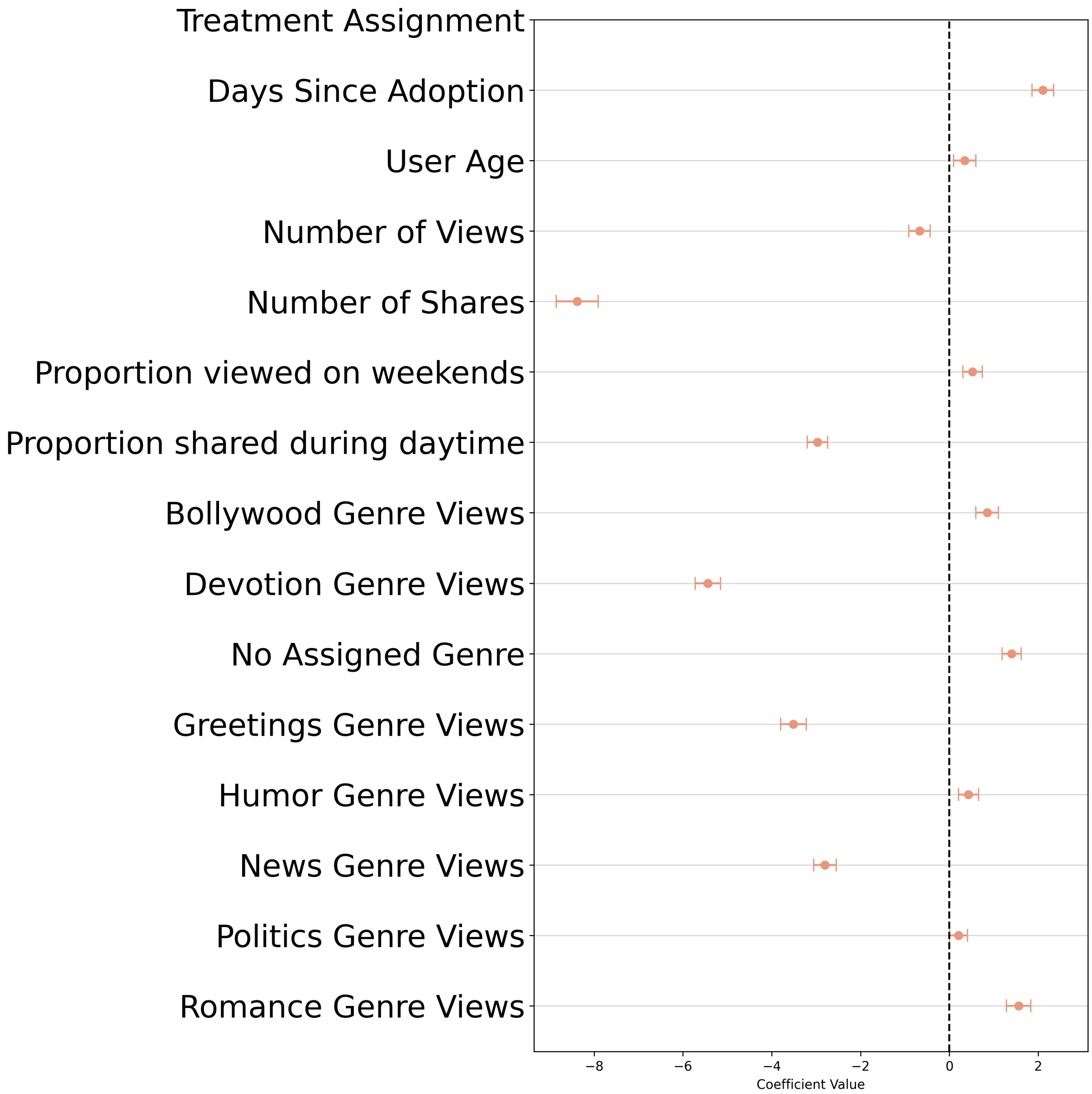

Notes: This Figure shows suggestive evidence on the mechanisms driving the treatment effects, by regressing the main outcome variable (proportion of shares that are toxic) on baseline user characteristics in the sample of treated users. Higher platform activity at baseline is associated with higher toxic sharing during intervention period, for all types of users. Users engaging with greetings content (or "good morning" posts) at baseline are the least likely to share toxic content during the intervention period. All variables were standardized as z-scores to get comparable magnitudes, and robust standard errors were computed.

Figure E.20: Treatment intensity with respect to Greetings content

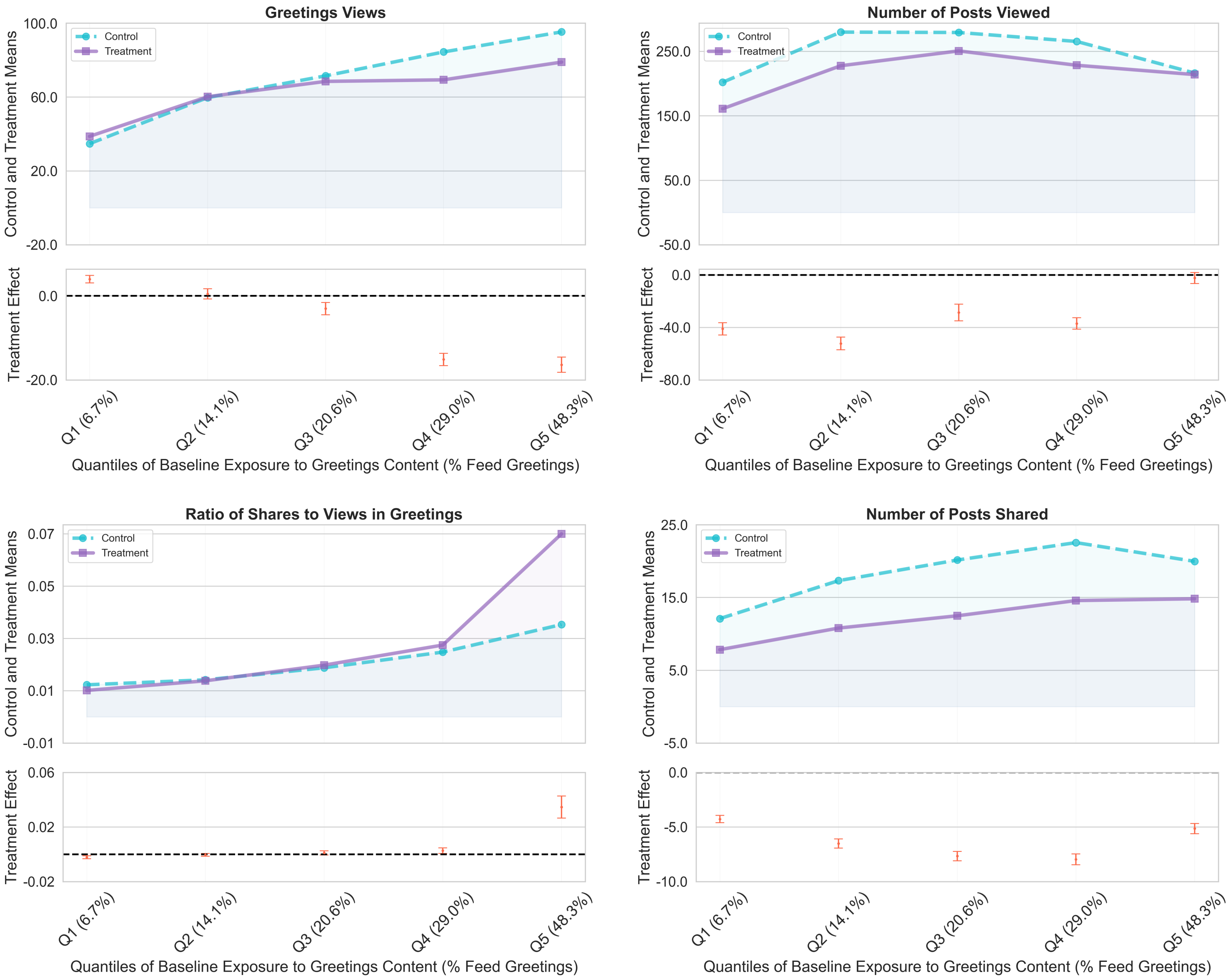

Notes: This Figure shows that the treatment effects on the number of Greetings related posts mimics the treatment intensity with respect to toxic content. The effects on the number of posts viewed in this genre, and the ratio of shares to views in the Greetings category, follow patterns similar to those observed for toxic content. However, the number of posts viewed (of any type) do not follow the same pattern as before. This is consistent with the explanation that users seek out content that they like, especially when Greetings content is not available on other platforms in India. Users are divided into quantiles based on their exposure to Greetings content at baseline, which is a proxy for their proclivity to such content.

# F Supplementary Tables

Table F.1: Regression results for all outcome variables

| | Num Logins | Time Spent (in hours) | Num Posts Viewed |
|---|---|---|---|
| Treatment | -1.270** | -2.531** | -35.497** |
| | (0.042) | (0.584) | (2.208) |
| Control Mean | 21.594** | 7.104** | 246.654** |
| | (0.021) | (0.583) | (1.361) |
| | **Time Spent per Post** | **Num Posts Shared** | **Shares to Views Ratio** |
| Treatment | -0.053** | -6.367** | -0.114** |
| | (0.002) | (0.206) | (0.007) |
| Control Mean | 0.127** | 18.396** | 0.261** |
| | (0.001) | (0.131) | (0.004) |
| | **Prop Activity on Weekends** | **Prop Activity during Daytime** | **Num Searches per Post Viewed** |
| Treatment | 0.010** | -0.035** | 0.016** |
| | (0.001) | (0.002) | (0.001) |
| Control Mean | 0.261** | 0.214** | 0.104** |
| | (0.001) | (0.001) | (0.001) |
| | **Prob Leaving Platform** | **Num Toxic Posts Viewed** | **Perc Toxic Posts Viewed** |
| Treatment | 0.006** | -5.024** | -0.641** |
| | (0.001) | (0.172) | (0.033) |
| Control Mean | 0.030** | 18.806** | 7.416** |
| | (0.000) | (0.129) | (0.018) |
| | **Num Toxic Posts Shared** | **Perc Toxic Posts Shared** | **Tox Share to Tox View Ratio** |
| Treatment | -0.093** | 0.120** | 0.007** |
| | (0.010) | (0.038) | (0.001) |
| Control Mean | 0.474** | 1.547** | 0.040** |
| | (0.006) | (0.018) | (0.001) |
| N | | 231814 | |

Notes: This table shows that the intervention caused disengagement with the platform, by showing negative and significant estimates of treatment effects on total number of posts viewed and shared, number of times users logged on, and total time spent. This table also shows that the average user viewed and shared fewer toxic posts, but the proportion of toxic posts shared increased. The intervention increased users' search costs of using the platform, as measured by the number of searches performed. Robust standard errors in parentheses. $p < 0.05^{*}, p < 0.01^{**}, p < 0.001^{***}$.

Table F.2: Regression results for different thresholds criterion for toxic content

| | (1) Continuous | (2) 0.2 Threshold | (3) 0.3 Threshold | (4) 0.4 Threshold |
|---|---|---|---|---|
| **Average Toxicity in Views** | | | | |
| Treatment | -0.824*** | -0.006*** | -0.00142*** | 0.0003*** |
| | (0.013) | (0.0003) | (0.000210) | (0.0001) |
| Control Mean | 5.930*** | 0.074*** | 0.030*** | 0.003*** |
| | (0.007) | (0.0002) | (0.0001) | (0.00003) |
| **Aberage Toxicity in Shares** | | | | |
| Treatment | 0.112*** | 0.001*** | 0.001*** | |
| | (0.015) | (0.0003) | | (0.0002) |
| Control Mean | 1.289*** | 0.015*** | 0.007*** | 0.0001** |
| | (0.008) | (0.0002) | (0.0001) | (0.0001) |
| $N$ | 231814 | 231814 | 231814 | 231814 |

Notes: This Table shows that the treatment effect on the proportion of toxic posts viewed and shared, when toxicity is defined using the continuous toxicity score (1), and different thresholds for the binary toxicity score (2-4). All the results are consistent with the main results on toxic exposure in Table F.1. Standard errors are robust at user level. $p < 0.05^{*}, p < 0.01^{**}, p < 0.001^{***}$.

Table F.3: User characteristics correlated with the probability of leaving the platform

| Variable | Coefficient | Interaction Coefficient |
|---|---|---|
| Treatment Effect | 0.014 | N/A |
| | (0.008096) | N/A |
| Number of Views (Baseline) | -0.000*** | 0.000 |
| | (0.000) | (0.000) |
| Number of Shares (Baseline) | -0.000 | -0.000 |
| | (0.000) | (0.000) |
| Toxic Shares (Baseline) | 0.000* | -0.000 |
| | (0.000) | (0.000) |
| Toxic Views (Baseline) | -0.000*** | 0.000 |
| | (0.000) | (0.000) |
| Male Gender | -0.004*** | -0.002 |
| | (0.001) | (0.002) |
| Days since account created | 0.000*** | -0.000 |
| | (0.000) | (0.000) |
| User Age | -0.000 | -0.000 |
| | (0.000) | (0.000) |
| Proportion content viewed on weekends | -0.002 | 0.003 |
| | (0.002) | (0.004) |
| Proportion content shared during daytime | 0.002 | -0.001 |
| | (0.001) | (0.003) |
| Share of views in Bollywood Genre | 0.039*** | 0.008 |
| | (0.005) | (0.011) |
| Share of views in Devotion Genre | 0.015*** | 0.006 |
| | (0.004) | (0.009) |
| No Assigned Genre | 0.049*** | 0.037 |
| | (0.011) | (0.023) |
| Share of views in Greetings Genre | 0.028*** | 0.001 |
| | (0.004) | (0.008) |
| Share of views in Humor Genre | 0.054*** | -0.018 |
| | (0.008) | (0.015) |
| Share of views in News Genre | 0.011 | -0.016 |
| | (0.007) | (0.013) |
| Share of views in Politics Genre | -0.014 | -0.011 |
| | (0.032) | (0.070) |
| Share of views in Romance Genre | 0.054*** | -0.002 |
| | (0.005) | (0.010) |

Notes: This Table shows that, conditional on observable user characteristics, treatment assignment is not correlated with the probability of leaving the platform. This also shows that the treatment does not differentially impact the probability of leaving the platform, for given observable user characteristics. This means that the treated leavers are not systematically different from the control leavers. These results are obtained by estimating the regression equation $\mathbb{1}_i(\text{left platform} = \text{ yes}) = \beta_0 + \beta_1 D_i + \sum_c \beta_c \mathbb{1}_i(\text{user characteristic} = c) + \sum_c \beta_{1c} D_i \mathbb{1}_i(\text{user characteristic} = c) + \varepsilon_i$, where $\mathbb{1}_i(\text{left platform} = \text{ yes})$ is an indicator taking value 1, when user $i$ leaves the platform. Column (1) reports estimated $\beta_c$'s, while column (2) reports estimated $\beta_{1c}$'s. Standard errors are robust at user level. $p < 0.05^{*}, p < 0.01^{**}, p < 0.001^{***}$.

Table F.4: Structural estimates using OLS regressions

| | (1)<br>Proportion Toxic Posts Shared<br>(Intervention - Baseline) |
|---|---|
| Proportion Toxic<br>Posts Viewed | -0.104$^{**}$<br>(0.037) |
| $N$ | 63041 |

Notes: This table shows that the structural estimates of $\theta$ obtained using an OLS regressions are biased downwards. Dependent variable is differences in differences between probability of sharing toxic and non-toxic content, between intervention period and baseline, for treated users only. The explanatory variables are constructed by averaging differences between proportion of toxic and non-toxic posts viewed by treated users. Robust standard errors in parentheses. $p < 0.05^{*}, p < 0.01^{**}, p < 0.001^{***}$.

# G Attrition

Table G.1 reports the estimated treatment effects of the intervention on various outcome variables. Throughout the paper I have maintained that the relevant value for users who stop coming to the platform, or leave it entirely, is zero. This is true for outcomes including the number of posts shared/ viewed, the number of toxic posts shared/ viewed, and the proportion of toxic posts shared/ viewed. Similarly, the time spent on the platform is zero for users who leave the platform.

Table G.1: Heterogeneous Treatment Effects

| Quantile | Number of Logins | | Number of Shares | | Number of Views | |
|---|---|---|---|---|---|---|
| | Control Mean | Effect (SE) | Control Mean | Effect (SE) | Control Mean | Effect (SE) |
| Q1 | 20.929 | -1.225 (0.114) | 22.838 | -4.752 (0.688) | 226.650 | 1.063 (5.82) |
| Q2 | 21.820 | -1.472 (0.113) | 24.629 | -8.165 (0.685) | 272.036 | -15.298 (5.819) |
| Q3 | 22.101 | -1.521 (0.112) | 22.386 | -7.435 (0.608) | 304.601 | -52.926 (6.521) |
| Q4 | 22.152 | -1.483 (0.111) | 19.508 | -6.994 (0.522) | 325.085 | -45.855 (7.785) |
| Q5 | 22.376 | -1.281 (0.111) | 14.311 | -5.456 (0.393) | 328.739 | -76.34 (6.155) |

| Quantile | Time Spent (in hours) | | Num of Toxic Shares | | % Toxic Shares | |
|---|---|---|---|---|---|---|
| | Control Mean | Effect (SE) | Control Mean | Effect (SE) | Control Mean | Effect (SE) |
| Q1 | 4.647 | -1.151 (0.089) | 0.329 | 0.009 (0.017) | 0.992 | 0.255 (0.081) |
| Q2 | 6.441 | -1.972 (0.564) | 0.462 | -0.042 (0.022) | 1.248 | 0.108 (0.081) |
| Q3 | 9.206 | -4.264 (2.592) | 0.584 | -0.092 (0.027) | 1.533 | 0.168 (0.092) |
| Q4 | 9.759 | -4.446 (2.613) | 0.695 | -0.186 (0.032) | 1.951 | 0.058 (0.104) |
| Q5 | 7.360 | -2.051 (0.137) | 0.722 | -0.246 (0.034) | 2.707 | -0.278 (0.126) |

| Quantile | Ratio of Toxic Share to View | | Num of Toxic Views | | % Toxic Views | |
|---|---|---|---|---|---|---|
| | Control Mean | Effect (SE) | Control Mean | Effect (SE) | Control Mean | Effect (SE) |
| Q1 | 0.043 | 0.003 (0.003) | 9.579 | 3.352 (0.299) | 4.998 | 1.105 (0.081) |
| Q2 | 0.042 | 0.005 (0.004) | 15.038 | 0.729 (0.36) | 6.000 | 0.38 (0.078) |
| Q3 | 0.041 | 0.009 (0.004) | 21.235 | -5.011 (0.483) | 7.043 | -0.312 (0.083) |
| Q4 | 0.039 | 0.004 (0.003) | 28.047 | -9.263 (0.608) | 8.319 | -1.349 (0.078) |
| Q5 | 0.034 | 0.016 (0.005) | 37.581 | -18.573 (0.641) | 10.775 | -2.989 (0.087) |

Notes: The table reports the estimated treatment effects of the intervention on the outcome variable, by the amount of toxicity user was exposed to at baseline, which is a proxy for their type. The treatment effect is estimated using a linear regression model, with the outcome variable as the dependent variable, and the treatment indicator as the independent variable, both aggregated at the user level. The treatment indicator is a dummy variable that takes the value of 1 if the user is treated, and 0 otherwise. The table also reports robust standard errors of the estimated treatment effects.

This may raise concerns that selective attrition could bias the estimated treatment effects, if the treated users who leave the platform are systematically different from those who stay. To test that treated leavers are not systematically different from treated stayers, I first estimated the treatment effect on the probability of leaving the platform. Although I find differential attrition by treatment status, controlling for various observable characteristics corrects for this bias. This means that upon controlling for user attributes that are correlated with the probability of leaving among the treated, there is no selection in the probability of leaving among treated users. This is seen in Table F.3.

Table G.2: Lee Bounds for Estimated Treatment Effects

| Outcome | Quantile | Treatment Effect | Standard Error | Lower Bound | Upper Bound |
|---|---|---|---|---|---|
| Num of Posts Viewed | Q1 | 1.063 | 5.820 | -1092.278 | 2.627 |
| | Q2 | -15.298 | 5.819 | -1412.702 | -14.473 |
| | Q3 | -52.926 | 6.521 | -1729.319 | -54.646 |
| | Q4 | -45.855 | 7.785 | -1844.840 | -46.813 |
| | Q5 | -76.340 | 6.155 | -1904.051 | -79.718 |
| Num of Toxic Posts Viewed | Q1 | 3.352 | 0.299 | -45.471 | 3.686 |
| | Q2 | 0.729 | 0.360 | -81.304 | 0.898 |
| | Q3 | -5.011 | 0.483 | -132.070 | -5.222 |
| | Q4 | -9.263 | 0.608 | -178.453 | -9.727 |
| | Q5 | -18.573 | 0.641 | -247.830 | -19.658 |
| Num of Posts Shared | Q1 | -4.752 | 0.688 | -163.773 | -4.978 |
| | Q2 | -8.165 | 0.685 | -190.771 | -8.591 |
| | Q3 | -7.435 | 0.608 | -176.153 | -7.815 |
| | Q4 | -6.994 | 0.522 | -158.755 | -7.358 |
| | Q5 | -5.456 | 0.393 | -123.217 | -5.753 |
| % Toxic Posts Viewed | Q1 | 1.105 | 0.081 | -15.512 | 1.228 |
| | Q2 | 0.380 | 0.078 | -15.785 | 0.454 |
| | Q3 | -0.312 | 0.083 | -16.999 | -0.283 |
| | Q4 | -1.349 | 0.078 | -17.639 | -1.384 |
| | Q5 | -2.989 | 0.087 | -21.795 | -3.132 |
| Num of Toxic Posts Shared | Q1 | 0.009 | 0.017 | -3.398 | 0.012 |
| | Q2 | -0.042 | 0.022 | -4.999 | -0.042 |
| | Q3 | -0.092 | 0.027 | -6.509 | -0.095 |
| | Q4 | -0.186 | 0.032 | -7.965 | -0.194 |
| | Q5 | -0.246 | 0.034 | -8.568 | -0.259 |
| % Toxic Posts Shared | Q1 | 0.255 | 0.081 | -11.275 | 0.283 |
| | Q2 | 0.108 | 0.081 | -14.687 | 0.126 |
| | Q3 | 0.168 | 0.092 | -17.297 | 0.191 |
| | Q4 | 0.058 | 0.104 | -21.491 | 0.078 |
| | Q5 | -0.278 | 0.126 | -30.113 | -0.280 |
| Ratio of Toxic Shares to Views | Q1 | 0.003 | 0.003 | -0.497 | 0.004 |
| | Q2 | 0.005 | 0.004 | -0.494 | 0.006 |
| | Q3 | 0.009 | 0.004 | -0.489 | 0.010 |
| | Q4 | 0.004 | 0.003 | -0.457 | 0.005 |
| | Q5 | 0.016 | 0.005 | -0.398 | 0.017 |

Notes: The table reports the Lee bounds for the estimated treatment effects of the intervention on the main outcome variables. The Lee bounds are constructed using the rate of attrition, which is computed using the inverse probability of logging on to the platform. The table shows that the Lee bounds for the treatment effects are tightly estimated.

However, there still may be concerns that the estimated treatment effects are biased due to selective attrition, if the treated users who leave the platform are systematically different from those who stay, on unobservable characteristics. To address this concern, I construct Lee bounds for the estimated treatment effects, with respect to all the outcome variables (Lee, 2009). The rate of attrition is computed using the inverse probability of logging on to the platform, and is used to construct the bounds. Table G.2 shows that the Lee bounds for negative treatment effects are tightly estimated.

# H Text Analysis

I begin describing the text data by translating from the original Hindi, and summarizing the most common words in the political posts in Figure H.1. This summary measure is based on more than 20 million posts that were viewed and shared by users in the baseline and intervention periods. The text analysis currently excludes a dozen other Indian regional languages in which users can consume content.

Figure H.1: Word clouds depicting words associated with highly toxic posts

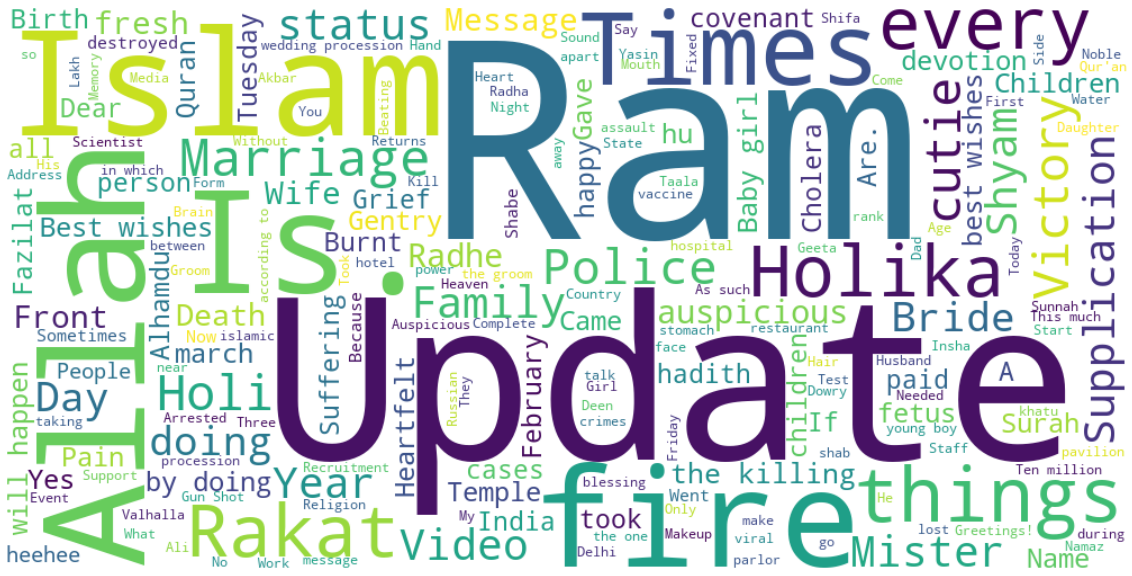

(a) High Toxicity

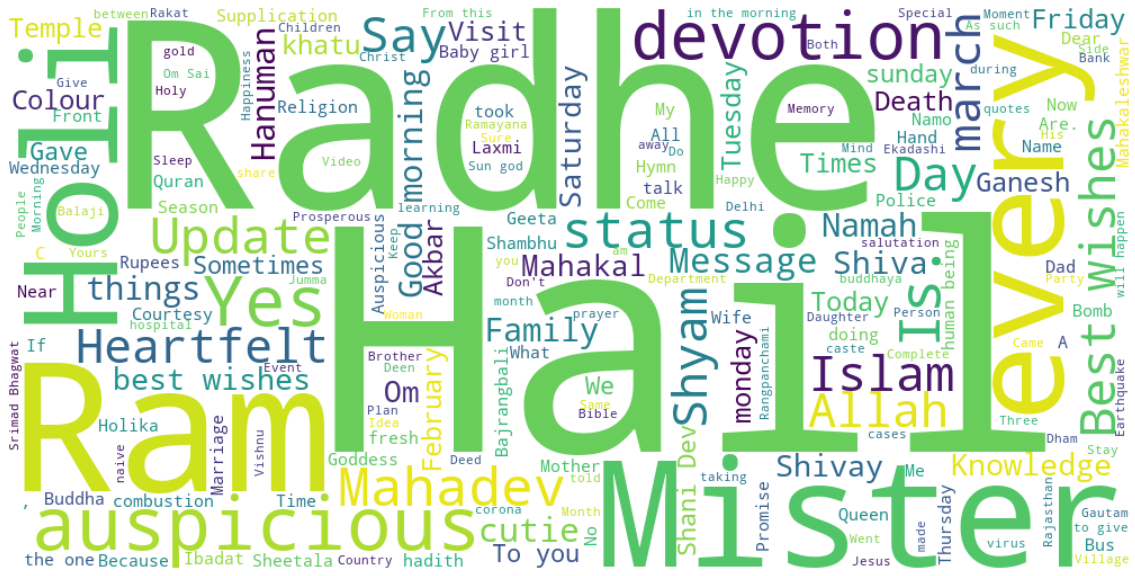

(b) Low Toxicity

Notes: This Figure shows word clouds constructed using the TF-IDF vectorizer, on posts classified into high and low toxicity categories respectively. Cut-off to classify posts into high and low toxicity categories is 0.2, based on the toxicity scores provided by Perspective API. The figure demonstrates overlap in words pertaining to religion in both categories, for example 'Islam' and the Hindu mythological god-king 'Ram,' who is also central to Hindu nation building agenda of the current ruling government. This highlights the need for contextual embeddings to characterize the text data.

Figure H.1 shows that the most common word in posts labelled as toxic is 'Ram,' which is a reference to legendary Hindu deity, who is said to have blessed the Hindu Nationalist project.[23] The Hindu nationalist project is a political ideology that is associated with the ruling party in India, that has been accused of promoting anti-minority sentiments, and even promoted outright calls for ethnic cleansing in extreme instances (Jaffrelot, 2021).

However, I find a significant overlap in the most common words across posts that were classified as toxic or not. For instance, the words 'Ram,' 'Islam,' 'Allah' are common in both toxic and non-toxic posts. This demonstrates that analyzing tokenized vector of words may lead to misleading conclusion. The text analysis must include sufficient information about the context in which the words are used. Therefore, I use semi-supervised Machine Learning methods that take contextual embeddings into account, while achieving a narrower objective: classifying posts as toxic or not. I provide examples to illustrate the toxicity classification algorithm in Table H.1.

[23]For instance, see Kalra (2021) for details on a coordinated campaign carried out in the name of Ram, that was aimed at inciting violence against Muslims in different parts of India.

Table H.1: Examples of text data (English translations) with toxicity scores

| Text | Toxicity Score | Toxicity Classification |
|---|---|---|
| Even p*s*i*s are opposing Modi these days, whose father has taken Rs. 12,000 from Modi government and has arranged for shitting and peeing in his house! | 0.643 | Toxic |
| Mohammed Shamim's disgusting act ! Lakhs of pilgrims kept trusting Mohammed Shamim... Mohammed Shamim used to make tea from urine water and sell it. Mohammed Shamim used to run a shop in Kerala's Sabarimala temple premises. | 0.479 | Toxic |
| Break those rocks Jai Shri Ram which are standing in the path of religion and shoot those criminals who have dirty intentions on the women of our country | 0.399 | Toxic |
| LIVE LATEST UPDATES 0.01% population wants 'Khalistan.' 18% want 'Ghazwa-e-Hind' and 80% want cheap onions and tomatoes. It is bitter but true. | 0.327 | Toxic |
| People travelling on "Bharat Jodo" route are now facing problem with the name "Bharat" instead of India. | 0.172 | Non-Toxic |
| 00 Death does not occur only when the soul leaves the body. He is also dead who remains silent even after seeing his religion and culture being attacked. 00 | 0.174 | Non-Toxic |
| Giqa Bihar wire procession of thieves (temple thief) (coal thief) (fodder four) (land thief) | 0.361 | Toxic |
| Don't make us jokers, when Christians being 2% do not celebrate Ramnavami, why do we Hindus being 80% celebrate Christmas, joke our children on 25th December, Jai Satya Sanatan | 0.361 | Toxic |
| Why has it been proved that sycophants are the biggest problem? Who is the master of sycophants? He is the biggest problem. | 0.061 | Non-Toxic |
| Bhajanlal Sharma will be the new Chief Minister of Rajasthan. | 0.008 | Non-Toxic |
| Why have Akhilesh, Rahul, Kejriwal, Mamta, Lalu, Sharad Pawar not spoken a single word against Islamic terrorists till date? India asks | 0.285 | Toxic |

Notes: The table shows examples of text data in English, with toxicity scores provided by the Perspective API. The toxicity score is a continuous measure that ranges from 0 to 1, with 0 indicating healthy contributions and 1 indicating very toxic content. The Perspective API uses a mix of supervised and semi-supervised machine learning methods, and is sensitive to context while assigning toxicity scores. The Perspective API is widely used in academic research and by publishers to identify and filter out abusive comments.